\documentclass[10pt]{article}
\pdfoutput=1
\usepackage{mathStyle}
\usepackage{amsmath,amssymb,amsthm,latexsym,bbm,calc,wasysym,mathtools,empheq, bbold, yfonts}
\usepackage[english]{babel}
\usepackage{relsize}
\usepackage{graphicx,color}
\usepackage[dvipsnames]{xcolor}
\usepackage[makeroom]{cancel}
\usepackage{xspace}
\usepackage{pifont,dsfont}
\usepackage{marvosym}
\usepackage{slashed}
\usepackage{subfigure}
\usepackage{sidecap}
\usepackage{stmaryrd}
\usepackage{empheq}

\usepackage[export]{adjustbox}
\mathtoolsset{showonlyrefs=false}
\usepackage{placeins}
\usepackage{enumitem}
\usepackage{verbatim}
\usepackage{amsthm}

\usepackage{tikz-cd}
\usepackage{tikz}
\usetikzlibrary{shapes.geometric}
\usetikzlibrary{arrows,matrix,calc,scopes,decorations.markings,intersections}
\usetikzlibrary{arrows.meta}
\usetikzlibrary{decorations.pathreplacing,angles,quotes}
\usetikzlibrary{external}

\newtheoremstyle{mydef}%
{0.9em} 
{0.7em}
{\hangindent=2em}
{1.8em}
{\scshape}
{.}
{.5em}
{}%
\theoremstyle{mydef}

\numberwithin{definition}{section}

\theoremstyle{mydef}

\numberwithin{lemma}{section}

\theoremstyle{mydef}

\numberwithin{theorem}{section}

\theoremstyle{mydef}

\numberwithin{example}{section}

\theoremstyle{mydef}

\numberwithin{convention}{section}

\theoremstyle{mydef}

\numberwithin{property}{section}

\newcommand{\nn}{\nonumber}
\newcommand{\snum}[1]{\text{\small $#1$}}
\newcommand{\act}{\triangleright}
\newcommand{\la}{\langle}
\newcommand{\ra}{\rangle}
\newcommand{\q}{\quad}
\newcommand{\sss}{\scriptstyle}
\newcommand{\bul}{{\sss \bullet}}
\newcommand{\mG}{\mathcal{G}}
\newcommand{\mC}{\mathcal{C}}
\newcommand{\mE}{\mathcal{E}}

\newcommand{\gx}{g_{\rm x}}
\newcommand{\gy}{g_{\rm y}}
\newcommand{\gz}{g_{\rm z}}
\newcommand{\hx}{h_{\hat{\rm x}}}
\newcommand{\hy}{h_{\hat{\rm y}}}
\newcommand{\hz}{h_{\hat{\rm z}}}
\newcommand{\tgx}{\tilde{g}_{\rm x}}
\newcommand{\tgy}{\tilde{g}_{\rm y}}
\newcommand{\tgz}{\tilde{g}_{\rm z}}
\newcommand{\bgy}{\bar{g}_{\rm y}}
\newcommand{\bgz}{\bar{g}_{\rm z}}
\newcommand{\thx}{\tilde{h}_{\hat{\rm x}}}
\newcommand{\bhx}{\bar{h}_{\hat{\rm x}}}
\newcommand{\thy}{\tilde{h}_{\hat{\rm y}}}
\newcommand{\thz}{\tilde{h}_{\hat{\rm z}}}
\newcommand{\pxi}{p_{{\rm x},i}}
\newcommand{\pxj}{p_{{\rm x},j}}

\newcommand{\pxk}{p_{{\rm x},k}}
\newcommand{\cyi}{c_{{\rm y},i}}
\newcommand{\czi}{c_{{\rm z},i}}
\newcommand{\dxi}{d_{\hat{\rm x},i}}
\newcommand{\cyj}{c_{{\rm y},j}}
\newcommand{\czj}{c_{{\rm z},j}}
\newcommand{\dxj}{d_{\hat{\rm x},j}}
\newcommand{\exi}{e_{{\rm x},i}}
\newcommand{\fyi}{f_{\hat{\rm y},i}}
\newcommand{\fzi}{f_{\hat{\rm z},i}}
\newcommand{\exOne}{e_{{\rm x},1}}
\newcommand{\fyOne}{f_{\hat{\rm y},1}}
\newcommand{\fzOne}{f_{\hat{\rm z},1}}
\newcommand{\texOne}{\tilde{e}_{{\rm x},1}}
\newcommand{\tfyOne}{\tilde{f}_{\hat{\rm y},1}}
\newcommand{\tfzOne}{\tilde{f}_{\hat{\rm z},1}}

\newcommand{\qyi}{q_{\hat{\rm y},i}}
\newcommand{\qyj}{q_{\hat{\rm y},j}}

\newcommand{\qzi}{q_{\hat{\rm z},i}}
\newcommand{\qzj}{q_{\hat{\rm z},j}}

\newcommand{\qzk}{q_{\hat{\rm z},k}}
\newcommand{\equivSpe}[3]{\raisebox{0.35em}{$\begin{array}{c} \text{\tiny (#1)} \\[-1.3em] \sim \\[-1.6em] {\sss #2,#3} \end{array}$}}

\newcommand{\ket}[1]{| #1 \ra}

\newcommand{\CC}{\mathbb{C}}
\newcommand{\summand}{\sum_{\substack{\gy,\gz\in G\\[-0.2em] \hx\in H\,|\,\gz=\gz^{\gy;\hx} \\[0.05em] \mE_{\gy,\gz} \in \mathfrak{B}_{\gy,\gz}}} \!\!\!\!}
\newcommand{\summandh}{\sum_{\substack{ \gy',\gz'\in G  \\[-0.3em] \hx'\in H \, |\, \gz'=\gz'^{\gy';\hx'}\\ \mE'_{\gy',\gz'} \in \mathfrak{B}_{\gy',\gz'}}} \!\!\!\!}
\newcommand{\g}{\tubeSavg{\hx}{\gy}{\gz}{\mE_{\gy,\gz}}{0.75}{1.25}}
\newcommand{\h}{\tubeSavg{\hx'}{\gy'}{\gz'}{\mE'_{\gy',\gz'}}{0.75}{1.25}}

\makeatletter
\newcommand{\xRrightarrow}[2][]{\ext@arrow 0359\Rrightarrowfill@{#1}{#2}}
\newcommand{\Rrightarrowfill@}{\arrowfill@\equiv\equiv\Rrightarrow}
\newcommand{\xLleftarrow}[2][]{\ext@arrow 3095\Lleftarrowfill@{#1}{#2}}
\newcommand{\Lleftarrowfill@}{\arrowfill@\Lleftarrow\equiv\equiv}
\makeatother

\hypersetup{%
	bookmarks=true,         
	bookmarksopen=true,
	unicode=true,           
	pdftoolbar=true,        
	pdfmenubar=true,        
	pdffitwindow=false,     
	pdfstartview={FitH},    
	pdftitle={Excitations in strict 2-group higher gauge models of topological phases},
	pdfauthor={},
	pdfsubject={},          
	pdfcreator={},          
	pdfproducer={pdfLaTeX}, 
	pdfkeywords={topological} {higher gauge theory} {excitations} {anyons},
	pdfnewwindow=true,      
	colorlinks,
	citecolor=BrickRed,
	filecolor=BrickRed,
	linkcolor=BlueViolet,
	urlcolor=BrickRed,
	linktocpage=true
}

\tikzset{->1-/.style={decoration={
			markings,
			mark=at position #1 with {\arrow{Triangle[fill=black, width=3.5pt, length=3.5pt]}}},postaction={decorate}}}
\tikzset{->2-/.style={decoration={
					markings,
					mark=at position #1 with {\arrow{Triangle[fill=white, width=5pt, length=5pt]}}},postaction={decorate}}}

\tikzset{
	every picture/.append style={
		line join=round,
		line cap=round,
		line width = 0.4pt
	}
}

\tikzstyle{dot}=[dotted]
\tikzstyle{op8}=[opacity=0.8]
\tikzstyle{op7}=[opacity=0.7]
\tikzstyle{op6}=[opacity=0.6]
\tikzstyle{op5}=[opacity=0.5]
\tikzstyle{op4}=[opacity=0.4]
\tikzstyle{op3}=[opacity=0.3]
\tikzstyle{op2}=[opacity=0.2]
\tikzstyle{op1}=[opacity=0.1]
\tikzstyle{sha}=[shorten >= 0.3em]
\tikzstyle{shb}=[shorten <= 0.3em]
\tikzstyle{shab}=[shorten >= 0.3em, shorten <= 0.3em]

\def\centerarc[#1](#2)(#3:#4:#5)
{ \draw[#1] ($(#2)+({#5*cos(#3)},{#5*sin(#3)})$) arc (#3:#4:#5); }

\newcommand{\plaq}[7]{
	\begin{tikzpicture}[scale=1,baseline=-0.3em]
	\matrix[matrix of math nodes,row sep =2.8em,column sep=3em, ampersand replacement=\&] (m) {
		{\sss 0} \& {\sss 2} \\
		{\sss 1} \& {\sss 3}
		\\
	};
	\path
	(m-1-1) edge[->] node[pos=0.5, above] {${\sss #2}$} (m-1-2)
	(m-2-1) edge[->] node[pos=0.5, below] {${\sss #3}$} (m-2-2)
	(m-2-1) edge[<-] node[pos=0.5, left] {${\sss #1}$} (m-1-1)
	(m-2-2) edge[<-] node[pos=0.5, right] {${\sss #4}$} (m-1-2)
	(m-#5) edge[double, double equal sign distance, -implies, line cap=butt] node[pos=0.5, sloped, above] {${\sss #7}$} (m-#6)
	;
	\end{tikzpicture}
}

\newcommand{\plaqTr}[7]{
	\begin{tikzpicture}[scale=1,baseline=-0.3em]
	\matrix[matrix of math nodes,row sep =2.8em,column sep=3em, ampersand replacement=\&] (m) {
		{\sss 0} \& {\sss 4} \\
		{\sss 1} \& {\sss 5}
		\\
	};
	\path
	(m-1-1) edge[->] node[pos=0.5, above] {${\sss #2}$} (m-1-2)
	(m-2-1) edge[->] node[pos=0.5, below] {${\sss #3}$} (m-2-2)
	(m-2-1) edge[<-] node[pos=0.5, left] {${\sss #1}$} (m-1-1)
	(m-2-2) edge[<-] node[pos=0.5, right] {${\sss #4}$} (m-1-2)
	(m-#5) edge[double, double equal sign distance, -implies, line cap=butt] node[pos=0.5, sloped, above] {${\sss #7}$} (m-#6)
	;
	\end{tikzpicture}
}

\newcommand{\plaqSpe}[5]{
	\begin{tikzpicture}[scale=1,baseline=-0.3em]
	\matrix[matrix of math nodes,row sep =2.8em,column sep=3em, ampersand replacement=\&] (m) {
		{\sss 0} \& {\sss 2} \\
		{\sss 1} \& {\sss 3}
		\\
	};
	\path
	(m-1-1) edge[->] node[pos=0.5, above] {${\sss #2}$} (m-1-2)
	(m-2-1) edge[->] node[pos=0.5, below] {${\sss #3}$} (m-2-2)
	(m-2-1) edge[<-, edge node={coordinate[pos=0.5] (sp1)}] node[pos=0.5, left] {${\sss #1}$} (m-1-1)
	(m-2-2) edge[<-, edge node={coordinate[pos=0.5] (sp2)}] node[pos=0.5, right] {${\sss #4}$} (m-1-2)
	(sp1) edge[double, double equal sign distance, -implies, line cap=butt, shab] node[pos=0.5, sloped, above] {${\sss #5}$} (sp2)
	;
	\end{tikzpicture}
}

\newcommand{\plaqDouble}[0]{
	\begin{tikzpicture}[scale=1,baseline=-0.3em]
	\matrix[matrix of math nodes,row sep =2.8em,column sep=3em, ampersand replacement=\&] (m) {
		{\sss 0} \& {\sss 2} \& {\sss 4} \\
		{\sss 1} \& {\sss 3} \& {\sss 5}
		\\
	};
	\path
	(m-1-1) edge[->] node[pos=0.5, above] {${\sss g_{02}}$} (m-1-2)
	(m-2-1) edge[->] node[pos=0.5, below] {${\sss g_{13}}$} (m-2-2)
	(m-2-1) edge[<-] node[pos=0.5, left] {${\sss g_{01}}$} (m-1-1)
	(m-2-2) edge[<-] node[pos=0.3, left] {${\sss g_{23}}$} (m-1-2)
	(m-1-2) edge[->] node[pos=0.5, above] {${\sss g_{24}}$}
	(m-1-3)
	(m-2-2) edge[->] node[pos=0.5, below] {${\sss g_{35}}$}
	(m-2-3)
	(m-2-3) edge[<-] node[pos=0.5, right] {${\sss g_{45}}$}
	(m-1-3)
	(m-1-2) edge[double, double equal sign distance, -implies, line cap=butt] node[pos=0.5, sloped, above] {${\sss h}$} (m-2-1)
	(m-1-3) edge[double, double equal sign distance, -implies, line cap=butt] node[pos=0.5, sloped, above] {${\sss h'}$} (m-2-2)
	;
	\end{tikzpicture}
}

\newcommand{\cubeSans}[0]{
	\begin{tikzpicture}[scale=1,baseline=-0.5em]
	\matrix[matrix of math nodes,row sep =0.35em,column sep=0.7em, ampersand replacement=\&] (m) {
		{} \& {} \& {\sss 0} \& {} \& {} \& {} \& {\sss 2}
		\\[0.2em]
		{\sss 4} \& {} \& {} \& {} \& {\sss 6} \& {} \& {}
		\\
		{} \& {} \& {} \& {} \& {} \& {} \& {}
		\\
		{} \& {} \& {} \& {} \& {} \& {} \& {}
		\\
		{} \& {} \& {} \& {} \& {} \& {} \& {}
		\\
		{} \& {} \& {\sss 1} \& {} \& {} \& {} \& {\sss 3}
		\\[0.2em]
		{\sss 5} \& {} \& {} \& {} \& {\sss 7} \& {} \& {}
		\\					
	};
	\path
	(m-1-3) edge[->] (m-1-7)
	(m-6-3) edge[sha] (intersection of m-6-3--m-6-7 and m-7-5--m-2-5)
	(intersection of m-6-3--m-6-7 and m-7-5--m-2-5) edge[shb, ->] (m-6-7)
	(m-1-3) edge[sha] (intersection of m-1-3--m-6-3 and m-2-1--m-2-5)
	(intersection of m-1-3--m-6-3 and m-2-1--m-2-5) edge[shb, ->] (m-6-3)
	
	(m-1-7) edge[->] (m-6-7)
	(m-7-1) edge[->] (m-7-5)
	(m-2-1) edge[->] (m-7-1)
	(m-1-3) edge[->] (m-2-1)
	(m-6-3) edge[->] (m-7-1)
	(m-1-7) edge[->] (m-2-5)
	(m-6-7) edge[->] (m-7-5)	
	
	(m-1-7) edge[double, double equal sign distance, -implies, line cap=butt] (m-6-3)
	
	(m-6-7) edge[double, double equal sign distance, -implies, line cap=butt,shorten <= 0.9em, shorten >= 0.9em] (m-7-1)
	
	(m-6-3) edge[double, double equal sign distance, -implies, line cap=butt, , shorten >=-0.2em] (m-2-1)
	
	(m-6-7) edge[double, double equal sign distance, -implies, line cap=butt, shorten >=-0.2em]  (m-2-5)
	
	(m-1-7) edge[double, double equal sign distance, -implies, line cap=butt, shorten <= 0.9em, shorten >= 0.9em]  (m-2-1)
	
	(m-2-5) edge[double, double equal sign distance, -implies, line cap=butt, shorten >= 0.2em, shorten <= 0.2em] (m-7-1)
	
	(m-2-1) edge[->]  (m-2-5)
	(m-2-5) edge[->]  (m-7-5)
	;
	\end{tikzpicture}
}

\newcommand{\cube}[6]{
	\begin{tikzpicture}[scale=1,baseline=-0.5em]
	\matrix[matrix of math nodes,row sep =0.35em,column sep=0.7em, ampersand replacement=\&] (m) {
		{} \& {} \& {\sss 0} \& {} \& {} \& {} \& {\sss 2}
		\\[0.2em]
		{\sss 4} \& {} \& {} \& {} \& {\sss 6} \& {} \& {}
		\\
		{} \& {} \& {} \& {} \& {} \& {} \& {}
		\\
		{} \& {} \& {} \& {} \& {} \& {} \& {}
		\\
		{} \& {} \& {} \& {} \& {} \& {} \& {}
		\\
		{} \& {} \& {\sss 1} \& {} \& {} \& {} \& {\sss 3}
		\\[0.2em]
		{\sss 5} \& {} \& {} \& {} \& {\sss 7} \& {} \& {}
		\\					
	};
	\path
	(m-1-3) edge[->] (m-1-7)
	(m-6-3) edge[sha] (intersection of m-6-3--m-6-7 and m-7-5--m-2-5)
	(intersection of m-6-3--m-6-7 and m-7-5--m-2-5) edge[shb, ->] (m-6-7)
	(m-1-3) edge[sha] (intersection of m-1-3--m-6-3 and m-2-1--m-2-5)
	(intersection of m-1-3--m-6-3 and m-2-1--m-2-5) edge[shb, ->] (m-6-3)
	
	(m-1-7) edge[->]  (m-6-7)
	(m-7-1) edge[->] node[pos=0.5, below] {${\sss #1}$} (m-7-5)
	(m-2-1) edge[->] node[pos=0.5, right] {${\sss #2}$} (m-7-1)
	(m-1-3) edge[->] node[pos=0.2, left, xshift=-0.2em, yshift=0.1em] {${\sss #3}$} (m-2-1)
	(m-6-3) edge[->]  (m-7-1)
	(m-1-7) edge[->] (m-2-5)
	(m-6-7) edge[->]  (m-7-5)	
	
	(m-1-7) edge[double, double equal sign distance, -implies, line cap=butt] (m-6-3)
	
	(m-6-7) edge[double, double equal sign distance, -implies, line cap=butt,shorten <= 0.9em, shorten >= 0.9em] node[pos=0.58, above, sloped, yshift=-0.1em] {${\sss #5}$} (m-7-1)
	
	(m-6-3) edge[double, double equal sign distance, -implies, line cap=butt, , shorten >=-0.2em] node[pos=0.55, above, sloped] {${\sss #4}$} (m-2-1)
	
	(m-6-7) edge[double, double equal sign distance, -implies, line cap=butt, shorten >=-0.2em]  (m-2-5)
	
	(m-1-7) edge[double, double equal sign distance, -implies, line cap=butt, shorten <= 0.9em, shorten >= 0.9em]  (m-2-1)
		
	(m-2-5) edge[double, double equal sign distance, -implies, line cap=butt, shorten >= 0.2em, shorten <= 0.2em] node[pos=0.2, above, sloped] {${\sss #6}$} (m-7-1)
	
	(m-2-1) edge[->]  (m-2-5)
	(m-2-5) edge[->]  (m-7-5)
	;
	\end{tikzpicture}
}

\newcommand{\cubeb}[6]{
	\begin{tikzpicture}[scale=1,baseline=-0.5em]
	\matrix[matrix of math nodes,row sep =0.35em,column sep=0.7em, ampersand replacement=\&] (m) {
		{} \& {} \& {\sss \phantom{5}} \& {} \& {} \& {} \& {\sss \phantom{5}}
		\\[0.2em]
		{\sss \phantom{5}} \& {} \& {} \& {} \& {\sss \phantom{5}} \& {} \& {}
		\\
		{} \& {} \& {} \& {} \& {} \& {} \& {}
		\\
		{} \& {} \& {} \& {} \& {} \& {} \& {}
		\\
		{} \& {} \& {} \& {} \& {} \& {} \& {}
		\\
		{} \& {} \& {\sss \phantom{5}} \& {} \& {} \& {} \& {\sss \phantom{5}}
		\\[0.2em]
		{\sss \phantom{5}} \& {} \& {} \& {} \& {\sss \phantom{5}} \& {} \& {}
		\\					
	};
	\node[draw=black,circle,scale=0.4,fill=black!80] at (m-1-3) {};
	\node[draw=black,circle,scale=0.4,fill=black!80] at (m-2-1) {};
	\node[draw=black,circle,scale=0.4,fill=black!80] at (m-6-3) {};
	\node[draw=black,circle,scale=0.4,fill=black!80] at (m-7-1) {};
	\node[draw=black,circle,scale=0.4,fill=gray!60] at (m-1-7) {};
	\node[draw=black,circle,scale=0.4,fill=gray!60] at (m-2-5) {};
	\node[draw=black,circle,scale=0.4,fill=gray!60] at (m-6-7) {};
	\node[draw=black,circle,scale=0.4,fill=gray!60] at (m-7-5) {};
	
	\path
	(m-1-3) edge[->] (m-1-7)
	(m-6-3) edge[sha] (intersection of m-6-3--m-6-7 and m-7-5--m-2-5)
	(intersection of m-6-3--m-6-7 and m-7-5--m-2-5) edge[shb, ->] (m-6-7)
	(m-1-3) edge[sha] (intersection of m-1-3--m-6-3 and m-2-1--m-2-5)
	(intersection of m-1-3--m-6-3 and m-2-1--m-2-5) edge[shb, ->] (m-6-3)
	
	(m-1-7) edge[->]  (m-6-7)
	(m-7-1) edge[->] node[pos=0.5, below] {${\sss #1}$} (m-7-5)
	(m-2-1) edge[->] node[pos=0.5, right] {${\sss #2}$} (m-7-1)
	(m-1-3) edge[->] node[pos=0.2, left, xshift=-0.2em, yshift=0.1em] {${\sss #3}$} (m-2-1)
	(m-6-3) edge[->]  (m-7-1)
	(m-1-7) edge[->] (m-2-5)
	(m-6-7) edge[->]  (m-7-5)	
	
	(m-1-7) edge[double, double equal sign distance, -implies, line cap=butt] (m-6-3)
	
	(m-6-7) edge[double, double equal sign distance, -implies, line cap=butt,shorten <= 0.9em, shorten >= 0.9em] node[pos=0.58, above, sloped, yshift=-0.1em] {${\sss #5}$} (m-7-1)
	
	(m-6-3) edge[double, double equal sign distance, -implies, line cap=butt, , shorten >=-0.2em] node[pos=0.55, above, sloped] {${\sss #4}$} (m-2-1)
	
	(m-6-7) edge[double, double equal sign distance, -implies, line cap=butt, shorten >=-0.2em]  (m-2-5)
	
	(m-1-7) edge[double, double equal sign distance, -implies, line cap=butt, shorten <= 0.9em, shorten >= 0.9em]  (m-2-1)
	
	(m-2-5) edge[double, double equal sign distance, -implies, line cap=butt, shorten >= 0.2em, shorten <= 0.2em] node[pos=0.2, above, sloped] {${\sss #6}$} (m-7-1)
	
	(m-2-1) edge[->]  (m-2-5)
	(m-2-5) edge[->]  (m-7-5)
	;
	\end{tikzpicture}
}

\newcommand{\cubep}[6]{
	\begin{tikzpicture}[scale=1,baseline=-0.5em]
	\matrix[matrix of math nodes,row sep =0.35em,column sep=0.7em, ampersand replacement=\&] (m) {
		{} \& {} \& {\sss \phantom{5}} \& {} \& {} \& {} \& {\sss \phantom{5}}
		\\[0.2em]
		{\sss \phantom{5}} \& {} \& {} \& {} \& {\sss \phantom{5}} \& {} \& {}
		\\
		{} \& {} \& {} \& {} \& {} \& {} \& {}
		\\
		{} \& {} \& {} \& {} \& {} \& {} \& {}
		\\
		{} \& {} \& {} \& {} \& {} \& {} \& {}
		\\
		{} \& {} \& {\sss \phantom{5}} \& {} \& {} \& {} \& {\sss \phantom{5}}
		\\[0.2em]
		{\sss \phantom{5}} \& {} \& {} \& {} \& {\sss \phantom{5}} \& {} \& {}
		\\					
	};
	\node[draw=black,circle,scale=0.4,fill=gray!60] at (m-1-3) {};
	\node[draw=black,circle,scale=0.4,fill=gray!60] at (m-2-1) {};
	\node[draw=black,circle,scale=0.4,fill=gray!60] at (m-6-3) {};
	\node[draw=black,circle,scale=0.4,fill=gray!60] at (m-7-1) {};
	\node[draw=black,circle,scale=0.4,fill=white] at (m-1-7) {};
	\node[draw=black,circle,scale=0.4,fill=white] at (m-2-5) {};
	\node[draw=black,circle,scale=0.4,fill=white] at (m-6-7) {};
	\node[draw=black,circle,scale=0.4,fill=white] at (m-7-5) {};

	\path
	(m-1-3) edge[->] (m-1-7)
	(m-6-3) edge[sha, shorten <= 0.55em] (intersection of m-6-3--m-6-7 and m-7-5--m-2-5)
	(intersection of m-6-3--m-6-7 and m-7-5--m-2-5) edge[shb, ->] (m-6-7)
	(m-1-3) edge[sha] (intersection of m-1-3--m-6-3 and m-2-1--m-2-5)
	(intersection of m-1-3--m-6-3 and m-2-1--m-2-5) edge[shb, ->] (m-6-3)
	
	(m-1-7) edge[->]  (m-6-7)
	(m-7-1) edge[->] node[pos=0.5, below] {${\sss #1}$} (m-7-5)
	(m-2-1) edge[->] node[pos=0.5, right] {${\sss #2}$} (m-7-1)
	(m-1-3) edge[->] node[pos=0.2, left, xshift=-0.2em, yshift=0.1em] {${\sss #3}$} (m-2-1)
	(m-6-3) edge[->]  (m-7-1)
	(m-1-7) edge[->] (m-2-5)
	(m-6-7) edge[->]  (m-7-5)	
	
	(m-1-7) edge[double, double equal sign distance, -implies, line cap=butt] (m-6-3)
	
	(m-6-7) edge[double, double equal sign distance, -implies, line cap=butt,shorten <= 0.9em, shorten >= 0.9em] node[pos=0.58, above, sloped, yshift=-0.1em] {${\sss #5}$} (m-7-1)
	
	(m-6-3) edge[double, double equal sign distance, -implies, line cap=butt, , shorten >=-0.2em] node[pos=0.55, above, sloped] {${\sss #4}$} (m-2-1)
	
	(m-6-7) edge[double, double equal sign distance, -implies, line cap=butt, shorten >=-0.2em]  (m-2-5)
	
	(m-1-7) edge[double, double equal sign distance, -implies, line cap=butt, shorten <= 0.9em, shorten >= 0.9em]  (m-2-1)
	
	(m-2-5) edge[double, double equal sign distance, -implies, line cap=butt, shorten >= 0.2em, shorten <= 0.2em] node[pos=0.2, above, sloped] {${\sss #6}$} (m-7-1)
	
	(m-2-1) edge[->]  (m-2-5)
	(m-2-5) edge[->]  (m-7-5)
	;
	\end{tikzpicture}
}

\newcommand{\cubeSpe}[6]{
	\begin{tikzpicture}[scale=1,baseline=-0.5em]
	\matrix[matrix of math nodes,row sep =0.35em,column sep=0.7em, ampersand replacement=\&] (m) {
		{} \& {} \& {\sss \phantom{5}} \& {} \& {} \& {} \& {\sss \phantom{5}}
		\\[0.2em]
		{\sss \phantom{5}} \& {} \& {} \& {} \& {\sss \phantom{5}} \& {} \& {}
		\\
		{} \& {} \& {} \& {} \& {} \& {} \& {}
		\\
		{} \& {} \& {} \& {} \& {} \& {} \& {}
		\\
		{} \& {} \& {} \& {} \& {} \& {} \& {}
		\\
		{} \& {} \& {\sss \phantom{5}} \& {} \& {} \& {} \& {\sss \phantom{5}}
		\\[0.2em]
		{\sss \phantom{5}} \& {} \& {} \& {} \& {\sss \phantom{5}} \& {} \& {}
		\\					
	};
	\node[draw=black,circle,scale=0.4,fill=black!80] at (m-1-3) {};
	\node[draw=black,circle,scale=0.4,fill=black!80] at (m-2-1) {};
	\node[draw=black,circle,scale=0.4,fill=black!80] at (m-6-3) {};
	\node[draw=black,circle,scale=0.4,fill=black!80] at (m-7-1) {};
	\node[draw=black,circle,scale=0.4,fill=white] at (m-1-7) {};
	\node[draw=black,circle,scale=0.4,fill=white] at (m-2-5) {};
	\node[draw=black,circle,scale=0.4,fill=white] at (m-6-7) {};
	\node[draw=black,circle,scale=0.4,fill=white] at (m-7-5) {};
	
	\path
	(m-1-3) edge[->] (m-1-7)
	(m-6-3) edge[sha, shorten <= 1.75em] (intersection of m-6-3--m-6-7 and m-7-5--m-2-5)
	(intersection of m-6-3--m-6-7 and m-7-5--m-2-5) edge[shb, ->] (m-6-7)
	(m-1-3) edge[sha] (intersection of m-1-3--m-6-3 and m-2-1--m-2-5)
	(intersection of m-1-3--m-6-3 and m-2-1--m-2-5) edge[shb, ->] (m-6-3)
	
	(m-1-7) edge[->]  (m-6-7)
	(m-7-1) edge[->] node[pos=0.5, below] {${\sss #1}$} (m-7-5)
	(m-2-1) edge[->] node[pos=0.5, right] {${\sss #2}$} (m-7-1)
	(m-1-3) edge[->] node[pos=0.2, left, xshift=-0.2em, yshift=0.1em] {${\sss #3}$} (m-2-1)
	(m-6-3) edge[->, shorten <= 1em]  (m-7-1)
	(m-6-3) edge[shorten >= 2em]  (m-7-1)
	(m-1-7) edge[->] (m-2-5)
	(m-6-7) edge[->]  (m-7-5)	
	
	(m-1-7) edge[double, double equal sign distance, -implies, line cap=butt] (m-6-3)
	
	(m-6-7) edge[double, double equal sign distance, -implies, line cap=butt,shorten <= 0.9em, shorten >= 0.9em] node[pos=0.62, above, sloped, yshift=-0.2em] {${\sss #5}$} (m-7-1)
	
	(m-6-3) edge[double, double equal sign distance, -implies, line cap=butt, , shorten >=-0.2em] node[pos=0.55, above, sloped] {${\sss #4}$} (m-2-1)
	
	(m-6-7) edge[double, double equal sign distance, -implies, line cap=butt, shorten >=-0.2em]  (m-2-5)
	
	(m-1-7) edge[double, double equal sign distance, -implies, line cap=butt, shorten <= 0.9em, shorten >= 0.9em]  (m-2-1)
	
	(m-2-5) edge[double, double equal sign distance, -implies, line cap=butt, shorten >= 0.2em, shorten <= 0.2em] node[pos=0.3, above, sloped, rectangle, fill=white, inner sep=0pt, yshift=0.3em] {${\sss #6}$} (m-7-1)
	
	(m-2-1) edge[->]  (m-2-5)
	(m-2-5) edge[->]  (m-7-5)
	;
	\end{tikzpicture}
}

\newcommand{\cubeDouble}[0]{
	\begin{tikzpicture}[scale=1,baseline=-0.5em]
	\matrix[matrix of math nodes,row sep =0.35em,column sep=0.7em, ampersand replacement=\&] (m) {
		{} \& {} \& {\sss \phantom{5}} \& {} \& {} \& {} \& {\sss \phantom{5}} \& {} \& {} \& {} \& {\sss \phantom{5}}
		\\[0.2em]
		{\sss \phantom{5}} \& {} \& {} \& {} \& {\sss \phantom{5}} \& {} \& {} \& {} \& {\sss \phantom{5}} \& {} \& {}
		\\
		{} \& {} \& {} \& {} \& {} \& {} \& {} \& {} \& {} \& {} \& {}
		\\
		{} \& {} \& {} \& {} \& {} \& {} \& {} \& {} \& {} \& {} \& {}
		\\
		{} \& {} \& {} \& {} \& {} \& {} \& {} \& {} \& {} \& {} \& {}
		\\
		{} \& {} \& {\sss \phantom{5}} \& {} \& {} \& {} \& {\sss \phantom{5}} \& {} \& {} \& {} \& {\sss \phantom{5}}
		\\[0.2em]
		{\sss \phantom{5}} \& {} \& {} \& {} \& {\sss \phantom{5}} \& {} \& {} \& {} \& {\sss \phantom{5}} \& {} \& {}
		\\					
	};

	\node[draw=black,circle,scale=0.4,fill=black!80] at (m-1-3) {};
	\node[draw=black,circle,scale=0.4,fill=black!80] at (m-2-1) {};
	\node[draw=black,circle,scale=0.4,fill=black!80] at (m-6-3) {};
	\node[draw=black,circle,scale=0.4,fill=black!80] at (m-7-1) {};
	\node[draw=black,circle,scale=0.4,fill=gray!60] at (m-1-7) {};
	\node[draw=black,circle,scale=0.4,fill=gray!60] at (m-2-5) {};
	\node[draw=black,circle,scale=0.4,fill=gray!60] at (m-6-7) {};
	\node[draw=black,circle,scale=0.4,fill=gray!60] at (m-7-5) {};
	\node[draw=black,circle,scale=0.4,fill=white] at (m-1-11) {};
	\node[draw=black,circle,scale=0.4,fill=white] at (m-2-9) {};
	\node[draw=black,circle,scale=0.4,fill=white] at (m-6-11) {};
	\node[draw=black,circle,scale=0.4,fill=white] at (m-7-9) {};

	\path
	(m-1-3) edge[->] (m-1-7)
	(m-1-7) edge[->] (m-1-11)
	(m-6-3) edge[sha] (intersection of m-6-3--m-6-7 and m-7-5--m-2-5)
	(intersection of m-6-3--m-6-7 and m-7-5--m-2-5) edge[shb, ->] (m-6-7)
	(m-6-7) edge[sha, shorten <= 0.55em] (intersection of m-6-7--m-6-11 and m-7-9--m-2-9)
	(intersection of m-6-7--m-6-11 and m-7-9--m-2-9) edge[shb, ->] (m-6-11)
	(m-1-3) edge[sha] (intersection of m-1-3--m-6-3 and m-2-1--m-2-5)
	(intersection of m-1-3--m-6-3 and m-2-1--m-2-5) edge[shb, ->] (m-6-3)
	
	(m-1-7) edge[sha] (intersection of m-1-7--m-6-7 and m-2-5--m-2-9)
	(intersection of m-1-7--m-6-7 and m-2-5--m-2-9) edge[shb, ->] (m-6-7)
	(m-1-11) edge[->] (m-6-11)
	(m-7-1) edge[->] node[pos=0.5, below] {${\sss \gx}$} (m-7-5)
	(m-7-5) edge[->] node[pos=0.5, below] {${\sss \gx'}$} (m-7-9)
	(m-2-1) edge[->] node[pos=0.5, right] {${\sss \gy}$} (m-7-1)
	(m-1-3) edge[->] node[pos=0.2, left, xshift=-0.2em, yshift=0.1em] {${\sss \gz}$} (m-2-1)
	(m-6-3) edge[->]  (m-7-1)
	(m-1-7) edge[->] (m-2-5)
	(m-1-11) edge[->] (m-2-9)
	(m-6-7) edge[->]  (m-7-5)
	(m-6-11) edge[->] (m-7-9)	
	
	(m-1-7) edge[double, double equal sign distance, -implies, line cap=butt] (m-6-3)
	
	(m-6-7) edge[double, double equal sign distance, -implies, line cap=butt,shorten <= 0.9em, shorten >= 0.9em] node[pos=0.58, above, sloped, yshift=-0.1em] {${\sss \hy}$} (m-7-1)
	
	(m-6-3) edge[double, double equal sign distance, -implies, line cap=butt, , shorten >=-0.2em] node[pos=0.55, above, sloped] {${\sss \hx}$} (m-2-1)
	
	(m-6-7) edge[double, double equal sign distance, -implies, line cap=butt, shorten >=-0.2em]  (m-2-5)
	
	(m-1-7) edge[double, double equal sign distance, -implies, line cap=butt, shorten <= 0.9em, shorten >= 0.9em]  (m-2-1)
	
	(m-2-5) edge[double, double equal sign distance, -implies, line cap=butt, shorten >= 0.2em, shorten <= 0.2em] node[pos=0.2, above, sloped] {${\sss \hz}$} (m-7-1)

	(m-1-11) edge[double, double equal sign distance, -implies, line cap=butt] (m-6-7)
	
	(m-6-11) edge[double, double equal sign distance, -implies, line cap=butt,shorten <= 0.9em, shorten >= 0.9em] node[pos=0.58, above, sloped, yshift=-0.1em] {${\sss \hy'}$} (m-7-5)
	
	(m-1-11) edge[double, double equal sign distance, -implies, line cap=butt, shorten <= 0.9em, shorten >= 0.9em]  (m-2-5)
	
	(m-2-9) edge[double, double equal sign distance, -implies, line cap=butt, shorten >= 0.2em, shorten <= 0.2em] node[pos=0.2, above, sloped] {${\sss \hz'}$} (m-7-5)
	
	(m-6-11) edge[double, double equal sign distance, -implies, line cap=butt, , shorten >=-0.2em] (m-2-9)
	
	(m-2-1) edge[->] (m-2-5)
	(m-2-5) edge[->] (m-7-5)
	(m-2-5) edge[->] (m-2-9)
	(m-2-9) edge[->] (m-7-9)
	;
	\end{tikzpicture}
}

\newcommand{\cubeDoubleAvg}[0]{
	\begin{tikzpicture}[scale=1,baseline=-0.5em]
	\matrix[matrix of math nodes,row sep =0.35em,column sep=0.7em, ampersand replacement=\&] (m) {
		{} \& {} \& {\sss \phantom{5}} \& {} \& {} \& {} \& {\sss \phantom{5}} \& {} \& {} \& {} \& {\sss \phantom{5}}
		\\[0.2em]
		{\sss \phantom{5}} \& {} \& {} \& {} \& {\sss \phantom{5}} \& {} \& {} \& {} \& {\sss \phantom{5}} \& {} \& {}
		\\
		{} \& {} \& {} \& {} \& {} \& {} \& {} \& {} \& {} \& {} \& {}
		\\
		{} \& {} \& {} \& {} \& {} \& {} \& {} \& {} \& {} \& {} \& {}
		\\
		{} \& {} \& {} \& {} \& {} \& {} \& {} \& {} \& {} \& {} \& {}
		\\
		{} \& {} \& {\sss \phantom{5}} \& {} \& {} \& {} \& {\sss \phantom{5}} \& {} \& {} \& {} \& {\sss \phantom{5}}
		\\[0.2em]
		{\sss \phantom{5}} \& {} \& {} \& {} \& {\sss \phantom{5}} \& {} \& {} \& {} \& {\sss \phantom{5}} \& {} \& {}
		\\					
	};
	
	\node[draw=black,circle,scale=0.4,fill=black!80] at (m-1-3) {};
	\node[draw=black,circle,scale=0.4,fill=black!80] at (m-2-1) {};
	\node[draw=black,circle,scale=0.4,fill=black!80] at (m-6-3) {};
	\node[draw=black,circle,scale=0.4,fill=black!80] at (m-7-1) {};
	\node[draw=black,circle,scale=0.4,fill=gray!60] at (m-1-7) {};
	\node[draw=black,circle,scale=0.4,fill=gray!60] at (m-2-5) {};
	\node[draw=black,circle,scale=0.4,fill=gray!60] at (m-6-7) {};
	\node[draw=black,circle,scale=0.4,fill=gray!60] at (m-7-5) {};
	\node[draw=black,circle,scale=0.4,fill=white] at (m-1-11) {};
	\node[draw=black,circle,scale=0.4,fill=white] at (m-2-9) {};
	\node[draw=black,circle,scale=0.4,fill=white] at (m-6-11) {};
	\node[draw=black,circle,scale=0.4,fill=white] at (m-7-9) {};
	
	\path
	(m-1-3) edge[->] (m-1-7)
	(m-1-7) edge[->] (m-1-11)
	(m-6-3) edge[shorten >= 2em] (intersection of m-6-3--m-6-7 and m-7-5--m-2-5)
	(intersection of m-6-3--m-6-7 and m-7-5--m-2-5) edge[shb, ->] (m-6-7)
	(m-6-7) edge[sha, shorten <= 1.7em] (intersection of m-6-7--m-6-11 and m-7-9--m-2-9)
	(intersection of m-6-7--m-6-11 and m-7-9--m-2-9) edge[shb, ->] (m-6-11)
	(m-1-3) edge[sha] (intersection of m-1-3--m-6-3 and m-2-1--m-2-5)
	(intersection of m-1-3--m-6-3 and m-2-1--m-2-5) edge[shb, ->] (m-6-3)
	
	(m-1-7) edge[sha] (intersection of m-1-7--m-6-7 and m-2-5--m-2-9)
	(intersection of m-1-7--m-6-7 and m-2-5--m-2-9) edge[shb, ->] (m-6-7)
	(m-1-11) edge[->] (m-6-11)
	(m-7-1) edge[->] node[pos=0.5, below] {${\sss \gx k^{-1}}$} (m-7-5)
	(m-7-5) edge[->] node[pos=0.5, below] {${\sss k\gx'}$} (m-7-9)
	(m-2-1) edge[->] node[pos=0.5, right] {${\sss \gy}$} (m-7-1)
	(m-1-3) edge[->] node[pos=0.2, left, xshift=-0.2em, yshift=0.1em] {${\sss \gz}$} (m-2-1)
	(m-6-3) edge[->, shorten <= 1em]  (m-7-1)
	(m-6-3) edge[shorten >= 2em]  (m-7-1)
	(m-1-7) edge[->] (m-2-5)
	(m-1-11) edge[->] (m-2-9)
	(m-6-7) edge[->, shorten <= 0.4em]  (m-7-5)
	(m-6-11) edge[->] (m-7-9)	
	
	(m-1-7) edge[double, double equal sign distance, -implies, line cap=butt] (m-6-3)
	
	(m-6-7) edge[double, double equal sign distance, -implies, line cap=butt,shorten <= 0.9em, shorten >= 0.9em] node[pos=0.58, above, sloped, yshift=-0.1em] {${\sss \hy(\gx \act \eta^{-1})}$} (m-7-1)
	
	(m-6-3) edge[double, double equal sign distance, -implies, line cap=butt, , shorten >=-0.2em] node[pos=0.55, above, sloped] {${\sss \hx}$} (m-2-1)
	
	(m-6-7) edge[double, double equal sign distance, -implies, line cap=butt, shorten >=-0.2em]  (m-2-5)
	
	(m-1-7) edge[double, double equal sign distance, -implies, line cap=butt, shorten <= 0.9em, shorten >= 0.9em]  (m-2-1)
	
	(m-2-5) edge[double, double equal sign distance, -implies, line cap=butt, shorten >= 0.2em, shorten <= 0.2em] node[rectangle, fill=white, inner sep=0pt, pos=0.4, above, sloped, yshift=0.3em] {${\sss \hz(\gx\act \eta'^{-1})}$} (m-7-1)
	
	(m-1-11) edge[double, double equal sign distance, -implies, line cap=butt] (m-6-7)
	
	(m-6-11) edge[double, double equal sign distance, -implies, line cap=butt,shorten <= 0.9em, shorten >= 0.9em] node[pos=0.58, above, sloped, yshift=-0.2em] {${\sss k\act (\eta \hy')}$} (m-7-5)
	
	(m-1-11) edge[double, double equal sign distance, -implies, line cap=butt, shorten <= 0.9em, shorten >= 0.9em]  (m-2-5)
	
	(m-2-9) edge[double, double equal sign distance, -implies, line cap=butt, shorten >= 0.2em, shorten <= 0.2em] node[rectangle, fill=white, inner sep=0pt, pos=0.3, above, sloped, yshift=0.3em] {${\sss k \act (\eta' \hz')}$} (m-7-5)
	
	(m-6-11) edge[double, double equal sign distance, -implies, line cap=butt, , shorten >=-0.2em] (m-2-9)
	
	(m-2-1) edge[->] (m-2-5)
	(m-2-5) edge[->] (m-7-5)
	(m-2-5) edge[->] (m-2-9)
	(m-2-9) edge[->] (m-7-9)
	;
	\end{tikzpicture}
}

\newcommand{\interpretation}[1]{
	\begin{tikzpicture}[scale=#1, baseline=0em, line width= 0.8pt]	
	\coordinate (0) at (1.5,0);
	
	\coordinate (1) at (0,0);
	\coordinate (2) at (0.8,0.1);
	\coordinate (3) at (2.2,0.1);
	\coordinate (4) at (3,0);
	
	\coordinate (5) at (0.8,2);
	\coordinate (6) at (2.2,2);
	
	\coordinate (7) at (0.8,-2);
	\coordinate (8) at (2.2,-2);
	
	\centerarc[](0)(0:360:2.2)
	
	\coordinate (10) at (0.6,-1);
	
	\draw[op6, name path = P14d] (1)  to[out=-90, in=270, looseness=0.8] (4);
	\path[name path = P14u] (1) to[out=90, in=-270, looseness=0.8](4);
	\draw[op6, name path = P23d] (2) to[out=-90, in=270, looseness=0.5, edge node={coordinate[pos=0.2] (sp1)}, edge node={coordinate[pos=0.8] (sp2)}](3);
	\path[name path = Psp1sp2] (sp1) to[out=90, in=-270, looseness=0.5] (sp2);
	
	\draw[name path = P56] (5) to[out=-90, in=270, looseness=0.5, edge node={coordinate[pos=0.2] (sp3)}, edge node={coordinate[pos=0.8] (sp4)}] (6);
	
	\draw[op4, name path = P78] (7) to[out=90, in=-270, looseness=0.5, edge node={coordinate[pos=0.2] (sp5)}, edge node={coordinate[pos=0.8] (sp6)}] (8);

	\path[name path = Psp3sp5] (sp3)  to[out=-20, in=20, looseness=0.3] (sp5);
	\path[name path = Psp4sp6] (sp4)  to[out=200, in=-200, looseness=0.3] (sp6);
	
	\begin{scope}
		\path [name intersections={of=Psp3sp5 and P23d,by={sp11}}];
		\path (sp3) ++ (-0.1,0) coordinate (spa);
		\path (sp11) ++ (0.1,0) coordinate (spb);
		\clip (spa) rectangle (spb);
		\draw[op6] (sp3)  to[out=-20, in=20, looseness=0.3] (sp5);
	\end{scope}
	
	\begin{scope}
		\path [name intersections={of=Psp3sp5 and P23d,by={sp11}}];
		\path [name intersections={of=Psp3sp5 and P14d,by={sp13}}];
		\path (sp11) ++ (0.1,0) coordinate (spa);
		\path (sp13) ++ (-0.1,0) coordinate (spb);
		\clip (spa) rectangle (spb);
		\draw[op3] (sp3)  to[out=-20, in=20, looseness=0.3] (sp5);
	\end{scope}
	
	\begin{scope}
		\path [name intersections={of=Psp3sp5 and P14d,by={sp13}}];
		\path (sp13) ++ (0.1,0) coordinate (spa);
		\path (sp5) ++ (-0.1,0) coordinate (spb);
		\clip (spa) rectangle (spb);
		\draw[op6] (sp3)  to[out=-20, in=20, looseness=0.3] (sp5);
	\end{scope}
	
	\begin{scope}
		\path [name intersections={of=Psp4sp6 and P23d,by={sp12}}];
		\path (sp4) ++ (0.1,0) coordinate (spa);
		\path (sp12) ++ (-0.1,0) coordinate (spb);
		\clip (spa) rectangle (spb);
		\draw[op6] (sp4) to[out=200, in=-200, looseness=0.3] (sp6);
	\end{scope}
	
	\begin{scope}
		\path [name intersections={of=Psp4sp6 and P23d,by={sp12}}];
		\path [name intersections={of=Psp4sp6 and P14d,by={sp14}}];
		\path (sp12) ++ (-0.1,0) coordinate (spa);
		\path (sp14) ++ (0.1,0) coordinate (spb);
		\clip (spa) rectangle (spb);
		\draw[op3] (sp4) to[out=200, in=-200, looseness=0.3] (sp6);
	\end{scope}
	
	\begin{scope}
		\path [name intersections={of=Psp4sp6 and P14d,by={sp14}}];
		\path (sp14) ++ (-0.1,0) coordinate (spa);
		\path (sp6) ++ (0.1,0) coordinate (spb);
		\clip (spa) rectangle (spb);
		\draw[op6] (sp4) to[out=200, in=-200, looseness=0.3] (sp6);
	\end{scope}
	
	\begin{scope}
		\path [name intersections={of=Psp3sp5 and P14u,by={sp7}}];
		\path (1) ++ (-0.1,-0.1) coordinate (spa);
		\path (sp7) ++ (0,0.1) coordinate (spb);
		\clip (spa) rectangle (spb);
		\draw[op6] (1) to[out=90, in=-270, looseness=0.8](4);
	\end{scope}
	
	\begin{scope}
		\path [name intersections={of=Psp4sp6 and P14u,by={sp8}}];
		\path (4) ++ (0.1,-0.1) coordinate (spa);
		\path (sp8) ++ (0,0.1) coordinate (spb);
		\clip (spa) rectangle (spb);
		\draw[op6] (1) to[out=90, in=-270, looseness=0.8](4);
	\end{scope}
	
	\begin{scope}
		\path [name intersections={of=Psp3sp5 and P14u,by={sp7}}];
		\path [name intersections={of=Psp4sp6 and P14u,by={sp8}}];
		\path (sp7) ++ (0,-0.1) coordinate (spa);
		\path (sp8) ++ (0,0.1) coordinate (spb);
		\clip (spa) rectangle (spb);
		\draw[op3] (1) to[out=90, in=-270, looseness=0.8](4);
	\end{scope}
	
	\begin{scope}
		\path [name intersections={of=Psp3sp5 and Psp1sp2,by={sp11}}];
		\path (1) ++ (-0.1,-0.1) coordinate (spa);
		\path (sp11) ++ (0,0.1) coordinate (spb);
		\clip (spa) rectangle (spb);
		\draw[op6] (sp1) to[out=90, in=-270, looseness=0.5] (sp2);
	\end{scope}
	
	\begin{scope}
		\path [name intersections={of=Psp4sp6 and Psp1sp2,by={sp12}}];
		\path (4) ++ (0.1,-0.1) coordinate (spa);
		\path (sp12) ++ (0,0.1) coordinate (spb);
		\clip (spa) rectangle (spb);
		\draw[op6] (sp1) to[out=90, in=-270, looseness=0.5] (sp2);
	\end{scope}
	
	\begin{scope}
		\path [name intersections={of=Psp3sp5 and Psp1sp2,by={sp11}}];
		\path [name intersections={of=Psp4sp6 and Psp1sp2,by={sp12}}];
		\path (sp11) ++ (0,-0.1) coordinate (spa);
		\path (sp12) ++ (0,0.1) coordinate (spb);
		\clip (spa) rectangle (spb);
		\draw[op3] (sp1) to[out=90, in=-270, looseness=0.5] (sp2);
	\end{scope}
	
	\node[circle, fill = black, inner sep=1.1pt, op6] at (10) {};
	\path[name path = P1010, ->, line width = 0.4pt, shab] (10) to[out=40, in=110, looseness=2, min distance = 5.5em] (10);
	
	\node[] at (0.3,0.1) {${\sss c_{{\rm y},1}}$};
	\node[] at (2.25,-1) {${\sss c_{{\rm z},1}}$};
	
	\begin{scope}
		\path [name intersections={of=P14d and P1010,by={spa}}];
		\path (spa) ++ (0.1,0) coordinate (spa);
		\clip (10) rectangle (spa);
		\draw[line width = 0.4pt, shab] (10) to[out=40, in=110, looseness=2, min distance = 5.5em] (10);
	\end{scope}
	
	\begin{scope}
		\path [name intersections={of=P14d and P1010,by={spa}}];
		\path [name intersections={of=P23d and P1010,by={spb}}];
		\path (spa) ++ (-0.1,0) coordinate (spa);
		\path (spb) ++ (0.1,0) coordinate (spb);
		\clip (spa) rectangle (spb);
		\draw[dot, line width = 0.4pt, shab] (10) to[out=40, in=110, looseness=2, min distance = 5.5em] (10);
	\end{scope}
	
	\begin{scope}
		\path [name intersections={of=P23d and P1010,by={spa}}];
		\clip (10) -- (spa) -- (1.5,-0.1) -- (-0.3,1) -- (0.4,-1) -- (10);
		\draw[->, line width = 0.4pt, shab] (10) to[out=40, in=110, looseness=2, min distance = 5.5em] (10);
	\end{scope}
	
	\path[name path = P1010b, <-, line width = 0.4pt, shab] (10) to[out=-20, in=20, looseness=2, min distance = 5.5em] (10);
	
	\begin{scope}
		\path [name intersections={of=Psp3sp5 and P1010b,by={spa1,spa2}}];
		\path [name intersections={of=Psp4sp6 and P1010b,by={spb1,spb2}}];
		\clip (10) -- (0.6,0) -- (spa2) -- (spb2) -- (3,0) -- (3,-1.5) -- (0.6,-1.5) -- (10);
		\draw[<-, line width = 0.4pt, shab] (10) to[out=-20, in=20, looseness=2, min distance = 5.5em] (10);
	\end{scope}
	
	\begin{scope}
		\path [name intersections={of=Psp3sp5 and P1010b,by={spa1,spa2}}];
		\path [name intersections={of=Psp4sp6 and P1010b,by={spb1,spb2}}];
		\path (spa2) ++ (0,-0.1) coordinate (spa2);
		\path (spb2) ++ (0,0.1) coordinate (spb2);
		\clip (spa2) rectangle (spb2);
		\draw[<-, line width = 0.4pt, shab, dot] (10) to[out=-20, in=20, looseness=2, min distance = 5.5em] (10);
	\end{scope}
	
	\end{tikzpicture}
}

\newcommand{\skippingRope}[1]{
	\begin{tikzpicture}[scale=#1, baseline=0em, line width= 0.8pt]
	\coordinate (0) at (0,0);
	\coordinate (1) at (-1.5,0);
	\coordinate (2) at (1.5,0);
	\coordinate (3) at (0,-1.5);
	\coordinate (4) at (0,1.5);
	
	\begin{scope}
		\clip (-1.5,1.5) rectangle (1.5,-1.5);
		\draw[line width=0.4pt] (3) edge[double, double equal sign distance, out=135, in=-135, looseness=0.8, line cap=butt, shorten <= -0.1em, shorten >= -0.1em, op6] (4);
		\draw[->, line width = 0.4pt, shab] (1) to[out=35, in=145, looseness=0.8,  edge node={coordinate[pos=0.64] (sp1)}] (2);
		\draw[line width=0.4pt] (sp1) edge[double, double equal sign distance, out=-85, in=45, looseness=0.8, line cap=butt, shorten <= 0.3em, shorten >= -0.1em] node[pos=0.5, right] {${\sss h}$} (3);
		\draw[line width=0.4pt] (4) edge[double, double equal sign distance, -implies, out=-45, in=95, looseness=0.8, line cap=butt, shorten <= -0.1em, shorten >= 0.3em] (sp1);
	\end{scope}	
	\centerarc[](0)(0:360:1.5) 
	
	\node[circle, fill = black, inner sep=1pt] at (1) {};
	\node[circle, fill = black, inner sep=1pt] at (2) {};
	\end{tikzpicture}
}

\newcommand{\tubeS}[6]{
	\begin{tikzpicture}[scale=1,baseline=-0.25em, line width=0.4pt]
	\centerarc[->](0,0.25)(-50:230:0.25)
	\centerarc[<-](0,-0.25)(50:-230:0.25)
	\draw[->] (0.3,0) -- (1,0);
	\node[] at (0,0) {${\sss #1 }$};
	\node[circle, fill = white, inner sep=0pt] at (0,0.45) {${\sss #2}$};
	\node[circle, fill = white, inner sep=0pt] at (0,-0.45) {${\sss #3}$};
	\node[circle, fill = white, inner sep=0pt] at (0.65,0) {${\sss #4}$};
	\node[] at (0.65,0.31) {${\sss #5}$};
	\node[] at (0.65,-0.31) {${\sss #6}$};
	\end{tikzpicture}
}

\newcommand{\tubeSavg}[6]{
	\begin{tikzpicture}[scale=1,baseline=-0.25em, line width=0.4pt]
	\centerarc[->](0,0.25)(-50:230:0.25)
	\centerarc[<-](0,-0.25)(50:-230:0.25)
	\draw[->] (0.3,0) -- (#6,0);
	\node[] at (0,0) {${\sss #1 }$};
	\node[circle, fill = white, inner sep=0pt] at (0,0.45) {${\sss #2}$};
	\node[circle, fill = white, inner sep=0pt] at (0,-0.45) {${\sss #3}$};
	\node[rectangle, fill = white, inner sep=0pt] at (#5,0.22) {${\sss #4}$};
	\end{tikzpicture}
}

\newcommand{\tubeSext}[8]{
	\begin{tikzpicture}[scale=1,baseline=-0.25em, line width=0.4pt]
	\centerarc[->](0,0.25)(-50:230:0.25)
	\centerarc[<-](0,-0.25)(50:-230:0.25)
	\draw[->] (0.3,0) -- (#8,0);
	\node[] at (0,0) {${\sss #1 }$};
	\node[circle, fill = white, inner sep=0pt] at (0,0.45) {${\sss #2}$};
	\node[circle, fill = white, inner sep=0pt] at (0,-0.45) {${\sss #3}$};
	\node[circle, fill = white, inner sep=0pt] at (#7,0) {${\sss #4}$};
	\node[] at (#7,0.31) {${\sss #5}$};
	\node[] at (#7,-0.31) {${\sss #6}$};
	\end{tikzpicture}
}

\newcommand{\tubeSsphere}[4]{
	\begin{tikzpicture}[scale=1,baseline=-0.25em, line width=0.4pt]
	\draw[->] (0.15,0) -- (#4,0);
	\node[] at (0,0) {${\sss #1 }$};
	\node[circle, fill = white, inner sep=0pt] at (#3,0) {${\sss #2}$};
	\end{tikzpicture}
}

\newcommand{\tubeSsphereavg}[4]{
	\begin{tikzpicture}[scale=1,baseline=-0.25em, line width=0.4pt]
	\draw[->] (0.15,0) -- (#4,0);
	\node[] at (0,0) {${\sss #1 }$};
	\node[rectangle, fill = white, inner sep=0pt] at (#3,0.19) {${\sss #2}$};
	\end{tikzpicture}
}

\title{Excitations in strict 2-group higher gauge models \\ of topological phases}

\author[\Square]{Alex Bullivant,}
\author[\pentagon, \hexagon]{Clement Delcamp}

\affiliation[\Square]{Department of Pure Mathematics \\ University of Leeds, Leeds, LS2 9JT, UK}

\affiliation[\pentagon]{Max-Planck-Institut f{\"u}r Quantenoptik \\ Hans-Kopfermann-Str. 1, 85748 Garching, Germany}
\affiliation[\hexagon]{Munich Center for Quantum Science and Technology (MCQST)\\ Schellingstr. 4, D-80799 M{\"u}nchen, Germany}

\emailAdd{a.l.bullivant@leeds.ac.uk}
\emailAdd{clement.delcamp@mpq.mpg.de}

\abstract{\\~\\
	We consider an exactly solvable model for topological phases in (3+1)d whose input data is a strict 2-group. This model, which has a higher gauge theory interpretation, provides a lattice Hamiltonian realisation of the Yetter homotopy 2-type topological quantum field theory. The Hamiltonian yields bulk flux and charge composite excitations that are either point-like or loop-like. Applying a generalised tube algebra approach, we reveal the algebraic structure underlying these excitations and derive the irreducible modules of this algebra, which in turn classify the elementary excitations of the model. As a further application of the tube algebra approach, we demonstrate that the ground state subspace of the three-torus is described by the central subalgebra of the tube algebra for torus boundary, demonstrating the ground state degeneracy is given by the number of elementary loop-like excitations.
}

\begin{document} 
	\vspace*{-2em}
	\maketitle
	\flushbottom
	\newpage

\section{Introduction\label{sec:intro}} 
\emph{Higher symmetries} refer to symmetries whose charge excitations have support on higher dimensional manifolds, in contrast to ordinary symmetries whose charge excitations are all \emph{point-like}. These generalized symmetries have been a very active topic of research within the physics community over the past few years. For instance, $q$-form \emph{global} symmetries together with their higher \emph{anomalies} have been thoroughly investigated in the quantum field theory context \cite{gaiotto2015generalized, Hsin:2018vcg, Cordova:2018cvg, Benini:2018reh, tachikawa2017gauging}, where a (higher) $q$-form global symmetry is a global symmetry whose symmetry operators are all $q$-dimensional. A theory displaying a non-anomalous $q$-form global symmetry can be gauged by coupling it to a ($q$+1)-form background connection, then resulting in a ($q$+1)-form \emph{gauge} theory.  Moreover, some theories display gauge fields of different degrees that interact in a non-trivial way. Such theories are  typically referred to as \emph{higher gauge theories}  \cite{bartels2004higher, baez2004higher, baez2011invitation, kapustin2014coupling} and are characterized by the ability of defining \emph{higher holonomies} which encode the parallel transport of higher-dimensional objects. A prototypical example of this scenario are \emph{2-group} gauge theories, which combine 0-form and 1-form symmetries.

Recently, \emph{gapped phases of matter} described by topological theories that have a higher gauge theory have been under scrutiny \cite{mackaay2000finite, Kapustin:2013uxa, kapustin2014topological, Thorngren:2015gtw,  2018arXiv180907325R, Wen:2018zux, Wan:2018bns, Wan:2018zql,  Wan:2018djl, Wan:2019oyr, zhu2018topological, Cheng:2017ftw,  2018arXiv180809394Z}. Examples of such theories are provided by sigma models whose target spaces are given by \emph{Postnikov towers} \cite{hatcher2002algebraic} built as fibrations of \emph{Eilenberg-MacLane spaces} \cite{eilenberg1953groups, eilenberg1954groups}, and the construction of lattice Hamiltonian realizations of such models was considered in \cite{Williamson:2016evv, delcamp2018gauge, Delcamp2019}. In general, we define a Hamiltonian realization on a $d$-dimensional hypersurface $\Sigma$ as a sum of mutually commuting projectors such that the ground state subspace is equal to the image of the corresponding ($d$+1)-dimensional partition function on the manifold $\Sigma \times [0,1]$, where the partition function is thought as an Hermitian projector. In this manuscript, we are interested in lattice Hamiltonian models that correspond to \emph{Yetter's homotopy 2-type} topological quantum field theories \cite{Yetter:1993dh, porter1998topological, martins2007yetter, Bullivant:2017sjz, Bullivant:2016clk}, whose input data are so-called (finite) \emph{strict 2-groups}. 

A strict 2-group can be presented in many equivalent ways. Most succinctly, it can be defined as a group object in the category of categories, i.e a \emph{strict monoidal category} where every object and every (1-)morphism is invertible. In practice, a more pedestrian definition in terms of \emph{crossed modules} is often used. Naturally, there is also a notion of \emph{weak 2-groups}, i.e. monoidal categories where every (1-)morphism is invertible and every object is \emph{weakly} invertible, but we focus on the strict version in this manuscript and postpone the weak case to a companion paper \cite{MCG}. Note furthermore that we could include an equivalence class in the cohomology of the classifying space of the strict 2-group as input of our model, but we choose not to in order to focus on the specificity of dealing with a higher gauge model. In the absence of cohomological twist, the model can be conveniently defined on \emph{cubulations} instead of \emph{triangulations}, which has the advantage of making the computations more readable.

\medskip \noindent
The main focus of this manuscript is the study of \emph{excitations} of strict 2-group gauge models, and more specifically their classification, where by excitation we mean a connected submanifold for which the energy is higher than the one of the ground state. Several equivalent approaches exist to tackle the question of classifying bulk excitations of a given topological model. The tube algebra approach, which is a generalization of Ocneanu's tube algebra \cite{ocneanu1994chirality, ocneanu2001operator}, is particularly intuitive and has proven very successful \cite{koenig2010quantum, Lan:2013wia, Aasen:2017ubm, DDR1,Delcamp:2016eya, bultinck2017anyons, Delcamp:2017pcw, Delcamp:2018efi, Bullivant:2019fmk}. More specifically, the authors showed in \cite{Bullivant:2019fmk} that for lattice Hamiltonian realizations of Dijkgraaf-Witten theory, it was possible to apply the \emph{tube algebra} approach in any dimensions in order to classify the excitations of the corresponding model. For instance, when applied to the case of bulk point-like excitations in (2+1)d and bulk loop-like excitations in (3+1)d, we can confirm that elementary excitations are classified by the simple modules of the \emph{twisted quantum double} algebra and the \emph{twisted quantum triple} algebra, respectively. In this manuscript, we propose a generalization of this approach to the case of strict 2-group gauge model in (3+1)d. 

In general, the tube algebra approach relies on two keys ideas: $(i)$ Properties of an excitation associated with a given submanifold are encoded into the \emph{boundary conditions} of its complementary open submanifold $\Sigma^{\rm o}$. $(ii)$ There exists an orientation-preserving diffeomorphism such that $(\partial \Sigma^{\rm o} \times [0,1]) \cup_{\partial \Sigma^{\rm o}} (\partial \Sigma^{\rm o} \times [0,1]) \simeq \partial \Sigma^{\rm o} \times [0,1]$. Crucially, it is always possible to extend this \emph{gluing} operation to a map on the ground state subspace of $\partial \Sigma^{\rm o} \times [0,1]$. This map in turn endows the corresponding Hilbert space with a finite-dimensional algebraic structure, whose simple modules classify the boundary conditions of $\Sigma^{\rm o}$, and thus the corresponding elementary excitations. We focus on the case of loop-like excitations in (3+1)d, in which case the relevant manifold to consider is the one obtained by cutting open the three-torus along one direction, i.e. $\mathbb T^2 \times [0,1]$. Note however that the strategy presented here is valid for more complicated excitations, associated with higher-genus boundary manifolds, and in higher dimensions. We comment on these more general scenarios at the end of the manuscript.

The derivation and study of the tube algebra for higher gauge models rely on similar techniques to conventional gauge models. However, there is a key distinguishing feature that can be appreciated via a simple geometrical remark: Given a manifold of the form $\partial \Sigma^{\rm o} \times [0,1]$, it is always possible to find a discretisation such that there are no vertices in the bulk, whereas the bulk must always contain at least one edge. Gauge models display a 0-form gauge invariance enforced at every vertex in the bulk of the discretised manifold. This implies that ground states on manifolds of the form $\partial \Sigma^{\rm o} \times [0,1]$ are simply labelled by flat (1-form) connections. In contrast, a strict 2-group gauge model displays a 0-form and a 1-form gauge invariance enforced at every vertex and edge in the bulk of the discretised manifold, respectively. In this case, ground states on manifolds of the form $\partial \Sigma^{\rm o} \times [0,1]$ are not simply labelled by strict 2-group flat connections but rather by equivalence classes of such connections. Physically, this translates into a confinement mechanism for the point-like charge excitations of the 0-form symmetry that are not invariant under the additional 1-form symmetry.

In (2+1)d, it is well-known that the number of elementary point-like excitations is equal to the ground state degeneracy on the two-torus. Similarly, for gauge models in (3+1)d, it can be shown that the number of elementary loop-like excitations equals the ground state degeneracy on the three-torus. This result can be established via a direct computation, or more elegantly, by demonstrating that the ground state subspace of the three torus is described by the central subalgebra of the tube algebra for torus boundary. We show that this statement generalizes to strict 2-group higher gauge models. 

\begin{center}
	\textbf{Organisation of the paper}
\end{center}
\noindent
In sec.~\ref{sec:Ham}, we define a lattice Hamiltonian model in (3+1)d whose input data is a strict 2-group. This model, which has a higher gauge theory interpretation, displays loop-like excitations. These loop-like excitations can be studied using the tube algebra approach. After briefly reviewing the general framework, we present in sec.~\ref{sec:tube} the explicit computation of the tube algebra for torus boundaries. The simple modules of the tube algebra are derived in sec.~\ref{sec:simple}. We then elucidate the physical interpretation of these simple modules as a classifying tool for the elementary bulk loop-like excitations of the higher gauge model.
In sec.~\ref{sec:torigs}, we utilise the tube algebra associated to loop-like excitations, demonstrating that the ground state subspace of the three-torus can be described by  the central subalgebra. We then deduce the ground state degeneracy corresponds to the number of elementary loop-like excitations.
The manuscript contains several appendices where technical details and proofs are relegated.

\newpage
\section{Higher gauge model\label{sec:Ham}}
\emph{In this section we introduce the model of interest in this manuscript. The input of this model, which has a higher gauge theory interpretation, is a strict 2-group. We first define the notion of strict 2-group connections on a cubulation, and then we construct the lattice Hamiltonian as a sum of mutually commuting projectors.}

\subsection{Strict 2-group connections}
Let $\Sigma$ be a closed oriented three-manifold endowed with a \emph{cubulation} $\Sigma_\boxempty$, which is a \emph{CW-complex} whose geometric realisation is homeomorphic to $\Sigma$. We require $\Sigma_\boxempty$ to be equipped with a complete ordering of the vertices. It follows that the one-skeleton of $\Sigma_\boxempty$ has the structure of a \emph{directed} graph such that each edge is oriented from the lowest to the highest vertex. 

The input for the model is a \emph{strict 2-group} $\mathcal{G}$. Succinctly, a strict 2-group can be defined as a \emph{strict monoidal category} where every object and every morphism is invertible, or via \emph{delooping} as a one-object \emph{2-groupoid} (see app.~\ref{sec:app_groupoid} for details). A more pedestrian definition is given in terms of  crossed modules \cite{whitehead1949combinatorial}: A \emph{crossed module} is a quadruple $\mathcal{G} \equiv (G, H, \partial, \act)$ which consists of two groups $G$ and $H$, a group homomorphism $\partial : H \to G$, and a group action of $G$ on $H$ by automorphisms $\act: G \times H \to H$ such that the so-called \emph{Peiffer identities} hold
\begin{align}
	\label{eq:Peiffer1}
	\partial(g \act h) &= g \partial(h)g^{-1} \\
	\label{eq:Peiffer2}
	\partial(h) \act h' &= hh'h^{-1} \; ,
\end{align}
for all $g \in G$ and $h,h' \in H$. Note furthermore that the action $\act$ fulfils the usual axioms
\begin{equation}
	\label{eq:axioms}
	\mathbb{1}_G \act h = h \; , \q g \act (g' \act h) = (gg') \act h \; , \q g \act (hh') = (g \act h)(g \act h') \; ,
\end{equation}
for all $g,g' \in G$ and $h,h' \in H$, where $\mathbb 1_G$ is the identity element in $G$. At this point it is worth mentioning that (\ref{eq:Peiffer1}--\ref{eq:Peiffer2}) imply that if either the action $\act$ is trivial or if $G$ is trivial, then $H$ must be \emph{abelian}.

 We define a strict 2-group \emph{connection} on $\Sigma_\boxempty$ as follows: To every directed edge $\mathsf{e} \subset \Sigma_\boxempty$ is assigned a $G$-valued {1-holonomy} $g_{\mathsf{e}}$. Every 1-holonomy has a source and a target vertex denoted by ${\rm s}(\mathsf{e})$ and ${\rm t}(\mathsf{e})$, respectively, such that 1-holonomies that have a matching source or target vertex can be composed. More specifically, given a 1-path of $\Sigma_\boxempty$, the corresponding 1-holonomy is defined as the oriented product from left to right of the 1-holonomies associated with the oriented edges along the path. Every plaquette $\mathsf{p} \subset \Sigma_\boxempty$ is assigned an $H$-valued {2-holonomy} $h_{\mathsf{p}}$. Every 2-holonomy has a source and a target 1-path denoted by ${\rm s}(\mathsf{p})$ and ${\rm t}(\mathsf{p})$, respectively. Crucially, 1- and 2-holonomies interact in a non-trivial way. Indeed, given a plaquette $\mathsf{p} \subset \Sigma_\boxempty$ decorated with a 2-holonomy $h_\mathsf{p}$, we assign to it the following 1-holonomy
\begin{equation}
	\label{eq:trivHol1}
	{\rm hol}_1(\mathsf{p}):= \partial(h_{\mathsf{p}})g_{{\rm s}(\mathsf{p})}g_{{\rm t}(\mathsf{p})}^{-1} \stackrel{!}{=} \mathbbm{1}_G\; ,
\end{equation}
which is required to be trivial, hence resulting in a constraint between 1- and 2-holonomies. Henceforth, this constraint will be enforced at every plaquette and will be referred to as the \emph{fake-flatness constraint}.

Both the source ${\rm s}(\mathsf{p})$ and the target ${\rm t}(\mathsf{p})$ 1-paths of a given 2-holonomy share the same source vertex referred to as the \emph{basepoint} ${\rm bp}(\mathsf{p})$ of the 2-holonomy.
For instance, let us consider the plaquette $\mathsf{p} \equiv \snum{(0123)} \subset \Sigma_\boxempty$ with the following assignment of 1- and 2-holonomies:
\begin{equation*}
	\plaq{g_{01}}{g_{02}}{g_{13}}{g_{23}}{1-2}{2-1}{h} \; ,
\end{equation*}
where the double arrow `$\Rightarrow$' is here to keep track of the source and target 1-paths of the 2-holonomy.
In this case, we have ${\rm bp}(\mathsf{p})=\snum{(0)}$, ${\rm s}(\mathsf{p}) = \snum{(02)} \cup \snum{(23)}$ and ${\rm t}(\mathsf{p}) = \snum{(01)} \cup \snum{(13)}$ such that the following fake-flatness constraint
\begin{equation}
	\label{eq:fake1}
	{\rm hol}_1(\mathsf{p})= \partial(h)g_{02}g_{23}(g_{01}g_{13})^{-1} \stackrel{!}{=} \mathbbm{1}_G
\end{equation}
is enforced. 

Such an assignment of 1- and 2-holonomies to the edges and the plaquettes of $\Sigma_\boxempty$ is referred to as a \emph{$\mathcal{G}$-labelling}, and it defines a local description of a strict 2-group $\mathcal{G}$-connection. Given a $\mathcal{G}$-labelling $\mathfrak{g} \equiv  (g,h)$, we notate by $g_{ab} \equiv \mathfrak{g}[ab]$ the restriction of $\mathfrak{g}$ to the edge $(ab) \subset \Sigma_\boxempty$ and by $h_{abcd} \equiv \mathfrak{g}[abcd]$ the restriction of $\mathfrak{g}$ to the plaquette $(abcd) \subset \Sigma_\boxempty$. When no confusion is possible, the subscripts of the 2-holonomies will often be omitted as in the diagram above.

Similarly to 1-holonomies, 2-holonomies can be composed. Given an ordered set of 2-holonomies such that the target 1-path of one coincides with the source 1-path of the following one, it is possible to compose them to define the 2-holonomy associated with the corresponding 2-path. More generally, given an arbitrary 2-path, there is a well-defined 2-holonomy associated to it. This follows from the fact that since condition \eqref{eq:trivHol1} is enforced, it is always possible to simultaneously change the source and target 1-paths of a plaquette 2-holonomy as well as its basepoint, as long as the corresponding  $H$-labelling is modified accordingly. For instance, one has the following relations
\begin{equation}
	\label{eq:whisker}
	\plaq{g_{01}}{g_{02}}{g_{13}}{g_{23}}{1-2}{2-1}{h}\; = \; \plaq{g_{01}}{g_{02}}{g_{13}}{g_{23}}{2-1}{1-2}{h^{-1}} \; = \;  
	\plaq{g_{01}}{g_{02}}{g_{13}}{g_{23}}{1-1}{2-2}{g_{01}^{-1} \act h} \; = \; \plaqSpe{g_{01}}{g_{02}}{g_{13}}{g_{23}}{g_{01}^{-1} \act h} \; ,
\end{equation}
where the basepoint in the first two diagrams and in the last two diagrams is $\snum{(0)}$ and $\snum{(1)}$, respectively. Relations of this kind are usually referred to as the \emph{whiskering rules}. We can check explicitly that these rules are such that \eqref{eq:trivHol1} is always preserved. For instance, the plaquette 1-holonomy in the third diagram reads
\begin{equation} 
	\partial(g_{01}^{-1}\act h)g_{01}^{-1}g_{02}(g_{13}g_{23}^{-1})^{-1} \stackrel{\eqref{eq:Peiffer1}}{=} g_{01}^{-1}\partial(h)g_{02}(g_{13}g_{23}^{-1})^{-1}
\end{equation}
which must be trivial given that the fake-flatness constraint \eqref{eq:fake1} is enforced. Applying the whiskering rules, it is always possible to compose 2-holonomies associated with adjacent plaquettes, and it was shown in \cite{Bullivant:2017sjz} that given a 2-path the corresponding 2-holonomy is uniquely defined. Let us for instance consider the cube $\mathsf{c} \equiv \snum{(01234567)} \subset \Sigma_\boxempty$ depicted below
\begin{equation*}
	 \cubeSans{} \!\! ,
\end{equation*}
we can show by means of the whiskering rules that the (closed) 2-holonomy associated with its boundary 2-path reads
\begin{align}
	\label{eq:cube2Hol}
	{\rm hol}_2(\mathsf{c}) =h_{0145} \, (g_{01} \act h_{1357}) \, h_{0123} \, (g_{02} \act h_{2367}^{-1}) \, h_{0246}^{-1} \, (g_{04} \act h_{4567}^{-1}) \; ,
\end{align}
where we made the choice that 2-holonomies multiply from right to left by convention. 

By defining strict 2-groups as one-object 2-groupoids, the definition of strict 2-group connections proposed above can be neatly recast as functors from the path 2-groupoid to the strict 2-group. We present this alternative approach in app.~\ref{sec:app_groupoid}.

\subsection{Lattice Hamiltonian}
Let us now define the higher gauge model. More details can be found in \cite{Bullivant:2016clk, Bullivant:2017sjz}. The microscopic Hilbert space $\mathcal{H}^\mathcal{G}[\Sigma_\boxempty]$ is spanned by graph-states $| \mathfrak{g} \ra$ where $\mathfrak{g}$ is a $\mathcal{G}$-labelling of $\Sigma_\boxempty$ as defined earlier. The lattice Hamiltonian is obtained as a sum of mutually commuting operators that come in three distinct classes: To every cube $\mathsf{c} \subset \Sigma_\boxempty$, we assign an operator $\mathbb{B}_{\mathsf{c}}$ whose action on a graph-state $| \mathfrak{g} \ra \in \mathcal{H}^\mathcal{G}[\Sigma_\boxempty] $ reads
\begin{equation}
	\label{eq:trivHol2}
	\mathbb{B}_\mathsf{c}|\mathfrak{g} \ra = \delta \big({\rm hol}_2(\mathsf{c}), \mathbbm{1}_H \big) |\mathfrak{g}  \ra \; ,
\end{equation}
where $\mathbb{1}_H$ is the identity element in $H$ and ${\rm hol}_2(\mathsf{c})$ is computed as in \eqref{eq:cube2Hol}. Such $\mathbb{B}$-operators penalise $\mathcal{G}$-labellings for which the 2-holonomies associated with every cube are not trivial. The corresponding constraint is known as the \emph{2-flatness constraint} and we refer to a $\mathcal{G}$-labelling that satisfies the 2-flatness constraint at every cube as a $\mathcal{G}$\emph{-colouring}. Such a $\mathcal{G}$-colouring constitutes a local description of a flat strict 2-group connection. The set of $\mathcal{G}$-colourings on $\Sigma_\boxempty$ is denoted by ${\rm Col}(\Sigma_\boxempty,\mathcal{G})$.\footnote{A flat strict 2-group connection can also be concisely defined as a homotopy $\gamma : \Sigma \to B\mG$ from $\Sigma$ to the classifying space of the 2-group as defined in \cite{brown2011nonabelian, martins2007yetter}. The classifying space $B \mG$ is such that only its first and second homotopy groups are non-vanishing such that non-trivial 1- and 2-holonomies can be found along non-contractible 1- and 2-cycles only.}

To every vertex $\mathsf{v} \subset \Sigma_\boxempty$, we assign an operator $\mathbb{A}_\mathsf{v} = 1/|G|\sum_{k \in G}\mathbb{A}^k_\mathsf{v}$ which enforces invariance under so-called \emph{0-form gauge transformations} via\footnote{Note that the first two terms are identical to the ones entering the definition of the gauge operator in Dijkgraaf-Witten models.}
\begin{equation}
	\label{eq:opAvertex}
	\mathbb{A}^k_\mathsf{v} = 
	\bigg(\bigotimes_{\mathsf{e}:{\rm t}(\mathsf{e})=\mathsf{v}}R_\mathsf{e}^k \bigg)
	\otimes\bigg(\bigotimes_{\mathsf{e}:{\rm s}(\mathsf{e})=\mathsf{v}}L_\mathsf{e}^k \bigg)
	\otimes 
	\bigg(\bigotimes_{\mathsf{p}:{\rm bp}(\mathsf{p})=\mathsf{v}}A_\mathsf{p}^k \bigg) \; ,
\end{equation} 
where $R^k_\mathsf{e}:g \mapsto gk^{-1}$, $L^k_{\mathsf{e}}:g \mapsto kg$ and $A^k_\mathsf{p}: h\mapsto k \act h$. The last term is here to ensure that the fake-flatness constraint \eqref{eq:trivHol1} commutes with the action of $\mathbb{A}_\mathsf{v}$. Applying definition \eqref{eq:opAvertex}, we have for instance
\begin{equation*}
	\mathbb{A}^k_{(0)} 
	\Bigg| \! \plaq{g_{01}}{g_{02}}{g_{13}}{g_{23}}{1-2}{2-1}{h} \!\! \Bigg\rangle =
	\Bigg| \! \plaq{kg_{01}}{kg_{02}}{g_{13}}{g_{23}}{1-2}{2-1}{k \act h} \!\! \Bigg\rangle \; .
\end{equation*}
We find that the plaquette 1-holonomy ${\rm hol}_1\snum{(0123)}$ transforms under the action of $\mathbb{A}^k_{(0)}$ as
\begin{align*}
	{\rm hol}_1\snum{(0123)} \stackrel{!}{=} \mathbb{1}_G \;  \mapsto \;
	&\partial(k \act h)kg_{02}g_{23}(kg_{01}g_{13})^{-1} \\
	& \stackrel{\eqref{eq:Peiffer1}}{=} k \partial(h)g_{02}g_{23}(g_{01}g_{13})k^{-1} = \mathbbm{1}_G 
\end{align*}
so that the fake-flatness constraint \eqref{eq:fake1} remains satisfied as expected.

Finally, to every edge $\mathsf{e} \subset \Sigma_\boxempty$, we assign an operator $\mathbb{A}_\mathsf{e} = 1/|H|\sum_{\lambda \in H}\mathbb{A}^\lambda_\mathsf{e}$ which enforces invariance under so-called \emph{1-form gauge transformations} via
\begin{equation}
	\label{eq:opAedge}
	\mathbb{A}^\lambda_\mathsf{e} = 
	\bigg(\bigotimes_{\mathsf{p}:{\rm s}(\mathsf{p}) \supset \mathsf{e}}R_\mathsf{p}^\lambda \bigg)
	\otimes\bigg(\bigotimes_{\mathsf{p}:{\rm t}(\mathsf{p}) \supset \mathsf{e}}L_\mathsf{p}^\lambda \bigg)
	\otimes 
	L_\mathsf{e}^{\partial(\lambda)} \; ,
\end{equation} 
where $R_\mathsf{p}^\lambda : h \mapsto h(g_{{\rm bp}(\mathsf{p}){\rm s}(\mathsf{e})} \act \lambda^{-1})$ and  $L_\mathsf{p}^\lambda : h \mapsto (g_{{\rm bp}(\mathsf{p}){\rm s}(\mathsf{e})} \act \lambda)h$, such that $g_{{\rm bp}(\mathsf{p}){\rm bp}(\mathsf{p})}= \mathbbm{1}_G$.
For instance, we have
\begin{align*}
	\mathbb{A}^\lambda_{(02)} 
	\Bigg| \! \plaq{g_{01}}{g_{02}}{g_{13}}{g_{23}}{1-2}{2-1}{h} \!\! \Bigg\rangle =
	\Bigg| \! \plaq{g_{01}}{\partial(\lambda)g_{02}}{g_{13}}{g_{23}}{1-2}{2-1}{h \lambda^{-1}} \!\! \Bigg\rangle
\end{align*}
and
\begin{align*}
	\mathbb{A}^\lambda_{(13)} \Bigg| \! \plaq{g_{01}}{g_{02}}{g_{13}}{g_{23}}{1-2}{2-1}{h} \!\! \Bigg\rangle =
	\Bigg| \! \plaq{g_{01}}{g_{02}}{\partial(\lambda)g_{13}}{g_{23}}{1-2}{2-1}{(g_{01}\act \lambda)h} \!\! \Bigg\rangle \; .
\end{align*}
As for the 0-form gauge transformations, we can check that the fake-flatness constraints remain satisfied under the action of $\mathbb{A}_{\mathsf e}$. For instance, the plaquette 1-holonomy ${\rm hol}_1\snum{(0123)}$ transforms under the action of $\mathbb{A}^\lambda_{(13)}$ as
\begin{align*}
	{\rm hol}_1\snum{(0123)} \stackrel{!}{=} \mathbb{1}_G \;  \mapsto \;
	&\partial \big((g_{01}\act \lambda)h\big)
	g_{02}g_{23}(g_{01}\partial(\lambda)g_{13})^{-1} \\ &\stackrel{\eqref{eq:Peiffer1}}{=}  g_{01}\partial(\lambda)g_{01}^{-1}\partial(h)g_{02}g_{23}g_{13}^{-1}\partial(\lambda^{-1})g_{01}^{-1} = \mathbbm{1}_G 
\end{align*}
so that condition \eqref{eq:trivHol1} still holds as expected.

It was shown in \cite{Bullivant:2017sjz} that all the operators commute and the lattice Hamiltonian reads
\begin{equation}
	\label{eq:ham}
	\mathbb{H}^\mathcal{G}[\Sigma_\boxempty] = - \sum_{\mathsf{v}\subset \Sigma_\boxempty}\mathbb{A}_\mathsf{v}- \sum_{\mathsf{e}\subset \Sigma_\boxempty}\mathbb{A}_\mathsf{e}
	- \sum_{\mathsf{c}\subset \Sigma_\boxempty}\mathbb{B}_\mathsf{c} \; ,
\end{equation}
where the sums run over all the vertices, edges and cubes in $\Sigma_\boxempty$, respectively.

\subsection{Ground state subspace\label{sec:ground}}

The ground state subspace $\mathcal{V}^\mathcal{G}[\Sigma_\boxempty]$ of the lattice Hamiltonian $\mathbb{H}^\mathcal{G}[\Sigma_\boxempty]$ defined in \eqref{eq:ham} is spanned by linear combinations of $\mathcal{G}$-labelled graph-states on $\Sigma_\boxempty$ that satisfy the \emph{stabiliser constraints} $\mathbb{B}_\mathsf{c}|\psi \ra = | \psi \ra$, $\mathbb{A}_\mathsf{v}|\psi \ra = | \psi \ra$ and $\mathbb{A}_\mathsf{e}|\psi \ra = | \psi \ra$ for every $\mathsf{c}, \mathsf{v}, \mathsf{e} \subset \Sigma_\boxempty$. In the following, we will need the corresponding ground state projector, namely
\begin{equation}
	\label{eq:proj}
	\mathbb{P}^\mathcal{G}[\Sigma_\boxempty] = \bigg( \bigotimes_{\mathsf{v}\subset \Sigma_\boxempty}\mathbb{A}_\mathsf{v}\bigg) \otimes \bigg(\bigotimes_{\mathsf{e}\subset \Sigma_\boxempty}\mathbb{A}_\mathsf{e}\bigg)
	\otimes \bigg(\bigotimes_{\mathsf{c}\subset \Sigma_\boxempty}\mathbb{B}_\mathsf{c}\bigg) 
\end{equation}
such that ${\rm Im} \, \mathbb P ^\mathcal{G}[\Sigma_\boxempty] = \mathcal{V}^\mathcal{G}[\Sigma_\boxempty]$.

It was demonstrated in \cite{Bullivant:2016clk} that the model in question corresponds to the Hamiltonian realisation of the \emph{Yetter homotopy 2-type topological theory} \cite{yetter1993tqft}. In particular, this relation is realised by the observation that the ground state projector for a given discretised three-manifold $\Sigma_{\boxempty}$ can be identified with the Yetter topological partition function applied to the space-time manifold $\Sigma_{\boxempty}\times\mathbb{I}$. As direct consequence, we can identify the ground state subspace $\mathcal{V}^{\mG}[\Sigma_{\boxempty}]$ as defined by the model with the Hilbert space the partition function assigns to $\Sigma_{\boxempty}$. 

The fact that the ground state subspace $\mathcal{V}^\mathcal{G}[\Sigma_\boxempty]$ is described by a topological field theory manifests itself upon performing changes of cubulations. Indeed, given two cubulations $\Sigma_\boxempty$ and $\Sigma_{\boxempty'}$ of $\Sigma$, the corresponding ground state subspaces are isomorphic, i.e. $\mathcal{V}^\mathcal{G}[\Sigma_\boxempty] \simeq \mathcal{V}^\mathcal{G}[\Sigma_{\boxempty'}]$. This signifies that it is possible to perform local changes of the cubulation while remaining in the ground state sector. Such modifications of the underlying cubulation are performed by local unitary transformations that are discrete implementations of a gap-preserving adiabatic evolution \cite{Chen:2010gda}. This means for instance that a ground state defined on a cubulation made of two adjacent plaquettes is isomorphic to another ground state defined on the `merger' of these two plaquettes, i.e.
\begin{equation}
	\label{eq:iso}
	\Bigg| \! \plaqDouble{} \!\! \Bigg\rangle
	\simeq \frac{1}{|G||H|^\frac{1}{2}}
	\Bigg| \! \plaqTr{g_{01}}{g_{02}g_{24}}{g_{13}g_{35}}{g_{45}}{1-2}{2-1}{h(g_{02}\act h')} \!\! \Bigg\rangle \; ,
\end{equation}
where the factor $|G||H|^{\frac{1}{2}}$ ensures that the isomorphism preserves the normalisation of states.\footnote{Such factors can be induced by the requirement that the normalisation of the states is preserved under such isomorphisms, or equally by considering the isomorphism as a cobordism operator in the corresponding Yetter homotopy 2-type topological theory.}
Since isomorphisms of this form play a crucial role in the following, let us explain it in more detail: The 1-holonomies associated with the edges $\snum{(04)}$ and $\snum{(15)}$ are provided by the oriented product of the 1-holonomies along the 1-paths $\snum{(02)} \cup \snum{(24)}$ and $\snum{(13)} \cup \snum{(35)}$, respectively, namely $g_{02}g_{24}$ and $g_{13}g_{35}$. Similarly, the 2-holonomy that labels the plaquette $\snum{(0145)}$ is obtained as the composition (from right to left) of the 2-holonomies associated with the initial plaquettes $\snum{(0123)}$ and $\snum{(2345)}$ such that ${\rm bp}\snum{(0123)} = \snum{(0)}$, ${\rm s}\snum{(0123)} = \snum{(02)} \cup \snum{(23)}$, ${\rm t}\snum{(0123)} = \snum{(01)} \cup \snum{(13)}$ and ${\rm bp}\snum{(2345)} = \snum{(2)}$, ${\rm s}\snum{(2345)} = \snum{(24)} \cup \snum{(45)}$, ${\rm t}\snum{(2345)} = \snum{(23)} \cup \snum{(35)}$. The two 2-holonomies not having the same basepoint, they cannot be composed right away. It is thus necessary to make use of the whiskering rules \eqref{eq:whisker} so that the 2-holonomy on the right has vertex $\snum{(0)}$ as basepoint. This requires modifying the corresponding $H$-labelling by action of $g_{02}$. It remains to modify simultaneously the source and target 1-paths of the 2-holonomy on the left to $\snum{(02)} \cup \snum{(23)} \cup \snum{(35)}$ and $\snum{(01)} \cup \snum{(13)} \cup \snum{(35)}$, respectively, which does not require a modification of the corresponding $H$-labelling. At this point, the two 2-holonomies have matching source and target 1-paths so that they can be composed. The resulting 2-holonomy is labelled by $h(g_{02} \act h')$ and is such that ${\rm bp}\snum{(0145)} = \snum{(0)}$, ${\rm s}\snum{(0145)} = \snum{(04)} \cup \snum{(45)}$ and ${\rm t}\snum{(0145)} = \snum{(01)} \cup \snum{(15)}$.

\newpage
\section{Tube algebra for loop-like excitations\label{sec:tube}}

\emph{In this section, we derive the algebraic structure underlying the loop-like excitations of the higher gauge model following the tube algebra approach.}

\subsection{Formal definition}
Given a closed oriented three-manifold $\Sigma$ endowed with a cubulation $\Sigma_\boxempty$, we defined in \eqref{eq:ham} the lattice Hamiltonian $\mathbb{H}^\mG[\Sigma_\boxempty]$ whose ground state subspace is denoted by $\mathcal{V}^\mG[\Sigma_\boxempty]$.  An \emph{excitation} in such model is defined as a connected subcomplex of $\Sigma_\boxempty$ for which some of the stabiliser constraints are violated so that it has an overall \emph{energy density} higher than that of the ground state. There are several equivalent approaches to study such excitations in a systematic way. In this paper, we follow the so-called \emph{tube algebra approach}. This approach relies on the following key concept: Properties of an excitation associated with a given subcomplex are encoded into the \emph{boundary conditions} of its complementary open subcomplex. This signifies that the classification of boundary conditions induces a classification of the excitation content of the model. 

Let us consider an excitation associated with a given subcomplex of $\Sigma_\boxempty$. We denote by $\Sigma^{\rm o}_\boxempty$ the open manifold obtained by removing this subcomplex. We are interested in the lattice Hamiltonian $\mathbb{H}^\mG[\Sigma^{\rm o}_\boxempty \backslash \partial \Sigma^{\rm o}_\boxempty]$ as defined in \eqref{eq:ham} where the sums now run over all the vertices, edges and cubes in the \emph{interior} of $\Sigma^{\rm o}_\boxempty$. This lattice Hamiltonian presents so-called \emph{open boundary conditions} since graph-states with different $\mathcal G$-colourings on $\partial \Sigma^{\rm o}_\boxempty$ are not mixed. In this case, the corresponding ground state subspace admits a decomposition in terms of boundary $\mG$-colourings:
\begin{equation}
	\label{eq:boundDecomposition}
	\mathcal{V}^\mG[\Sigma^{\rm o}_\boxempty] = \bigoplus_{\mathfrak{a} \in {\rm Col}(\partial \Sigma^{\rm o}_\boxempty,\mG)}\mathcal{V}^\mG[\Sigma^{\rm o}_\boxempty ]_\mathfrak{a}
\end{equation}
where $\mathcal{V}^\mG[\Sigma^{\rm o}_\boxempty]_{\mathfrak a}$ is the ground state subspace spanned by graph-states with boundary $\mG$-colouring ${\mathfrak a}$. It follows that a state with a given boundary colouring defines a specific excitation, which is a superposition of so-called \emph{elementary} excitations. The elementary excitations can then be found as the irreducible modules of the corresponding \emph{tube algebra}. 

Given an open manifold $\Sigma^{\rm o}_\boxempty$, we define by $\mathfrak{T}[\partial \Sigma_\boxempty^{\rm o}]$ a cubulation of $\partial \Sigma_\boxempty^{\rm o} \times \mathbb I \equiv \partial \Sigma_\boxempty^{\rm o} \times \mathbb [0,1]$ such that $\partial \mathfrak{T}[\partial \Sigma^{\rm o}_\boxempty] = \overline{\partial \Sigma^{\rm o}_\boxempty} \sqcup \partial \Sigma^{\rm o}_\boxempty$. Naturally, we can always glue a copy of $\mathfrak{T}[\partial \Sigma_\boxempty^{\rm o}]$ to $\Sigma_\boxempty^{\rm o}$ along $\partial \Sigma_\boxempty^{\rm o}$ without affecting its topology, i.e. $\mathfrak{T}[\partial \Sigma^{\rm o}_\boxempty] \cup \Sigma^{\rm o}_\boxempty \simeq \Sigma^{\rm o}_\boxempty$. As shown in \cite{Bullivant:2019fmk}, this gluing operation can be extended to a \emph{symmetry} of the ground state subspace $\mathcal{V}^\mG[\Sigma^{\rm o}_\boxempty]$. It follows from the discussion in sec.~\ref{sec:ground} that it is always possible to perform cubulation changes so as to find a representative $\mathcal{V}^\mG[\Sigma^{\rm o}_{\boxempty '}]$ isomorphic to $\mathcal{V}^\mG[\Sigma^{\rm o}_{\boxempty}]$ whereby a local neighbourhood of $\partial \Sigma^{\rm o}_{\boxempty}$ is of the form $\mathfrak{T}[\partial \Sigma^{\rm o}_\boxempty]$. This can then be used to localise the action of this ground state subspace symmetry in such way that it only involves degrees of freedom contained within $\mathfrak{T}[\partial \Sigma^{\rm o}_\boxempty]$. Boundary configurations for $\partial \Sigma^{\rm o}_\boxempty$ can then be classified by the irreducible modules of the symmetry map  on $\mathcal{V}^\mG[\mathfrak{T}[\partial \Sigma^{\rm o}_\boxempty]]$ associated with the gluing operation $\mathfrak{T}[\partial \Sigma^{\rm o}_\boxempty] \cup \mathfrak{T}[\partial \Sigma^{\rm o}_\boxempty] \simeq \mathfrak{T}[\partial \Sigma^{\rm o}_\boxempty]$. Let us now construct this map. 

We are interested in classifying loop-like excitations for the higher gauge model \eqref{eq:ham}, where by loop-like excitations we mean an excitation whose topology is given by the circle $\mathbb{S}^1$. Given such a loop embedded in a three-manifold, its regular neighbourhood is provided by a solid two-torus $\mathbb{D}^2 \times \mathbb S ^1$, so that loop-like excitations can be classified in terms of boundary conditions of the torus $\mathbb{T}^2 = \partial(\mathbb D^2 \times \mathbb S^1)$. We shall therefore construct the map on $\mathcal{V}^\mG[\mathfrak{T}[\mathbb T^2_\boxempty]]$ associated with the gluing operation $\mathfrak{T}[\mathbb T^2_\boxempty] \cup \mathfrak{T}[\mathbb T^2_\boxempty] \simeq \mathfrak{T}[\mathbb T^2_\boxempty]$.

It follows from \eqref{eq:boundDecomposition} that the ground state subspace on $\mathfrak{T}[\mathbb T^2_\boxempty]$ satisfies
\begin{equation*}
	\mathcal{V}^\mG[\mathfrak T[\mathbb T^2_\boxempty]] = \bigoplus_{\substack{\mathfrak a \in {\rm Col}(\mathbb T^2_\boxempty \times \{0\}, \mG) \\ \mathfrak b \in {\rm Col}(\mathbb T^2_\boxempty \times \{1\}, \mG)}} 	\mathcal{V}^\mG[\mathfrak T[\mathbb T^2_\boxempty]]_{\mathfrak a,\mathfrak b} \; .
\end{equation*}
We want to construct a \emph{gluing} map for two states of such ground state subspace. Firstly, we define a map that identifies boundary conditions along the gluing interface:
\begin{equation*}
	\mathfrak{G} : 
	\mathcal{V}^\mG[\mathfrak{T}[\mathbb T^2_\boxempty]] \otimes
	\mathcal{V}^\mG[\mathfrak{T}[\mathbb T^2_\boxempty]]
	\to 
	\bigoplus_{\substack{\mathfrak{a},\mathfrak{a}' \in {\rm Col}(\mathbb T^2_\boxempty \times \{0\}, \mathcal G) \\ \mathfrak{b},\mathfrak{b}' \in {\rm Col}(\mathbb T^2_\boxempty \times \{1\}, \mathcal G)}} 
	\mathcal{V}^\mG[\mathfrak{T}[\mathbb T^2_\boxempty]]_{\mathfrak a, \mathfrak b} \otimes
	\mathcal{V}^\mG[\mathfrak{T}[\mathbb T^2_\boxempty]]_{\mathfrak a', \mathfrak b'} 
	\subseteq   
	\mathcal{H}^\mG[\mathfrak{T}[\mathbb T^2_\boxempty] \cup \mathfrak{T}[\mathbb T^2_\boxempty]]
\end{equation*}
such that
\begin{equation*}
\begin{array}{cccccc}
	\mathfrak{G} : &
	\mathcal{V}^\mG[\mathfrak{T}[\mathbb T^2_\boxempty]]_{\mathfrak a,\mathfrak b} \otimes
	\mathcal{V}^\mG[\mathfrak{T}[\mathbb T^2_\boxempty]]_{\mathfrak a', \mathfrak b'} & \rightarrow &  
	\mathcal{V}^\mG[\mathfrak{T}[\mathbb T^2_\boxempty]]_{\mathfrak a,\mathfrak b} \otimes
	\mathcal{V}^\mG[\mathfrak{T}[\mathbb T^2_\boxempty]]_{\mathfrak a', \mathfrak b'} \\
	{} & | \psi_{\mathfrak a,\mathfrak b} \ra \otimes | \varphi_{\mathfrak a',\mathfrak b'} \ra & \mapsto &  \hspace{-1.5em} \delta_{\mathfrak b,\mathfrak a'}|\psi_{\mathfrak a,\mathfrak b} \ra \otimes | \varphi_{\mathfrak b,\mathfrak b'} \ra \; .
\end{array}
\end{equation*}
Crucially, the image of $\mathfrak{G}$ differs from the ground state subspace $\mathcal{V}^\mG[\mathfrak{T}[\mathbb T^2_\boxempty] \cup \mathfrak{T}[\mathbb T^2_\boxempty]]$ because some stabiliser constraints may be violated along the \emph{gluing interface}. However, these constraints can be enforced by means of the projection operator $\mathbb{P}^\mG[\mathfrak{T}[\mathbb T^2_\boxempty] \cup \mathfrak{T}[\mathbb T^2_\boxempty]]$ as defined in \eqref{eq:proj}. Once all the constraints are enforced, it is possible to find a cubulation changing unitary isomorphism between $\mathcal{V}^\mG[\mathfrak{T}[\mathbb T^2_\boxempty] \cup \mathfrak{T}[\mathbb T^2_\boxempty]]$ and $\mathcal{V}^\mG[\mathfrak{T}[\mathbb T^2_\boxempty]]$. Putting everything together, we define the gluing map
\begin{equation*}
	\star : 
	\mathcal{V}^\mG[\mathfrak{T}[\mathbb T^2_\boxempty]] \otimes
	\mathcal{V}^\mG[\mathfrak{T}[\mathbb T^2_\boxempty]]
	\xrightarrow{\, \mathfrak{G} \,} 
	\mathcal{H}^\mG[\mathfrak{T}[\mathbb T^2_\boxempty] \cup \mathfrak{T}[\mathbb T^2_\boxempty]]
	\xrightarrow{\mathbb{P}^\mG[\mathfrak{T}[\mathbb T^2_\boxempty] \cup \mathfrak{T}[\mathbb T^2_\boxempty]]}
	\mathcal{V}^\mG[\mathfrak{T}[\mathbb T^2_\boxempty] \cup \mathfrak{T}[\mathbb T^2_\boxempty]]
	\xrightarrow{\, \sim \,} 	\mathcal{V}^\mG[\mathfrak{T}[\mathbb T^2_\boxempty]]  \; ,
\end{equation*}
which endows $\mathcal{V}^\mG[\mathbb T^2_\boxempty]$ with a finite-dimensional algebra structure denoted by ${\rm Tube}^\mG[\mathbb T^2_\boxempty]$. This algebra can be shown to be an \emph{associative semi-simple $*$-algebra}. It follows from the discussion above that the irreducible modules of ${\rm Tube}^\mG[\mathbb T^2_\boxempty]$ classify the elementary loop-like excitations of the model. 

\subsection{Ground states of the tube\label{sec:gsTube}}

We explained above that the elementary loop-like excitations of the model \eqref{eq:ham} can be classified by the simple modules of the tube algebra ${\rm Tube}^\mG[\mathbb T^2_\boxempty]$. Crucially, the choice of discretisation for the manifold $\mathbb T^2 \times \mathbb I$ does not matter. More precisely, given a cubulation $\mathbb T^2_\boxempty$ of $\mathbb T^2$, different choices for $\mathfrak{T}[\mathbb T^2_\boxempty]$ yield isomorphic algebras, while different  choices of boundary cubulations yield \emph{Morita equivalent} algebras. It follows from the definition of isomorphic algebras and Morita equivalent algebras that the classification of simple modules is independent of the choice of both bulk and boundary cubulations. Therefore, we shall make a choice that makes carrying-out the computations explicitly as straightforward as possible.

We choose to cubulate $\mathbb T^2$ as a plaquette with opposite edges identified. A cubulation $\mathfrak{T}[\mathbb T^2_\boxempty]$ of $\mathbb T^2_\boxempty \times \mathbb I$ is then obtained as a cube whose opposite edges and opposite faces are identified. We then consider the space of  $\mG$-coloured graph-states on $\mathfrak{T}[\mathbb T^2_\boxempty]$ of the form
\begin{align}
	\label{eq:colouredTube}
	{\rm Span}_{\mathbb C} 
	\Bigg\{ \Bigg| \!\! \cube{\gx}{\gy}{\gz}{\hx}{\hy}{\hz} \! \Bigg\rangle \Bigg\}_{\substack{\hspace{-2.5em} \forall \, \gx,\gy,\gz \in G \\ \hspace{-3.3em} \forall \, \hy, \hz \in H \\[-0.25em] \forall \, \hx \in H \, | \, \gz = \gz^{\gy \,;\, \hx}}} \!\!\!\!
	 =:  {\rm Span}_{\mathbb C} \Big\{ \Big| \,
	\tubeS{\hx}{\gy}{\gz}{\gx}{\hz}{\hy}
	\Big\ra \Big\}_{\substack{\hspace{-2.5em} \forall \, \gx,\gy,\gz \in G \\ \hspace{-3.3em} \forall \, \hy, \hz \in H \\[-0.25em] \forall \, \hx \in H \, | \, \gz = \gz^{\gy \,;\, \hx}}}  ,
\end{align}
where we introduced the shorthand notation 
\begin{equation*}
	g^{a \, ; \,b} := a^{-1}\partial(b^{-1})ga \; .
\end{equation*} 
In the definition above, we made the following identifications:
\begin{align*}
		&\mathfrak{g}[\snum{02}] \equiv \mathfrak{g}[\snum{46}] \equiv \mathfrak{g}[\snum{13}] \equiv \mathfrak{g}[\snum{57}] = \gx
		\; , \q
		\mathfrak{g}[\snum{01}] \equiv \mathfrak{g}[\snum{45}] = \gy
		\; , \q
		\mathfrak{g}[\snum{04}] \equiv \mathfrak{g}[\snum{15}] = \gz
		\; , \\
		&\mathfrak{g}[\snum{23}] \equiv \mathfrak{g}[\snum{67}] =  \gx^{-1}\partial(\hz^{-1})\gy\gx
		\; , \q
		\mathfrak{g}[\snum{26}] \equiv \mathfrak{g}[\snum{37}] =  \gx^{-1}\partial(\hy^{-1})\gz\gx \; ,
\end{align*}
where the last two $G$-labellings  are deduced from the fake-flatness constraints, and
\begin{align*}
		&\mathfrak{g}[\snum{0145}] = \hx \; , \q
		\mathfrak{g}[\snum{0123}] \equiv \mathfrak{g}[\snum{4567}] = \hy \; , \q \mathfrak{g}[\snum{0246}] \equiv \mathfrak{g}[\snum{1357}] = \hz \; , \\
		&\mathfrak{g}[\snum{4567}] = \gx^{-1} \act [\hy^{-1} \, (g_z \act \hz^{-1}) \, \hx \, (\gy \act \hy) \, \hz] \; ,
\end{align*}
where the last $H$-labelling is deduced from the 2-flatness constraint. Henceforth, we refer to $(\gy,\gz,\hx) \in G^2 \times H$ such that $\gz = \gz^{\gy \, ; \, \hx}$ as a \emph{boundary} $\mG$-colouring,  and $(\gx,\hy,\hz) \in G \ltimes H^2$ as a \emph{bulk} $\mG$-colouring of $\mathfrak{T}[\mathbb T^2_\boxempty]$. 

In order to obtain the ground states on $\mathfrak{T}[\mathbb T^2_\boxempty]$, we are left to enforce the 1-form gauge invariance along $\snum{(02)} \equiv \snum{(13)} \equiv \snum{(57)} \equiv \snum{(46)}$ via the projector $\mathbb{A}_\mathsf{e}$ as defined in \eqref{eq:opAedge}. However, in order for the ground states to be normalised to unity, we need to analyse beforehand the sets of bulk and boundary $\mG$-colourings. Let us first consider the set of boundary $\mG$-colourings, namely
\begin{equation*}
	{\rm Col}(\mathbb T^2_\boxempty \times \{0\}, \mG) = \big\{(\gy ,\gz , \hx) \in G^2 \times H  \; \big| \; \partial(\hx)\gy \gz (\gz \gy)^{-1} = \mathbbm{1}_G \big\} \; .
\end{equation*}
We define an equivalence relation on ${\rm Col}(\mathbb T^2_\boxempty \times \{0\}, \mG)$ given by
\begin{equation}
	\label{eq:equivBdry}
	(\gy ,\gz , \hx) \equivSpe{I}{\gy}{\gz}(\tgy , \tgz , \thx)
\end{equation}	
if there exists $(a , b_1 , b_2) \in G \ltimes H^2$ such that
\begin{align*}
	(\tgy , \tgz , \thx) = (\gy^{a \,;\, b_2} , \gz^{a \,;\, b_1} ,\hx^{a,\gy,\gz \,;\, b_1,b_2}) \; ,
\end{align*} 
where we introduced the shorthand notation
\begin{equation*}
	h^{a_1,a_2,a_3 \,;\, b_1,b_2} := a_1^{-1} \act [b_1^{-1}(a_3 \act b_2^{-1})h(a_2 \act b_1)b_2] \; .
\end{equation*} 
Equivalence classes with respect to such relation define a partition of ${\rm Col}(\mathbb T^2_\boxempty \times \{0\}, \mG)$ into disjoint subsets of boundary $\mG$-colourings. Let $\mC$ be such an equivalence class. In the following, we notate the elements in $\mC$ by 
\begin{equation*}
	(\cyi , \czi , \dxi) \; , \q  i=1,\ldots,|\mC| 
\end{equation*}
and we call  $(c_{{\rm y},1} , c_{{\rm z},1} , d_{\hat{\rm x},1})$ the \emph{representative} element.

So far we have a partition of the set of boundary $\mG$-colourings into equivalence classes $\mC \subset {\rm Col}(\mathbb T^2_\boxempty \times \{0\}, \mG)$ with respect to \eqref{eq:equivBdry}. In light of this statement, let us now analyse the set of bulk $\mG$-colourings. Letting $(\gy,\gz,\hx)$ be a boundary $\mG$-colouring, the set of bulk $\mG$-colourings is $G \ltimes H^2$. But, the 1-form gauge invariance along the edge $\snum{(02)} \equiv \snum{(13)} \equiv \snum{(57)} \equiv \snum{(46)}$ defines an equivalence relation on $G \ltimes  H^2$ where
\begin{equation}
	\label{eq:equivTube}
	(\gx  ,  \hy  ,  \hz )  \equivSpe{II}{\gy}{\gz}	(\tgx  ,  \thy  , \thz )
\end{equation} 
if there exists $\lambda \in H$ such that 
\begin{align*}	
	(\tgx  ,  \thy  , \thz ) = \big(\partial(\lambda)\gx \, ,\, (\gz \act \lambda)\hy \lambda^{-1} \, , \,(\gy \act \lambda)\hz \lambda^{-1}\big) \; .
\end{align*}
Similarly to \eqref{eq:equivBdry}, which define a partition of the set of boundary $\mG$-colourings, the equivalence relation \eqref{eq:equivTube} defines a partition of $G \ltimes H^2$ into disjoint subsets of bulk $\mG$-colourings. The corresponding set of equivalence classes (or orbits) is denoted by
\begin{equation}
	\label{eq:defO}
	\mathfrak{B}_{\gy,\gz} := G \ltimes H^2 / \equivSpe{II}{\gy}{\gz} \; .
\end{equation}
Let $\mE_{\gy,\gz} \subset G \ltimes H^2$ be such an equivalence class. We notate the elements in $\mE_{\gy,\gz}$ by
\begin{equation}
	(\exi , \fyi , \fzi) \; , \q  i=1,\ldots,|\mE_{\gy,\gz}| 
\end{equation}
and we call  $(e_{{\rm x},1} , f_{\hat{\rm y},1} , f_{\hat{\rm z},1})$ the \emph{representative} element. We then define the corresponding \emph{stabiliser} as 
\begin{equation}
	\label{eq:defZO}
	Z_{\mE_{\gy,\gz}} := \big\{ \lambda \in H \; \big| \;
	(e_{{\rm x},1}  ,  f_{\hat{\rm y},1}  , f_{\hat{\rm z},1} ) = \big(\partial(\lambda)e_{{\rm x},1} \, ,\, (\gz \act \lambda)f_{\hat{\rm y},1} \lambda^{-1} \, , \,(\gy \act \lambda)f_{\hat{\rm z},1} \lambda^{-1}\big) \big\}  \; .
\end{equation}
Crucially, the size of this stabiliser only depends on the equivalence class $\mC \ni (\gy,\gz,\hx)$ of boundary $\mG$-colourings. More precisely, given an equivalence class  $\mC$ of boundary $\mG$-colourings, two elements $(\cyi,\czi,\dxi), (\cyj,\czj,\dxj) \in \mC$, and two equivalence classes $\mE_{\cyi,\czi}$, $\mE'_{\cyj,\czj} \subset G \ltimes H^2$ of bulk $\mG$-colourings whose representative elements are $(e_{{\rm x},1}, f_{\hat{\rm y},1} , f_{\hat{\rm z},1})$ and $(e'_{{\rm x},1}, f'_{\hat{\rm y},1} , f'_{\hat{\rm z},1})$, respectively, we have $|Z_{\mE_{\cyi,\czi}}| = |Z_{\mE'_{\cyj,\czj}}|$. 

The statement above can be proven by showing explicitly that the aforementioned centralisers are isomorphic. By definition of $\mC$, we know there exists $(a,b_1,b_2) \in G \ltimes H^2$ such that
\begin{equation*}
	\label{eq:aux1}
	\cyj = \cyi^{a \, ; \, b_2}  = a^{-1}\partial(b_2^{-1})\cyi a \q {\rm and} \q
	\czj = \czi^{a \, ; \, b_1} = a^{-1}\partial(b_1^{-1})\czi a \; .
\end{equation*}
Moreover, if $\lambda_i \in Z_{\mE_{\cyi,\czi}}$, then $\lambda_i \in H$ and
\begin{equation*}
	\label{eq:aux2}
	\partial(\lambda_i) e_{{\rm x},1} =  e_{{\rm x},1} \; , \q
	(\czi \act \lambda_i) f_{\hat{\rm y},1} \lambda_i^{-1} =  f_{\hat{\rm y},1} \; ,\q
	(\cyi \act \lambda_i) f_{\hat{\rm z},1} \lambda_i^{-1} =  f_{\hat{\rm z},1} \; . 
\end{equation*}
From the equations above, we can deduce several constraints on $\lambda_i$: The first equation informs us that $\lambda_i \in {\rm Ker}\, \partial_i$. But it follows from the second Peiffer identity, that any element in ${\rm Ker} \, \partial_i$ commutes with every element in $H$. Together with the remaining two conditions above, this implies that $\cyi \act \lambda_i = \lambda_i$ and $\czi \act \lambda_i = 
\lambda_i$. Conversely, it is easy to check that if $\lambda_i$ satisfies these constraints, then it belongs to $Z_{\mE_{\cyi,\czi}}$. Consequently, we already know that $Z_{\mE_{\cyi,\czi}}$ depends only on $\cyi, \czi$, and not a specific choice of equivalence class $\mE_{\cyi,\czi}$. Furthermore, given $\lambda_i \in Z_{\mE_{\cyi,\czi}}$, it is possible to construct a group element in $Z_{\mE'_{\cyj,\czj}}$. 
Defining $\lambda'_j := a^{-1} \act \lambda_i$, we easily check that
\begin{equation*}
	\partial(\lambda'_j)  e_{{\rm x},1}' = \partial(a^{-1} \act \lambda_i)  e'_{{\rm x},1} = a^{-1}\partial(\lambda_i)a e_{{\rm x},1}' = e_{{\rm x},1}'
\end{equation*}
and 
\begin{align*}
	(\czj \act \lambda'_j) f'_{\hat{\rm y},1} \lambda'^{-1}_j 
	& \; = \; \big( [a^{-1}\partial(b_1^{-1})\czi a] \act (a^{-1} \act \lambda_i)  \big)  f'_{\hat{\rm y},1} (a^{-1} \act \lambda_i^{-1}) \\
	&\stackrel{\mathmakebox[\widthof{\;=\;}]{\eqref{eq:axioms}}}{=} a^{-1} \act [\partial(b_1^{-1}) \act (\czi \act \lambda_i)]  f'_{\hat{\rm y},1} (a^{-1} \act \lambda_i^{-1}) \\
	&\stackrel{\mathmakebox[\widthof{\;=\;}]{\eqref{eq:Peiffer2}}}{=} a^{-1} \act (b_1^{-1}\lambda_i b_1) f'_{\hat{\rm y},1} (a^{-1} \act \lambda_i^{-1}) =  f'_{\hat{\rm y},1} \; ,
\end{align*}
where we used the fact that $\lambda_i \in {\rm Ker }\, \partial$ and  $\czi \act \lambda_i = \lambda_i$. Similarly we find $(\cyj \act \lambda'_j) f'_{\hat{\rm z},1} \lambda'^{-1}_j= f'_{\hat{\rm z},1}$. All the relations above are invertible and thus $Z_{\mE_{\cyi,\czi}} \simeq Z_{\mE'_{\cyj,\czj}}$ so that $|Z_{\mE_{\cyi,\czi}}| =| Z_{\mE'_{\cyj,\czj}} |$. In summary, given a boundary $\mG$-colouring $(\cyi,\czi,\dxi)$ and an equivalence class $\mE_{\cyi,\czi}$, the centralizer $Z_{\mE_{\cyi,\czi}}$ only depends on the equivalence class $\mC$ that contains $(\cyi,\czi,\dxi)$ so that
\begin{align*}
	 |Z_{\mE_{\cyi,\czi}}| =  |Z_{\mE_{c_{{\rm y},1},c_{{\rm z},1}}}|  =: |Z_{\mE_\mC}| \; &, \q  
	\forall \, i=1,\ldots,|\mC|  \\
	{\rm and} \q |Z_{\mE_\mC}| = |Z_{\mE_\mC'}| \; &, \q \forall \, \mE_\mC, \mE_\mC' \in \mathfrak{B}_{c_{{\rm y},1},c_{{\rm z},1}} \; .
\end{align*}
According to  the \emph{orbit-stabiliser theorem} we know that $|H| = |\mE_\mC| \cdot |Z_{\mE_\mC}|$ for every equivalence class $\mE_{\mC} \in \mathfrak{B}_\mC := \mathfrak{B}_{c_{{\rm y},1},c_{{\rm z},1}}$, so that every orbit in $\mathfrak{B}_{\mC}$ has the same size. Together with the fact that the set of such orbits forms a partition of $G \ltimes H^2$, it implies that the number of independent bulk $\mG$-colourings is
\begin{equation}
	\label{eq:sizeO}
	|\mathfrak{B}_\mC|
	\; = \; \sum_{\mE_{\mC} \in \mathfrak{B}_{\mC}} \! 1  
	\; = \;
	\sum_{\mE_\mC \in \mathfrak{B}_\mC}\sum_{(\exi,\fyi,\fzi) \in \mE_\mC} \!\!
	|\mE_\mC|^{-1} \; = \; |\mE_\mC^0|^{-1} \!\!\!\! \sum_{(\gx,\hy,\hz) \in G \ltimes H^2} \!\! 1
	\; = \; \frac{|G| \cdot |H|^2}{|\mE_\mC^0|} \; ,
\end{equation} 
where $\mE^0_\mC$ is any preferred equivalence class in $\mathfrak{B}_\mC$.
It is worth emphasizing that this formula does not follow straightforwardly from the definition \eqref{eq:defO} of the quotient set, but rather is a non-trivial consequence of the peculiar structure of the 1-form gauge transformations.

We are now ready to define the normalized ground states on $\mathfrak{T}[\mathbb T^2_\boxempty]$. The projection itself is performed via group averaging whereas the normalisation factor follows from the discussion above:
\begin{equation}
	\label{eq:avgTube}
	\Big| \,
	\tubeSavg{\hx}{\gy}{\gz}{\mE_{\gy,\gz}}{0.75}{1.25}
	\Big\ra
	:= \frac{1}{|H|^\frac{1}{2}|Z_{\mE_{\gy,\gz}}|^\frac{1}{2}}\sum_{\lambda \in H}
	\Big| \, \tubeSext{\hx}{\gy}{\gz}{\partial(\lambda)\exOne}{(\gy \act \lambda)\fzOne \lambda^{-1}}{(\gz \act \lambda)\fyOne \lambda^{-1}}{1.2}{2.1} \Big\ra \; ,
\end{equation}
where the notation makes explicit the fact that ground states only depend on equivalence classes of bulk $\mG$-colourings,
so that the ground state subspace on the tube $\mathfrak{T}[\mathbb T^2_\boxempty]$ explicitly reads
\begin{equation}
		\label{eq:gsTube}
		\mathcal{V}^\mG[\mathfrak{T}[\mathbb T^2_\boxempty]] =
		{\rm Span}_\mathbb{C} \Big\{ 	
		\Big| \, 		\tubeSavg{\hx}{\gy}{\gz}{\mE_{\gy,\gz}}{0.75}{1.25} \Big\ra \Big\}_{\substack{\hspace{-3.5em} \forall \, \gy,\gz \in G \\[-0.2em]  \forall \, \hx \in H \, | \, \gz = \gz^{\gy \,;\, \hx} \\ \hspace{-1.35em} \forall \, \mE_{\gy,\gz} \in \mathfrak{B}_{\gy,\gz}}} \; .
\end{equation}
Equipped with the inner product
\begin{align*}
	\Big\la \; 	
	\tubeSavg{\hx}{\gy}{\gz}{\mE_{\gy,\gz}}{0.75}{1.25} \; \Big| \, \tubeSavg{\thx}{\tgy}{\tgz}{\tilde{\mE}_{\tgy,\tgz}}{0.75}{1.25} \Big\ra
	= \delta\big(\gy , \tgy \big) \, \delta\big(\gz , \tgz\big) \, 
	\delta\big(\hx , \thx\big) \, 
	\delta\big( \mE_{\gy,\gz} , \tilde{\mE}_{\tgy,\tgz}\big) \; ,
\end{align*}
it defines the ground state Hilbert space on $\mathfrak{T}[\mathbb T^2_\boxempty]$.

\subsection{Computation of the tube algebra} 
Let us now derive the tube algebra for the ground states \eqref{eq:avgTube}. It suffices to apply successively the three operations entering the definition of the $\star$-product. However, we find it convenient to first define an auxiliary product $\odot$ whose definition is identical to the one of the $\star$-product but whose domain is the tensor product of two copies of the Hilbert space associated with $\eqref{eq:colouredTube}$. In other words, we first perform the computation omitting the group averaging \eqref{eq:avgTube}, which enforces the 1-form gauge invariance, and only in a second time reinstate it in order to obtain the final result in terms of the ground states. 

Firstly, boundary $\mG$-colourings are identified via the map $\mathfrak{G}$:
\begin{align}
	\label{eq:gluingDoubleTube}
	&\mathfrak{G} \Bigg( \,
	\Bigg| \!\! \cubeb{\gx}{\gy}{\gz}{\hx}{\hy}{\hz} \!
	\Bigg\ra
	\otimes \,
	\Bigg| \!\! \cubep{\gx'}{\gy'}{\gz'}{\hx'}{\hy'}{\hz'} \!
	\Bigg\ra\Bigg)
	\\[-2.5em]
	\nn
	& \q = 
	\delta \big(\hx', \hx^{\gx,\gy,\gz \,;\, \hy,\hz} \big) \,
	\delta \big(\gy', \gy^{\gx \,;\, \hz} \big) \, 
	\delta \big(\gz' , \gz^{\gx \,;\, \hy} \big)
	\Bigg| \!\! \cubeDouble \!
	\Bigg\ra ,
\end{align}
where we represented identified vertices with the same coloured dot.
Secondly, 0-form and 1-gauge invariance are enforced along the gluing interface via $\mathbb{P}^\mG[\mathfrak{T}[\mathbb T^2_\boxempty] \cup \mathfrak{T}[\mathbb T^2_\boxempty]]$ so that the state in $\mathcal{H}^\mG[\mathfrak{T}[\mathbb T^2_\boxempty] \cup \mathfrak{T}[\mathbb T^2_\boxempty]]$ is projected to 
\begin{align}
	\label{eq:avgDoubleTube}
	\frac{1}{|G||H|^2} 
	\sum_{\substack{k \in G \\ \eta,\eta' \in H}} 	
	\Bigg| \!\! \cubeDoubleAvg \!
	\Bigg\ra .
\end{align}
Thirdly, the following cubulation changing isomorphism is applied 
\begin{align}
	\label{eq:isoDoubleTube}
	\Bigg| \!\! \cubeDoubleAvg \!
	\Bigg\ra
	\simeq \; \frac{1}{|G|^\frac{1}{2}|H|}
	\Bigg| \!\! \cubeSpe{\gx\gx'}{\gy}{\gz}{\hx}{\hy(\gx \act \hy')}{\hz(\gx \act \hz')} \!
	\Bigg\ra ,
\end{align}
where we used axioms \eqref{eq:axioms} so that for instance $\hy ( \gx \act \eta^{-1} ) \big( ( \gx k^{-1} ) \act [k \act (\eta \hy')]\big) = \hy (\gx \act \hy')$. Note that under this isomorphism, the summation variables appearing in \eqref{eq:avgDoubleTube} cancel each other, so that the sums become trivial and thus compensate for the corresponding normalization factors. Putting everything together and using the symbolic notation defined in \eqref{eq:colouredTube}, we obtain 
\begin{align*}
	\Big| \, \tubeS{\hx}{\gy}{\gz}{\gx}{\hz}{\hy}
	\Big\ra \!
	\odot
	\Big| \, \tubeS{\hx'}{\gy'}{\gz'}{\gx'}{\hz'}{\hy'}
	\Big\ra\!
	= 
	 \frac{	\delta \big(\hx', \hx^{\gx,\gy,\gz \,;\, \hy,\hz} \big) 
	 	\delta \big(\gy',\gy^{\gx \,;\, \hz} \big) 
	 	\delta \big(\gz' , \gz^{\gx \,;\, \hy} \big)}{|G|^\frac{1}{2}|H|}
	\Big| \, \tubeSext{\hx}{\gy}{\gz}{\gx \gx'}{\hz(\gx \act \hz')}{\hy(\gx \act \hy')}{1.1}{1.9} \Big\ra  .
\end{align*}
Using this result, let us now include the group averaging to derive ${\rm Tube}^\mG[\mathbb T^2_\boxempty]$:
\begin{align*}
	\Big| \,
	\tubeSavg{\hx}{\gy}{\gz}{\mE_{\gy,\gz}}{0.75}{1.25}
	\Big\ra \!
	\star 
	\Big| \,
	\tubeSavg{\hx'}{\gy'}{\gz'}{\mE'_{\gy',\gz'}}{0.75}{1.25}
	\Big\ra
	& =
	\frac{1}{|H| |Z_{\mE_{\gy,\gz}}| |Z_{\mE'_{\gy',\gz'}}|}
	\sum_{\lambda, \lambda' \in H}
	\Big| \, \tubeSext{\hx}{\gy}{\gz}{\partial(\lambda)\exOne}{(\gy \act \lambda)\fzOne \lambda^{-1}}{(\gz \act \lambda)\fyOne \lambda^{-1}}{1.2}{2.1} \Big\ra 
	\odot 
	\Big| \, \tubeSext{\hx'}{\gy'}{\gz'}{\partial(\lambda')\exOne'}{(\gy' \act \lambda')\fzOne' \lambda'^{-1}}{(\gz' \act \lambda')\fyOne' \lambda'^{-1}}{1.3}{2.3} \Big\ra 
	\\[0.6em]
	& \simeq \delta \big(\hx', \hx^{\exOne,\gy,\gz \,;\, \fyOne,\fzOne} \big) 
	\delta \big(\gy',\gy^{\exOne \,;\, \fzOne} \big) 
	\delta \big(\gz' , \gz^{\exOne \,;\, \fyOne} \big) 
	\\[-0.7em]
	& \times \frac{1}{|G|^\frac{1}{2}|H|^2 |Z_{\mE_{\gy,\gz}}|}
	\sum_{\lambda, \lambda' \in H}
	\Big| \, \tubeSext{\hx}{\gy}{\gz}{\partial(\lambda)\exOne \partial(\lambda')\exOne'}{(\gy \act \lambda)\fzOne \lambda^{-1}\left([\partial(\lambda)\exOne] \act [(\gy' \act \lambda')\fzOne' \lambda'^{-1}]\right)}{(\gz \act \lambda)\fyOne \lambda^{-1}\left([\partial(\lambda)\exOne] \act [(\gz' \act \lambda')\fyOne' \lambda'^{-1}]\right)}{3.05}{5.8} \Big\ra  
	\\
	& =  \delta \big(\hx', \hx^{\exOne,\gy,\gz \,;\, \fyOne,\fzOne} \big) 
	\delta \big(\gy',\gy^{\exOne \,;\, \fzOne} \big) 
	\delta \big(\gz' , \gz^{\exOne \,;\, \fyOne} \big) 
	\\[-0.7em]
	& \times \frac{1}{|G|^\frac{1}{2}|H|^\frac{1}{2}|Z_{\mE_{\gy,\gz}}|^\frac{1}{2}}\! \cdot \! \frac{1}{|H|^\frac{1}{2}|Z_{\mE_{\gy,\gz}}|^\frac{1}{2}} \!
	\sum_{\mu \in H}
	\Big| \, \tubeSext{\hx}{\gy}{\gz}{\partial(\mu)\exOne \exOne'}{(\gy \act \mu)\fzOne (\exOne \act \fzOne')\mu^{-1}}{(\gz \act \mu)\fyOne (\exOne \act \fyOne')\mu^{-1}}{1.85}{3.4} \Big\ra  .  
\end{align*}
In the second step, we used the equality $|\mE_{\gy,\gz}| = |\mE'_{\gy',\gz'}|$, which is true since, in virtue of the delta functions, $(\gy,\gz\,\hx)$ and $(\gy',\gz',\hx')$ are in the same equivalence class with respect to \eqref{eq:equivBdry}. Moreover, we reproduced the $\odot$-product formula for $\mG$-coloured graph states derived above. In the third step, we made a shift of summation variable by defining $\mu := \lambda (\exOne \act \lambda')$, which makes one of the sums trivial, hence cancelling one of the $1/|H|$ factors. Indeed, using the delta functions in the derivation above, we check for instance that
\begin{equation*}
	\partial(\mu)\exOne\exOne' 
	\stackrel{\eqref{eq:Peiffer1}}{=}
	\partial(\lambda) \exOne \partial(\lambda') \exOne'  \; ,
\end{equation*}
and
\begin{align*}
	(\gz \act \mu)\fyOne (\exOne \act \fyOne') \mu^{-1} 
	&\; = \;
	\big(\gz \act [\lambda (\exOne \act \lambda')]\big)\fyOne (\exOne \act \fyOne')(\exOne \act \lambda'^{-1})\lambda^{-1} 
	\\
	&\stackrel{\mathmakebox[\widthof{\;=\;}]{\eqref{eq:axioms}}}{=}  (\gz \act \lambda) \big( [\partial(\fyOne)\exOne \gz'] \act \lambda'\big)\fyOne (\exOne \act \fyOne')(\exOne \act \lambda'^{-1})\lambda^{-1}
	\\
	&\stackrel{\mathmakebox[\widthof{\;=\;}]{\eqref{eq:Peiffer2}}}{=}  (\gz \act \lambda) \fyOne \big( \exOne \act [(\gz' \act \lambda')\fyOne' \lambda'^{-1}]\big)\lambda^{-1}
	\\
	&\stackrel{\mathmakebox[\widthof{\;=\;}]{\eqref{eq:Peiffer2}}}{=}  
	(\gz \act \lambda) \fyOne \lambda^{-1} \big( [\partial(\lambda)\exOne] \act [(\gz' \act \lambda')\fyOne' \lambda'^{-1}]\big) \; .
\end{align*}
In order to obtain the final result, we are left to simplify the normalization factors. Firstly, the factor $1/|H|^\frac{1}{2}|Z_{\mE_{\gy,\gz}}|^\frac{1}{2}$ enters the definition of the resulting ground state according to \eqref{eq:avgTube}. Secondly, the orbit-stabiliser theorem states that $|H| = |\mE_{\gy,\gz}| \cdot |Z_{\mE_{\gy,\gz}}|$ and thus
\begin{equation*}
	\frac{1}{|G|^\frac{1}{2}|H|^\frac{1}{2}|Z_{\mE_{\gy,\gz}}|^\frac{1}{2}} \; = \; \frac{|\mE_{\gy,\gz}|^\frac{1}{2}}{|G|^\frac{1}{2}|H|} \stackrel{\eqref{eq:sizeO}}{=}
	\frac{1}{|\mathfrak{B}_{\gy,\gz}|^\frac{1}{2}} \; .
\end{equation*}
Putting everything together, the tube algebra ${\rm Tube}^\mG[\mathbb T^2_\boxempty]$ finally reads 
\begin{align}
	\label{eq:tube}
	\Big| \,
	\tubeSavg{\hx}{\gy}{\gz}{\mE_{\gy,\gz}}{0.75}{1.25}
	\Big\ra \!
	\star 
	\Big| \,
	\tubeSavg{\hx'}{\gy'}{\gz'}{\mE'_{\gy',\gz'}}{0.75}{1.25}
	\Big\ra
	=  
	\frac{\delta \big(\hx', \hx^{\exOne,\gy,\gz \,;\, \fyOne,\fzOne} \big) 
		\delta \big(\gy',\gy^{\exOne \,;\, \fzOne} \big) 
		\delta \big(\gz' , \gz^{\exOne \,;\, \fyOne} \big)}{|\mathfrak{B}_{\gy,\gz}|^\frac{1}{2}}
	\Big| \, \tubeSavg{\hx}{\gy}{\gz}{(\mE \cdot \mE')_{\gy,\gz}}{1.05}{1.85} \Big\ra 
\end{align}
where $(\mE \cdot \mE')_{\gy,\gz}$ is the equivalence class in $\mathfrak{B}_{\gy,\gz}$ whose representative element reads 
\begin{equation*}
	\big(\exOne \exOne' \,,\,  \fyOne (\exOne \act \fyOne') \,,\, \fzOne (\exOne \act \fzOne') \big) \; .
\end{equation*}

\newpage
\section{Elementary loop-like excitations \label{sec:simple}}
\emph{In the previous section, we obtained the tube algebra associated with the gluing operation $\mathfrak{T}[\mathbb T^2_\boxempty] \cup \mathfrak{T}[\mathbb T^2_\boxempty] \simeq \mathfrak{T}[\mathbb T^2_\boxempty]$. We now derive the simple modules of this tube algebra. In the next section, we will elucidate the physical interpretation of these simple modules as a classifying tool for the elementary bulk loop-like excitations of the higher gauge model.}

\subsection{Simple modules of the tube algebra}

Let us derive the representation theory of the tube algebra ${\rm Tube}^\mG[\mathbb T^2_\boxempty]$ whose defining formula is \eqref{eq:tube}. In order to do so, we will first decompose the tube algebra into a direct sum of subalgebras. Recall that we defined earlier the equivalence relation \eqref{eq:equivBdry} on the set ${\rm Col}(\mathbb T^2_\boxempty \times \{0\}, \mG)$ of boundary $\mG$-colourings.
As stated earlier, equivalence classes with respect to this equivalence relation, which correspond to \emph{sets of boundary colourings invariant under the action of the tube algebra}, forms a partition of ${\rm Col}(\mathbb T^2_\boxempty \times \{0\}, \mG)$ into disjoint sets.  Furthermore, given two states in ${\rm Tube}^\mG[\mathbb T^2_\boxempty]$ whose boundary colourings belong to two disjoint equivalence classes, the algebra product vanishes. 
This induces that each equivalence class $\mC \subset {\rm Col}(\mathbb T^2_\boxempty \times \{0\}, \mG)$ defines a subalgebra ${\rm Tube}^\mG[\mathbb T^2_\boxempty]_\mC$, and since the set of equivalence classes forms a partition of the boundary colourings, one has
\begin{equation*}
	{\rm Tube}^\mG[\mathbb T^2_\boxempty] = \bigoplus_\mC {\rm Tube}^\mG[\mathbb T^2_\boxempty]_\mC \; .
\end{equation*}
We can therefore find the simple modules of ${\rm Tube}^\mG[\mathbb T^2_\boxempty]$ in terms of the ones of its subalgebras ${\rm Tube}^\mG[\mathbb T^2_\boxempty]_\mC$ for every $\mC \subset {\rm Col}(\mathbb T^2_\boxempty \times \{0\}, \mG) $. As explained in more detail further, the label $\mC$ corresponds to a magnetic \emph{flux} quantum number, while the simple modules of ${\rm Tube}^\mG[\mathbb T^2_\boxempty]_\mC$ provide the corresponding \emph{charge} components. 

Given an equivalence class $\mC$, let us now construct explicitly the simple modules of   ${\rm Tube}^\mG[\mathbb T^2_\boxempty]_\mC$. Recall that we notate elements in $\mC$ by 
\begin{equation*}
	(\cyi  , \czi , \dxi) \; , \q  i=1,\ldots,|\mC| 
\end{equation*}
such that $(c_{{\rm y},1}, c_{{\rm z},1} , d_{\hat{\rm x},1})$  is the \emph{representative} element. Let us introduce the set
\begin{equation*}
	Q_\mC = \big\{(\pxi , \qyi , \qzi)\big\}_{i=1,\ldots,|\mC|}
\end{equation*}	
such that each triplet is defined according to
\begin{align}
	\label{eq:defQ}
	(c_{{\rm y},1} , c_{{\rm z},1} , d_{\hat{\rm x},1}) = 
	(c_{{\rm y},i}^{p_{{\rm x},i} \, ; \, q_{\hat{\rm z},i}} ,
	c_{{\rm z},i}^{p_{{\rm x},i} \, ; \, q_{\hat{\rm y},i}} , d_{\hat{\rm x},i}^{p_{{\rm x},i}, c_{{\rm y},i},c_{{\rm z},i} \, ; \, q_{\hat{\rm y},i},q_{\hat{\rm z},i}})	
\end{align} 
with $(p_{{\rm x},1} , q_{\hat{\rm y},1} , q_{\hat{\rm z},1}) = (\mathbbm{1}_G , \mathbbm{1}_H , \mathbbm{1}_H)$. We then define the \emph{stabiliser group} 
\begin{align}
	\label{eq:defZC}
	Z_\mC &:= \big\{ \mE_\mC \in \mathfrak{B}_\mC \; \big| \;
	(c_{{\rm y},1} , c_{{\rm z},1} , d_{\hat{\rm x},1}) = 
	(c_{{\rm y},1}^{\exOne \, ; \,\fzOne} ,
	c_{{\rm z},1}^{\exOne \,;\, \fyOne} , d_{\hat{\rm x},1}^{\exOne, c_{{\rm y},1}, c_{{\rm z},1} \,;\, \fyOne,\fzOne})  \big\}  ,
\end{align} 
where the relevant notations regarding $\mE_\mC$ were introduced in the previous section.
The group algebra $\mathbb C[Z_\mC]$ is then defined as the algebra whose defining vector space is 
\begin{equation*}
	{\rm Span}_\mathbb{C} \Big\{ 	
	\big| \xrightarrow{\; \mE_\mC \;}\big\ra \Big\}_{\mE_\mC \in Z_\mC} 
\end{equation*}
and whose algebra product reads
\begin{equation}
	\label{eq:algZC}
	\big| \xrightarrow{\; \mE_\mC \;}\big\ra
	\star 	\big| \xrightarrow{\; \mE'_\mC \;}\big\ra = 
	\big| \xrightarrow{\; (\mE \cdot \mE')_\mC \;}\big\ra  ,
\end{equation}  
where $(\mE \cdot \mE')_\mC$ is the equivalence class in $\mathfrak{B}_\mC$ whose representative element reads
\begin{equation*}
	\big(\exOne \exOne' \,,\,  \fyOne (\exOne \act \fyOne') \,,\, \fzOne (\exOne \act \fzOne') \big) \; .
\end{equation*}

\bigskip \noindent
Given an irreducible representation $(D^R,V_R)$ of the centraliser $Z_\mC$, where $V_R$ is a complex vector space and $D^R:\mathbb{C}[Z_\mC] \to {\rm End}(V_R)$ an algebra homomorphism, we can now construct a simple representation of the tube algebra $ {\rm Tube}^\mG[\mathbb T^2_\boxempty]$ via a homomorphism $D^{\mC,R} : {\rm Tube}^\mG[\mathbb T^2_\boxempty]_\mC \to {\rm End}(V_{\mC,R})$ where the vector space $V_{\mC,R}$ is defined as
\begin{align*}
	V_{\mC,R} := {\rm Span}_\mathbb{C}\big\{ | c_{{\rm y},i} \,,\, c_{{\rm z},i} \,,\, d_{\hat{\rm x},i}\, , \, v_m \ra \big\}_{\substack{\hspace{-2.5em }\forall \, i =1,\ldots,|\mC| \\ \forall \, m=1,\ldots,{\rm dim}(V_R)}} \; .
\end{align*}
For $i,j \in \{1,\ldots,|\mC|\}$, $m,n \in \{1,\ldots, {\rm dim}(V_R)\}$, the matrix elements read
\begin{align}
	\nn
	D^{\mC,R}_{im,jn}\Big( 	\Big| \, \tubeSavg{\hx}{\gy}{\gz}{\mE_{\gy,\gz}}{0.75}{1.25} \Big\ra \Big)
	:=\; &\delta \big(\gy , c_{{\rm y},i}\big) \, 
	\delta \big(\gz , c_{{\rm z},i} \big) \, 
	\delta \big(\hx , d_{\hat{\rm x},i}\big) 
	\\[-0.5em]  
	\nn
	\times \; &\delta \big(\gy^{\exOne \,;\,\fzOne} ,\, c_{{\rm y},j}\big) \, \delta \big(\gz^{\exOne \, ; \, \fyOne} , c_{{\rm z},j}\big) \, \delta \big(\hx^{\exOne,\gy,\gz \, ; \, \fyOne,\fzOne} , d_{\hat{\rm x},j} \big)
	\\[-0.2em]
	\label{eq:irreps}
	\times \; &
	D^R_{mn}\big( 	\big| \xrightarrow{\; [\mE_\mC^{\rm stab.}]_{i,j} \;}\big\ra \big) \; ,
\end{align}
where $[\mE_\mC^{\rm stab.}]_{i,j}$ is the equivalence class in $\mathfrak{B}_\mC$ whose representative element reads
\begin{equation}
	\label{eq:defEstab}
	\big(\pxi^{-1}\exOne \pxj \,,\, 
	\pxi^{-1} \act [\qzi^{-1} \fzOne(\exOne \act \qzj)] \,,\,
	\pxi^{-1} \act [\qyi^{-1} \fyOne(\exOne \act \qyj)] \big) \; ,
\end{equation}
such that
\begin{equation*}
	D^{\mC,R}\Big( 	\Big| \, \tubeSavg{\hx}{\gy}{\gz}{\mE_{\gy,\gz}}{0.75}{1.25}
	\Big\ra \Big)\! := \frac{1}{|\mathfrak{B}_{\gy,\gz} \! |^\frac{1}{2}}
	\!\! \sum_{i,j=1}^{|\mC|} \!\! \sum_{m,n=1}^{{\rm dim}(V_R)} \!\!\! D^{\mC,R}_{im,jn}\Big( 	\Big| \, \tubeSavg{\hx}{\gy}{\gz}{\mE_{\gy,\gz}}{0.75}{1.25}
	\Big\ra \Big)
	|c_{{\rm y},i} , c_{{\rm z},i}, d_{\hat{\rm x},i}, v_m \ra \la c_{{\rm y},j} , c_{{\rm z},j}, d_{\hat{\rm x},j} , v_n |
	 .
\end{equation*}
Crucially, the delta functions in definition \eqref{eq:irreps} ensure that 
\begin{equation}
	\label{eq:repStab}
	[\mE_\mC^{\rm stab.}]_{i,j} \in Z_\mC  \; , \q  
	\forall \, \mE_{\gy,\gz} \in \mathfrak{B}_{\gy,\gz} \; , 
\end{equation}
which is checked explicitly in app.~\ref{sec:app_proofStab}.

It follows directly from the definition that the representation matrices above  realise an algebra homomorphism (see proof in app.~\ref{sec:app_proofLin}):
\begin{equation}
	\label{eq:Lin}
	\sum_{k=1}^{|\mC|}\sum_{o=1}^{{\rm dim}(V_R)}
	D^{\mC,R}_{im,ko}\Big( 	\Big| \, \tubeSavg{\hx}{\gy}{\gz}{\mE_{\gy,\gz}}{0.75}{1.25}
	\Big\ra \Big)
	D^{\mC,R}_{ko,jn} \Big(
	\Big| \, \tubeSavg{\hx'}{\gy'}{\gz'}{\mE'_{\gy,\gz}}{0.75}{1.25}
	\Big\ra\Big) = D^{\mC,R}_{im,jn}\Big( 	\Big| \,	\tubeSavg{\hx}{\gy}{\gz}{\mE_{\gy,\gz}}{0.75}{1.25}
	\Big\ra \!\star  \Big| \, \tubeSavg{\hx'}{\gy'}{\gz'}{\mE'_{\gy,\gz}}{0.75}{1.25}
	\Big\ra \Big) \; .
\end{equation}
Furthermore, the matrices satisfy the following \emph{orthogonality} and \emph{completeness} relations (see proofs in app.~\ref{sec:app_proofOrtho} and \ref{sec:app_proofComplete}): 
\begin{align}
	\label{eq:ortho}
	\!\!\!\!	\summand \!\!\!\!
	D^{\mC,R}_{im,jn}\Big( 	\Big| \, 	\tubeSavg{\hx}{\gy}{\gz}{\mE_{\gy,\gz}}{0.75}{1.25}
	\Big\ra \Big)
	\overline{D^{\mC',R'}_{i'm',j'n'}\Big( 	\Big| \, 	\tubeSavg{\hx}{\gy}{\gz}{\mE_{\gy,\gz}}{0.75}{1.25}
	\Big\ra \Big)} = \frac{|\mathfrak{B}_\mC |\delta_{\mC, \mC'}\delta_{R,R'}}{|\mC|{\rm dim}(V_R)}\delta_{i,i'}\delta_{j,j'}\delta_{m,m'}\delta_{n,n'}
	\\
	\label{eq:complete}
	\frac{1}{| \mathfrak{B}_{\gy,\gz}|}\sum_{\mC,R}\sum_{\substack{i,j \\ m ,n}}
	|\mC|{\rm dim}(V_R)
	D^{\mC,R}_{im,jn}\Big( 	\Big| \, 	\tubeSavg{\hx}{\gy}{\gz}{\mE_{\gy,\gz}}{0.75}{1.25}
	\Big\ra \Big)
	\overline{D^{\mC,R}_{im,jn}\Big( \Big| \, \tubeSavg{\thx}{\tgy}{\tgz}{\tilde{\mE}_{\tgy,\tgz}}{0.75}{1.25}
	\Big\ra \Big)} = 	
	\Big\la \; 	
	\tubeSavg{\hx}{\gy}{\gz}{\mE_{\gy,\gz}}{0.75}{1.25}\; \Big| \, \tubeSavg{\thx}{\tgy}{\tgz}{\tilde{\mE}_{\tgy,\tgz}}{0.75}{1.25}
	\Big\ra
\end{align}
where $\overline{\;\cdot \;}$ stands for complex conjugation.
These two conditions can be used to check \emph{a posteriori} that the set of simple modules is indeed indexed by pairs $(\mC,R)$.

\subsection{Physical interpretation}
We found above the simple modules of the tube algebra ${\rm Tube}^\mG[\mathbb T^2_\boxempty]$, and we showed that they are indexed by pairs $(\mC,R)$ such that equivalence classes $\mC$ represent sets of boundary $\mG$-colourings that are in the same orbit with respect to the action of the tube algebra, while the representations labelled by $R$ decompose the symmetries of a given boundary $\mG$-colouring under the action of the tube algebra. We now would like to interpret these simple modules in terms of elementary loop-like excitations of the higher gauge model. However, due to the tube algebra itself involving many degrees of freedom, deriving a consistent interpretation turns out to be a rather subtle task. In order to make progress in this direction, it is useful to consider limiting cases of ${\rm Tube}^\mG[\mathbb T^2_\boxempty]$ so as to isolate the different flux and charge components.

\medskip \noindent
Let us assume for now that the group $H$ is trivial, i.e. $\mG = (G, \{\mathbb 1_G\}, \partial: \mathbb 1_G \to \mathbb 1_G, {\rm id})$. Under this assumption, the model \eqref{eq:ham} reduces to a \emph{gauge} model, namely the Hamiltonian realization of \emph{Dijkgraaf-Witten} theory with trivial cohomology class in $H^4(BG, {\rm U}(1))$ \cite{dijkgraaf1990topological, Wan:2014woa}. The authors showed in \cite{Delcamp:2017pcw,Bullivant:2019fmk} that in this case the tube algebra for loop-like excitations is isomorphic to the (untwisted) quantum triple algebra, which we reproduce below for convenience:
\begin{align*}
	\Big| \, \tubeS{\hspace{0.1em}\mathbb 1_G}{\gy}{\gz}{\gx}{\hspace{0.1em}\mathbb 1_G}{\hspace{0.1em}\mathbb 1_G}
	\Big\ra 
	\star 
	\Big| \, \tubeS{\hspace{0.1em}\mathbb 1_G}{\gy'}{\gz'}{\gx'}{\hspace{0.1em}\mathbb 1_G}{\hspace{0.1em}\mathbb 1_G}
	\Big\ra 
	= 
	\frac{\delta \big(\gy', \gx^{-1}\gy \gx \big) \, 
	\delta \big(\gz' ,\gx^{-1}\gz \gx \big)}{|G|^\frac{1}{2}}
	\Big| \, \tubeSext{\hspace{0.1em}\mathbb 1_G}{\gy}{\gz}{\gx \gx'}{\hspace{0.1em}\mathbb 1_G}{\hspace{0.1em}\mathbb 1_G}{0.9}{1.5} \Big\ra \;  .
\end{align*}
The simple modules of this algebra are labelled by equivalence classes that correspond to sets of $G$-colourings of $\mathbb T^2 \times \{0\}$ related via simultaneous conjugation, and irreducible representations of the corresponding stabiliser groups. The physical interpretation of these simple modules goes as follows \cite{Bullivant:2019fmk}:  Given the three-disk $\mathbb D^3$, removing a solid torus $\mathbb D^2 \times \mathbb S^1$ from it creates a loop-like defect. After this operation, we can find a non-contractible 1-cycle, starting and ending at a given basepoint, that winds once around the hole left by the torus. The $G$-colouring assigns a non-trivial group variable to this non-contractible 1-cycle that is interpreted as a magnetic flux. This situation corresponds to the case where we consider equivalence classes whose representatives are of the form $(c_{{\rm y},1}, \mathbb 1_G , \mathbb 1_G)$ or $(\mathbb 1_G , c_{{\rm z},1}, \mathbb 1_G)$, in which case the tube algebra is isomorphic to the \emph{quantum double algebra} \cite{Drinfeld:1989st, Dijkgraaf1991} and the loop-like excitations are in one-to-one correspondence with the point-like anyonic particles of the (2+1)d Dijkgraaf-Witten model with trivial input 3-cocycle. Subsequently removing a solid cylinder that \emph{threads} the hole previously created and whose bounding circles are incident with the boundary of $\mathbb D^3$, we can find a second non-contractible 1-cycle, starting and ending at the same basepoint, that winds once around the hole left by the cylinder. The $G$-colouring assigns another non-trivial group variable to this non-contractible 1-cycle, which is interpreted as a magnetic flux threading the loop-like excitation. Composition of the 1-cycles is commutative and therefore the corresponding group variables must commute as well. Such sets of commuting group variables provide representatives $(c_{{\rm y},1}, c_{{\rm z},1}, \mathbb 1_G)$ for the equivalences classes appearing in the description of the simple modules of the quantum triple algebra. We can depict this situation as follows:
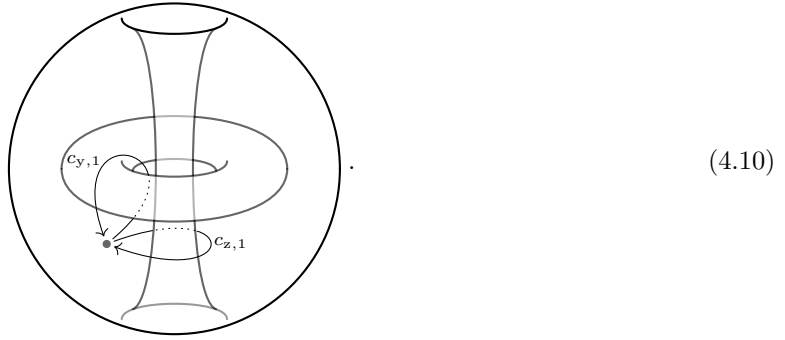
\begin{equation}
	\label{eq:interpretation}
	\hspace{-0.7em}\interpretation{1} \; .
\end{equation}	
The requirement that the two fluxes commute imply that the magnetic flux and electric charge quantum numbers labelling the loop-like excitation are constrained by the presence of the non-trivial threading flux.

In specifying $H = \{\mathbb 1_G\}$ as in the analysis above, we isolate the 1-form component of a strict 2-group flat connection. But it is well-known that flat $G$-connections on $\mathbb T^2$ are equivalent to group homomorphisms in ${\rm Hom}(\pi_1(\mathbb T^2),G)$, where $\pi_1(\mathbb T^2)$ denotes the \emph{fundamental group} group of $\mathbb T^2$, so that non-trivial (1-)holonomies can be assigned to non-contractible 1-cycles only. Furthermore, since $\pi_1(\mathbb T^2) = \mathbb Z \times \mathbb Z$, the corresponding group variables must commute, as explained above. Similarly, we can isolate the 2-form component of a strict 2-group flat connection by choosing a crossed module of the form $\mG = (\{\mathbb 1_H\},H, \partial : H \to \mathbb 1_H, {\rm id})$, in which case the second  Peiffer identity \eqref{eq:Peiffer2} imposes that $H$ must be abelian. Flat (2-form) $H$-connections on 2d surfaces $\Sigma$ correspond to group homomorphisms in ${\rm Hom}(\pi_2(\Sigma),H)$, where $\pi_2(\Sigma)$ denotes the \emph{second homotopy group} of $\Sigma$, so that non-trivial 2-holonomies can be assigned to non-contractible 2-cycles only.\footnote{Given a $d$-dimensional manifold, the number of non-contractible $i$-cycles is provided by the so-called $i$-th \emph{Betti number} denoted by $b_i$. For the two-torus, we have $b_0(\mathbb T^2)=1$, $b_1(\mathbb T^2)=2$, and $b_2(\mathbb T^2)=1$.} Similarly to the previous scenario, the corresponding group variables are interpreted as magnetic fluxes, but now with respect to the 2-flatness constraint. Under this assumption, the model \eqref{eq:ham} reduces to a so-called 2-form gauge model, namely the Hamiltonian realization of \emph{Crane-Yetter} theory for the braided fusion category of $H$-graded vector spaces with trivial cohomology class in $H^4(B^2H, {\rm U}(1))$ \cite{Walker:2011mda,Bullivant:2016clk, Delcamp2019}.\footnote{It was shown in \cite{Bullivant:2016clk} and \cite{Delcamp2019} that this model is equivalent to the Walker-Wang model in the untwisted and twisted cases, respectively.}  Since $H$ is abelian, the tube algebra simplifies considerably, i.e.
\begin{align*}
	\Big| \, \tubeS{\hx}{\mathbb 1_H}{\mathbb 1_H}{\mathbb 1_H}{\hz}{\hy}
	\Big\ra \!
	\star 
	\Big| \, \tubeS{\hx'}{\mathbb 1_H}{\mathbb 1_H}{\mathbb 1_H}{\hz'}{\hy'}
	\Big\ra
	= 
	\frac{\delta \big(\hx', \hx \big)}{|H|}
	\Big| \, \tubeSext{\hx}{\mathbb 1_H}{\mathbb 1_H}{\mathbb 1_H}{\hz+\hz'}{\hy+\hy'}{1.1}{1.9} \Big\ra \; ,
\end{align*}
and deriving its representation theory is immediate: Equivalence classes are in one-to-one correspondence with the group elements in $H$ and label point-like flux excitations with respect to the 2-flatness constraint, while we distinguish two independent representation labels which amount to string-like charge excitations with respect to the 1-form gauge invariance along the two 1-cycles of the torus.

So we have a good understanding of the simple modules of the tube algebra ${\rm Tube}^{\mathcal G}[\mathbb T^2_\boxempty]$ in the limiting cases where one of the groups entering the definition of the strict 2-group is trivial. Before tackling the interpretation of the elementary loop-like excitations in the general case, we are going to consider another limiting case, namely the tube algebra  ${\rm Tube}^\mG[\mathbb S^2_\boxempty]$ for spherical boundaries. In any dimensions and for any kind of models, the tube algebra associated with the $d$-sphere always yields the algebraic structure underlying the \emph{point-like} excitations of the model. Indeed, the regular neighbourhood of a point embedded in a $d$-dimensional manifold is a $d$-disk $\mathbb D^d$ which upon removal leaves an $\mathbb S^d$-boundary. The two-sphere contains one non-contractible 2-cycle. As explained above, the 2-form component of the strict 2-group connection can assign a non-trivial group element $h \in H$ to it. More precisely, this group variable amounts to the 2-holonomy associated with parallel-transporting a string with fixed endpoints around $\mathbb S^2$, i.e.
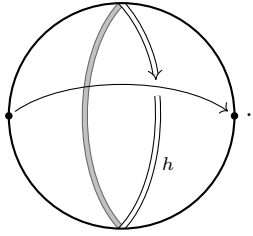
\begin{equation*}
	\skippingRope{1} \; .
\end{equation*}
This 2-holonomy is well-defined only when the fake-flatness constraint is satisfied, which in this case amounts to imposing that $\partial(h) = \mathbb 1_G$, i.e. $h \in {\rm Ker}\, \partial$. We can then compute ${\rm Tube}^\mG[\mathbb S^2]$ by discretising $\mathbb S^2$ as the two-disk $\mathbb D^2$ such that all the points in its boundary are identified to a unique vertex.  A discretisation of the interior of $\mathfrak{T}[\mathbb S^2]$ is then obtained as a single edge coloured by a group variable in $G$ that accounts for the violation of the 0-form gauge invariance at the boundary vertex. We thus consider $\mG$-coloured graph-states on $\mathfrak{T}[\mathbb S^2]$ of the form
\begin{align*}
	{\rm Span}_{\mathbb C} \big\{
	\big| \! \tubeSsphere{h}{g}{0.5}{0.85}
	\big\ra \big\}_{\substack{\hspace{-1.3em} \forall \, g \in G \\ \forall \, h \in {\rm Ker}\, \partial}} \; .
\end{align*}
It remains to impose the 1-form gauge invariance along the single edge in the interior of $\mathfrak{T}[\mathbb S^2]$ in order to obtain the corresponding ground states:
\begin{equation*}
	\big|\!  \tubeSsphereavg{h}{\mathcal E}{0.4}{0.65}
	\big\ra  := \frac{1}{|H|^\frac{1}{2}|{\rm Ker}\, \partial|^\frac{1}{2}} \sum_{\lambda \in H}
	\big| \! \tubeSsphere{h}{\partial(\lambda)\exOne}{0.85}{1.55}
	\big\ra \; ,
\end{equation*}
where the notation descends from the one used in sec.~\ref{sec:gsTube}. In this case, it turns out that the set of equivalence classes of bulk $\mG$-colourings is particularly simple. Indeed, it is equal to the co-kernel of $\partial$, i.e. ${\rm coKer} \, \partial := G/ {\rm Im} \, \partial$, which is well-defined since the Peiffer identities ensure that ${\rm Im} \, \partial$ is a \emph{normal} subgroup of $G$. The ground state subspace on $\mathfrak{T}[\mathbb S^2]$ therefore reads
\begin{align*}
	\mathcal{V}^\mG[\mathfrak{T}[\mathbb S^2]] = {\rm Span}_{\mathbb C} \big\{
	\big|\!  \tubeSsphereavg{h}{\mathcal E}{0.4}{0.65}
	\big\ra \big\}_{\substack{\hspace{-0.8em} \forall \, h \in {\rm Ker}\, \partial \\  \forall \, \mathcal{E} \in {\rm coKer}\, \partial }} \; ,
\end{align*}
and the tube algebra ${\rm Tube}^\mG[\mathbb S^2]$ is simply given by\footnote{We remark that the spherical boundary tube-algebra is Morita equivalent to the subalgebra of the torus boundary tube algebra defined by equivalence classes $\mC_h \subset {\rm Col}(\mathbb T^2_\boxempty \times \{0\}, \mG)$ whose representatives are of the form $(\mathbb 1_{G},\mathbb 1_{G},h)$, for all $h \in {\rm Ker} \, \partial$, i.e. 
	\begin{equation*}
	{\rm Tube}^\mG[\mathbb S^2]\,\,\,
	\substack{{\rm{Morita}}\\\sim} \!\!
	\bigoplus_{h\in {\rm Ker} \, \partial} {\rm Tube}^\mG[\mathbb T^2_\boxempty]_{\mC_h} \; .
	\end{equation*}
}
\begin{align*}
	\big|\!  \tubeSsphereavg{h}{\mathcal E}{0.4}{0.65}
	\big\ra 
	\star 
	\big| \! \tubeSsphereavg{h'}{\mathcal E'}{0.4}{0.65}
	\big\ra 
	= 
	\frac{\delta \big(h', \exOne^{-1} \act h \big)}{|{\rm coKer}\, \partial|^\frac{1}{2}}
	\big| \! \tubeSsphereavg{h}{\mathcal E \cdot \mathcal E'}{0.6}{1.05}\big\ra \;  .
\end{align*}
The simple modules of this tube algebra descend from the ones derived earlier so that flux point-like excitations are labelled by equivalence classes of ${\rm Ker}\, \partial$ where for $h,\tilde h \in {\rm Ker}\, \partial$ we have $h \sim \tilde h$ if there exists $a \in G$ such that $\tilde h = a^{-1} \act h$, and charge point-like excitations are labelled by irreducible representations of the centralizer
\begin{align*}
	\big\{ \mE \in {\rm coKer}\, \partial \; \big| \;
	h = \exOne^{-1} \act h \big\} \; .
\end{align*} 
So the spherical boundary case teaches us that for higher gauge models, elementary point-like excitations are labelled by both a flux quantum number with respect to the 2-flatness constraint and a charge quantum number with respect to the usual 0-form gauge invariance, whereby the charge label is constrained by the presence of the point-like flux. Let us emphasize that in the limiting case of the connected component of the spherical tube algebra with boundary colouring $h=\mathbb 1_{H}$, 0-form charge excitations are indexed by representations of ${\rm coKer}\, \partial$, and not representations of $G$ as it is the  case for the untwisted Dijkgraaf-Witten model. 
Indeed, we know that for the untwisted Dijkgraaf-Witten model pure charge excitations are indexed by representations $(D^R,V_R)$ of $G$. This can be appreciated from a string operator point of view: Define a path $\gamma$ of edges on the lattice connecting two vertices $\mathsf{v} \xrightarrow{\, \gamma \, }\mathsf{v}'$; Charges are then created in the states $|v_m \ra,|v_n \ra \in V_R$ at the vertices $\mathsf{v}$ and $\mathsf{v}'$, respectively, by multiplying each $G$-colouring by $D^R_{mn}(g_{\gamma})$, where $g_{\gamma}\in G$ is the holonomy assigned to such a path. This operator commutes with all vertex gauge operators on the lattice except at the end-points $\mathsf{v},\mathsf{v}'$ by the observation that such gauge operators do not change the resulting holonomy. Such excitations are call \emph{deconfined} as the energy cost of producing such a pair of charges is independent of their separation in the metric of the lattice.
Applying this construction to the higher gauge model, we realise that the previous string operator would fail to commute with the edge gauge operators along the length of the path. More precisely, the edge gauge operators perform  a transformation of the holonomy via $g_{\gamma}\mapsto \partial(\lambda)g_{\gamma}$ for some $\lambda\in H$. This observation demonstrates that the energy cost of such a charge excitation would be proportional to the length of the string, and we call such a pair of charge excitations \emph{confined} as the energetics of the model favour small separations of the charges. In order for such excitations to be deconfined, having energy cost at only the end-points, we must require that $D^R(\partial(\lambda)g_{\gamma})=D^R(g_{\gamma})$ for all $\lambda\in H$, which is equivalent to requiring that $D^R$ defines a representation of ${\rm coKer}\, \partial$, as expected.

Putting all the remarks above together, let us now propose a physical interpretation of the simple modules of the tube algebra ${\rm Tube}^{\mathcal G}[\mathbb T^2_\boxempty]$ for the strict 2-group higher gauge model. Firstly, we distinguish three types of flux excitations, which in terms of \eqref{eq:interpretation} can be interpreted as follows: The loop-like flux that corresponds to the 1-holonomy going around the hole left by the torus; the threading flux that corresponds to the second 1-holonomy going around the hole left by the cylinder; the point-like flux that corresponds to the 2-holonomy associated with the parallel-transport of the first loop along the second one. But this parallel-transport is well-defined only when the fake-flatness condition is imposed. Together, this implies that the magnetic flux quantum number labelling the loop-like excitation is constrained by the presence of both the threading flux and the 2-holonomy. Similarly, the charge quantum number accounts for the composite of three types of charge excitations: two string-like charges with respect to the 1-form gauge invariance along the non-contractible 1-cycles described above, one point-like charge with respect to the 0-form gauge invariance at the basepoint of the 1-cycles. Crucially, there is a non-trivial interplay between these charge excitations as they are constrained by the flux components via the flatness conditions but also because of the confinment mechanism put forward above in the spherical boundary case.

\newpage
\section{Ground states of the three-torus}\label{sec:torigs}

\emph{We derived in the previous section the simple modules of the tube algebra ${\rm Tube}^\mG[\mathbb T^2_\boxempty]$ classifying the elementary loop-like excitations of the higher gauge model. We now build upon this construction to derive a complete orthonormal ground state basis for the three-torus $\mathbb{T}^{3}$, demonstrating that such states are spanned by the central elements of ${\rm Tube}^\mG[\mathbb T^2_\boxempty]$. As a consequence, we find the ground state degeneracy of $\mathbb{T}^{3}$ to be given by the number of elementary loop-like excitations.}

\subsection{Canonical basis for ${\rm Tube}^\mG[\mathbb T^2_\boxempty]$}

We begin by introducing an alternative basis for the tube algebra ${\rm Tube}^\mG[\mathbb T^2_\boxempty]$ that we will refer to as the \emph{canonical basis}. The primary purpose of the canonical basis is to simplify the tube algebra product, rendering many calculations simpler than in the conventional basis.
Henceforth, given a simple module $(\mC,R)$, we use the shorthand notation $M,N$ for the basis indices $im,jn \in\{1,\ldots, {\rm dim}(V_{\mC,R}) \}$ introduced in sec.~\ref{sec:simple} and define $d_{\mC,R}:={\rm dim}(V_{\mC,R})=|\mC| \cdot {\rm dim}(V_R)$.

The canonical basis for ${\rm Tube}^\mG[\mathbb T^2_\boxempty]$ is defined by the set of elements $\ket{\mC,R \, ; MN}\in {\rm Tube}^\mG[\mathbb T^2_\boxempty]$ for each simple module $(\mC,R)$ and $M,N\in \{1,\ldots,d_{\mC,R} \}$ such that
\begin{align}\label{eq:canonicaltrans}
	&|\mC,R \, ;MN \ra:=
	\frac{d_{\mC,R}^\frac{1}{2}}{|\mathfrak{B}_{\mC}|^\frac{1}{2}} \!\!\!
	\summand
	\overline{
	D^{\mC,R}_{MN}\Big(\Big| \, \tubeSavg{\hx}{\gy}{\gz}{\mE_{\gy,\gz}}{0.75}{1.25}
	\Big\ra \Big)} \;
	\Big| \, \tubeSavg{\hx}{\gy}{\gz}{\mE_{\gy,\gz}}{0.75}{1.25}
	\Big\ra \;.
\end{align}
The transformation above defines an isomorphism between the two bases with inverse given by
\begin{align}\label{eq:canonicaltransinv}
	&\Big| \, \tubeSavg{\hx}{\gy}{\gz}{\mE_{\gy,\gz}}{0.75}{1.25}
	\Big\ra
	=\frac{1}{|\mathfrak{B}_{\gy,\gz}|^\frac{1}{2}}\sum_{\mC,R}d_{\mC,R}^\frac{1}{2}
	\sum_{M,N}
	D^{\mC,R}_{MN}\Big(\Big| \, \tubeSavg{\hx}{\gy}{\gz}{\mE_{\gy,\gz}}{0.75}{1.25}
	\Big\ra \Big) \; |\mC,R \, ;MN\ra\,.
\end{align}
	An immediate consequence of this definition is that the canonical basis is orthonormal, i.e.
\begin{align}
	\label{eq:canonicalortho}
	\la \mC',R';M'N'\, | \, \mC,R \, ;MN\ra = \delta_{\mC,\mC'} \, \delta_{R,R'} \, \delta_{M,M'} \,\delta_{N,N'}	\; ,
\end{align}
where the inner product is induced from the inner product in ${\rm Tube}^\mG[\mathbb T^2_\boxempty]$, and complete, i.e.
\begin{equation}
	\label{eq:canonicalcomp}
	\sum_{\mC,R}\sum_{M,N}\la \mC,R \, ;MN \, | \, \mC, R \, ; MN \ra = \big|{\rm Tube}^\mG[\mathbb T^2_\boxempty] \big|\; .
\end{equation}
These two statements are proven in app.~\ref{sec:app_canonicalortho} and \ref{sec:app_canonicalcomplete}, respectively. As desired, the $\star$-product in the canonical basis takes a particularly convenient form, namely
\begin{align}
	\label{eq:canonicalproduct}
	\ket{\mC,R \, ;MN}\star\ket{\mC',R';M'N'}
	=\frac{\delta_{N,M'} \, \delta_{\mC,\mC'} \, \delta_{R,R'}}{d_{\mC,R}^\frac{1}{2}}
	\, \ket{\mC,R \, ;MN'}\; ,
\end{align}
which is proven in app.~\ref{sec:app_canonicalalgebraprod}.
A useful corollary is the relations
\begin{align*}
	\Big| \, \g \Big\ra \star\ket{\mC,R \, ;MN}&=
	\frac{1}{|\mathfrak{B}_{\mC}|^\frac{1}{2}}\sum_{M'}D^{\mC,R}_{MM'}\Big( \Big| \, \g \Big\ra \Big) \ket{\mC,R \, ;M'N}
	\\
	\ket{\mC,R \, ;MN}\star\Big| \, \g \Big\ra &=\frac{1}{|\mathfrak{B}_{\mC}|^\frac{1}{2}}\sum_{N'}D^{\mC,R}_{N'N}\Big( \Big| \, \g \Big\ra \Big)\ket{\mC,R \, ;MN'} \; ,
\end{align*}
which follows from the definition of the $\star$-product and eq.~\eqref{eq:canonicaltransinv}.

\subsection{Centre of ${\rm Tube}^\mG[\mathbb{T}^2_\boxempty]$}\label{sec:centralbasis}

We now utilise the canonical basis defined in the previous part to define a natural basis for the central subalgebra
$Z({\rm Tube}^\mG[\mathbb T^2_\boxempty])\subset{\rm Tube}^\mG[\mathbb T^2_\boxempty]$, where $Z({\rm Tube}^\mG[\mathbb T^2_\boxempty])$ is defined as the subalgebra of ${\rm Tube}^\mG[\mathbb T^2_\boxempty]$ consisting of the set of all elements $|\psi \ra\in {\rm Tube}^\mG[\mathbb T^2_\boxempty]$ such that
\begin{align*}
	|\psi \ra
	\star
	\Big| \, \tubeSavg{\hx}{\gy}{\gz}{\mE_{\gy,\gz}}{0.75}{1.25}\Big\ra
	=
	\Big| \, \tubeSavg{\hx}{\gy}{\gz}{\mE_{\gy,\gz}}{0.75}{1.25}\Big\ra
	\star
	|\psi \ra
	\; , \q  
	\forall \, \Big| \, \tubeSavg{\hx}{\gy}{\gz}{\mE_{\gy,\gz}}{0.75}{1.25}\Big\ra\in {\rm Tube}^\mG[\mathbb T^2_\boxempty] \; .
\end{align*}
Building upon the canonical basis defined above, we can describe a complete and orthonormal basis for $Z({\rm Tube}^\mG[\mathbb T^2_\boxempty])$ as follows:
\begin{align*}
	Z({\rm Tube}^\mG[\mathbb T^2_\boxempty])={\rm Span}_{\CC}\big\{ |\mC,R \ra \big\}_{\forall \, \mC,R}
\end{align*}
where
\begin{align}\label{eq:centralcanonicalbasis}
	| \mC ,R \ra :=
	\frac{1}{d_{\mC,R}^\frac{1}{2}}\sum_{M}\ket{\mC,R \, ;MM} \; .
\end{align}
Orthonormality of these states follows directly from eq.~\eqref{eq:canonicalortho}, while
it is straightforward to verify such basis elements are indeed central, i.e.
\begin{align*}
	\ket{\mC,R} \star\Big| \, \g \Big\ra 
	&=\frac{1}{d_{\mC,R}^\frac{1}{2}}\sum_{M}\ket{\mC,R \, ;MM}\star\Big| \, \g \Big\ra 
	\nn
	\\
	\nn
	&=\frac{1}{d_{\mC,R}^\frac{1}{2}|\mathfrak{B}_{\mC}|^\frac{1}{2}}\sum_{M,N}\ket{\mC,R \, ;MN} D^{\mC,R}_{NM}\Big( \Big|  \;  \g \Big\rangle \Big) 
	\\
	&=\frac{1}{d_{\mC,R}^\frac{1}{2}|\mathfrak{B}_{\mC}|^\frac{1}{2}}
	\sum_{M,N}\ket{\mC,R \, ;NM}D^{\mC,R}_{MN}\Big( \Big| \, \g \Big\ra \Big)
	=\Big| \, \g \Big\ra \star\ket{\mC,R} \; .
\end{align*}
Completeness of the basis follows from the observation that any other element of ${\rm Tube}^\mG[\mathbb T^2_\boxempty]$ is either a sum of such elements or not central.

\subsection{Three-torus ground state basis}

Building upon the previous discussion, let us now show that the ground state subspace of the higher gauge model for the three-torus $\mathbb{T}^{3}$ is described by the centre $Z({\rm Tube}^\mG[\mathbb T^2_\boxempty])$ of the tube algebra ${\rm Tube}^\mG[\mathbb T^2_\boxempty]$.
We utilise a cubulation $\mathbb{T}^{3}_\boxempty$ of $\mathbb{T}^{3}$ induced from the tube $\mathfrak{T}[\mathbb T^2_\boxempty]$ defined in \eqref{eq:colouredTube} by further requiring the identifications $\snum{(0145)} \equiv \snum{(2367)}$, $\snum{(01)} \equiv \snum{(45)} \equiv \snum{(23)} \equiv \snum{(67)}$ and  $\snum{(04)} \equiv \snum{(15)} \equiv \snum{(26)} \equiv \snum{(37)}$. Applying such constraints, we can identify the space of $\mG$-coloured graph-states of $\mathbb{T}^{3}_\boxempty$ with a subspace of the space of $\mG$-coloured graph-states of $\mathfrak{T}[\mathbb T^2_\boxempty]$.
Specifically, a $\mG$-colouring of $\mathfrak{T}[\mathbb T^2_\boxempty]$ as defined in \eqref{eq:colouredTube} induces a  $\mG$-colouring of $\mathbb{T}^{3}_{\boxempty}$ if and only if
\begin{align*}
	(\gy,\gz,\hx)
	=
	(\gy^{\gx\,;\,\hz},\gz^{\gx\,;\, \hy},\hx^{\gx,\gy,\gz\,;\, \hy,\hz})\; ,
\end{align*}
and we notate $\mG$-coloured graph-states of $\mathfrak{T}[\mathbb T^2_\boxempty]$ that satisfy the above conditions, and thus define $\mG$-coloured graph-states of $\mathbb{T}^{3}_{\boxempty}$, as 
\begin{align*}
	\Big| \, \tubeS{\hx}{\gy}{\gz}{\gx}{\hz}{\hy} \Big\ra_{\mathbb{T}^{3}_{\boxempty}} .
\end{align*}
In order to obtain the ground states on $\mathbb T^3_\boxempty$, we are left to enforce the 0-form gauge invariance along the unique vertex $\snum{(0)} \equiv \snum{(1)} \equiv \ldots \equiv \snum{(7)}$ and the 1-form gauge invariance along the edges ${\rm x} := \snum{(02)} \equiv \snum{(13)} \equiv \snum{(46)} \equiv \snum{(57)}$, ${\rm y} := \snum{(01)} \equiv \snum{(23)} \equiv \snum{(45)} \equiv \snum{(67)}$ and ${\rm z} := \snum{(04)} \equiv \snum{(15)} \equiv \snum{(26)} \equiv \snum{(37)}$ via the projectors $\mathbb A_\mathsf{v}$ and $\mathbb{A}_\mathsf{e}$, respectively. The action of the gauge operators on the $\mG$-colourings of $\mathbb{T}^{3}_{\boxempty}$ reads
\begin{align*}
	\mathbb{A}^{k}_{(0)}
	\mathbb{A}^{\lambda_{\rm z}}_{\rm z}
	\mathbb{A}^{\lambda_{\rm y}}_{\rm y}
	\mathbb{A}^{\lambda_{\rm x}}_{\rm x}
	\Big| \,
	\tubeS{\hx}{\gy}{\gz}{\gx}{\hz}{\hy}
	\Big\ra_{\mathbb{T}^{3}_{\boxempty}}
	=
	\Big| \,
	\tubeS{\thx}{\tgy}{\tgz}{\tgx}{\thz}{\thy}
	\Big\ra_{\mathbb{T}^{3}_{\boxempty}}
\end{align*}
for all $k \in G$ and $\lambda_{\rm x},\lambda_{\rm y},\lambda_{\rm z}\in H$, where
\begin{align*}
	(\tgy, \tgz, \thx) &= 
	(\gy^{k^{-1}  ;\, \lambda^{-1}_{\rm y}} , 
	\gz^{k^{-1}  ;\, \lambda^{-1}_{\rm z}} , 
	\hx^{k^{-1} \! ,\gz,\gy \, ; \, \lambda^{-1}_{\rm y},\lambda^{-1}_{\rm z} })
	\\
	(\tgx, \thy, \thz)
	&=
	(\gx^{k^{-1}; \,\lambda^{-1}_{\rm x}},  
	\hy^{k^{-1} \! ,\gx,\gz \, ; \, \lambda^{-1}_{\rm z},\lambda^{-1}_{\rm x} } ,
	\hz^{k^{-1} \! ,\gx,\gy \, ; \,\lambda^{-1}_{\rm y},\lambda^{-1}_{\rm x} })
	\; .
\end{align*}
Using the above action of the gauge operators on the $\mG$-colourings of $\mathbb{T}^{3}_{\boxempty}$, we can explicitly define the ground state projector for $\mathbb{T}^{3}_{\boxempty}$ via
\begin{align}
	\label{eq:tori-gsproj}
	\mathbb{P}^{\mG}[\mathbb{T}^{3}_{\boxempty}]
	=
	\frac{1}{|G||H|^{3}} \! \!
	\sum_{
		\substack{
			\gx,\gy,\gz\in G\\
			\hx,\hy,\hz\in H
		}
	}
	\sum_{
	\substack{
		k \in G\\
		\lambda_{\rm x},\lambda_{\rm y},\lambda_{\rm z} \in H
	}
	} \!\!
	&\mathbb{A}^{k}_{(0)}
	\mathbb{A}^{\lambda_{\rm z}}_{\rm z}
	\mathbb{A}^{\lambda_{\rm y}}_{\rm y}
	\mathbb{A}^{\lambda_{\rm x}}_{\rm x}
	\Big| \,
	\tubeS{\hx}{\gy}{\gz}{\gx}{\hz}{\hy}
	\Big\ra
	\Big\la \,
	\tubeS{\hx}{\gy}{\gz}{\gx}{\hz}{\hy} \,
	\Big|
	\\[-0.6em] \nn
	\times \; &\delta(\hx,\hx^{\gx,\gy,\gz \,;\, \hy,\hz}) \,
	\delta(\gy,\gy^{\gx\,;\, \hz }) \,
	\delta(\gz,\gz^{\gx\,;\, \hy }) \,
	\delta(\gz,\gz^{\gy\,;\,\hx}) \; .
\end{align}
Having described the ground state projector $\mathbb{P}^{\mG}[\mathbb{T}^{3}_{\boxempty}]$, we are now able to construct the ground state subspace $\mathcal{V}^{\mG}[\mathbb{T}^{3}_{\boxempty}] :={\rm Im}\, \mathbb{P}^{\mG}[\mathbb{T}^{3}_{\boxempty}]$ of $\mathbb{T}^{3}_{\boxempty}$.
However, from \eqref{eq:tori-gsproj} alone, the form of the ground state subspace is relatively obtuse. In order to proceed with our discussion, it is instructive to observe that the ground state projector $\mathbb{P}^{\mG}[\mathbb{T}^{3}_{\boxempty}]$ can be equally expressed in terms of the basis elements of ${\rm Tube}^\mG[\mathbb T^2_\boxempty]$ as follows:
\begin{align}
	\label{eq:tori-gsprojalt}
	\mathbb{P}^{\mG}[\mathbb{T}^{3}_{\boxempty}]= \!\!\!\!\!\!\!
	\summand \;\; \summandh \!\!\!\!
	\bigg(
	\Big| \, 
	\tubeSavg{\hx'}{\gy'}{\gz'}{{\mE}'_{\gy',\gz'}}{0.75}{1.25}
	\Big\ra^{\!\! -1} \!\!\!\!\!
	\star
	\Big| \,
	\tubeSavg{\hx}{\gy}{\gz}{\mE_{\gy,\gz}}{0.75}{1.25}
	\Big\ra \!
	\star
	\Big| \,  
	\tubeSavg{\hx'}{\gy'}{\gz'}{\mE'_{\gy',\gz'}}{0.75}{1.25}
	\Big\ra
	\bigg)
	\Big\la \,
	\tubeSavg{\hx}{\gy}{\gz}{\mE_{\gy,\gz}}{0.75}{1.25} \,
	\Big| \, ,
\end{align}
where given an equivalence class $\mE_{\gy,\gz}$ whose representative element is $(\exOne, \fyOne, \fzOne)$ we have 
\begin{align}\label{eq:inverse}
	\Big| \, 
	\tubeSavg{\hx}{\gy}{\gz}{{\mE}_{\gy,\gz}}{0.75}{1.25}
	\Big\ra^{\!\! -1} \!\!\!
	:= \,\,
	\Big| \, 
	\tubeSavg{\bhx}{\bgy}{\bgz}{{\mE}^{-1}_{\bgy,\bgz}}{0.75}{1.25}
	\Big\ra
\end{align}
with $\bgy=\gy^{\exOne\,;\, \fzOne}, \bgz=\gz^{\exOne \,;\, \fyOne}$, $\bhx=\hx^{\exOne,\gy,\gz \,;\, \fyOne,\fzOne}$ and such that ${\mE}^{-1}_{\bgy,\bgz}\in\mathfrak{B}_{\bgy,\bgz}$ is the equivalence class with representative element  $(\exOne^{-1}, \exOne^{-1} \act \fyOne^{-1}, \exOne^{-1} \act \fzOne^{-1})$. The proof that both expressions do define the same operator is presented in app.~\ref{sec:app_gsprojector}. Furthermore, we can define the identity element of ${\rm Tube}^\mG[\mathbb T^2_\boxempty]$ via
\begin{align*}
	\mathbb{1}_{{\rm Tube}^\mG[\mathbb T^2_\boxempty]}:=
	\!\!\!\!\!\!\!
	\summand
	\Big| \,
	\tubeSavg{\hx}{\gy}{\gz}{{\mE}_{\gy,\gz}}{0.75}{1.25}
	\Big\rangle^{\!\! -1} \!\!\!\!\!
	\star
	\Big| \,
	\tubeSavg{\hx}{\gy}{\gz}{{\mE}_{\gy,\gz}}{0.75}{1.25}
	\Big\ra
\end{align*}
such that
\begin{align*}
	\mathbb{1}_{{\rm Tube}^\mG[\mathbb T^2_\boxempty]}
	\star
	\Big|\; 
	\tubeSavg{\hx}{\gy}{\gz}{\mE_{\gy,\gz}}{0.75}{1.25}
	\Big\rangle \!
	=
	\Big| \,
	\tubeSavg{\hx}{\gy}{\gz}{\mE_{\gy,\gz}}{0.75}{1.25}
	\Big\rangle \!
	=
	\Big| \,
	\tubeSavg{\hx}{\gy}{\gz}{\mE_{\gy,\gz}}{0.75}{1.25}
	\Big\rangle
	\star
	\mathbb{1}_{{\rm Tube}^\mG[\mathbb T^2_\boxempty]}
	\; , \q  
	\forall \, \Big| \, \tubeSavg{\hx}{\gy}{\gz}{\mE_{\gy,\gz}}{0.75}{1.25}\Big\ra\in {\rm Tube}^\mG[\mathbb T^2_\boxempty] \; .
\end{align*}
It then follows that the image of the ground state projector
consists of all elements $|\psi\ra\in {\rm Tube}^\mG[\mathbb T^2_\boxempty]$ satisfying the condition
\begin{align*}
	|\psi \ra
	\star
	\Big| \, \tubeSavg{\hx}{\gy}{\gz}{\mE_{\gy,\gz}}{0.75}{1.25}\Big\ra
	=
	\Big| \, \tubeSavg{\hx}{\gy}{\gz}{\mE_{\gy,\gz}}{0.75}{1.25}\Big\ra
	\star
	|\psi \ra
	\; , \q  
	\forall \, \Big| \, \tubeSavg{\hx}{\gy}{\gz}{\mE_{\gy,\gz}}{0.75}{1.25}\Big\ra\in {\rm Tube}^\mG[\mathbb T^2_\boxempty] \; .
\end{align*}
This expression is nothing else than the definition of
the central subalgebra $Z({\rm Tube}^\mG[\mathbb T^2_\boxempty])$ defined in sec.~\ref{sec:centralbasis}, and as such we can make the identifications
\begin{align*}
	\mathcal{V}^\mG[\mathbb{T}^{3}_{\boxempty}]
	:=
	{\rm Im} \, \mathbb{P}^{\mG}[\mathbb{T}^{3}_{\boxempty}]
	=
	Z({\rm Tube}^\mG[\mathbb T^2_\boxempty])
	=
	{\rm Span}_{\mathbb{C}}\big\{ | \mC ,R \ra \big\}_{\forall \, \mC,R}
	\; ,
\end{align*}
where the central elements $\ket{\mC,R}$ were defined in \eqref{eq:centralcanonicalbasis}. One immediate consequence of this result is that the ground state degeneracy of the three-torus in the higher gauge model is equal to the number of elementary loop-like excitations.

\bigskip
\section{Discussion\label{sec:discussion}}
Topological models with a higher gauge theory interpretation have recently been under much scrutiny. In this manuscript, we studied within the lattice Hamiltonian formalism the excitation content of higher gauge models whose input data are strict 2-groups. In order to accomplish this task, we generalized the tube algebra approach, which has been very successful in the study of gauge models, to higher gauge models. More precisely, we considered the tube algebra associated with the manifold $\mathbb T^2 \times [0,1]$ so that the corresponding simple modules classify the elementary loop-like excitations of the model. The methodology is exactly the same as the one followed to derive the elementary loop-like excitations of Dijkgraaf-Witten models. However, the derivations are considerably more subtle in the higher gauge theory case due to the presence of both 1-form and 2-form degrees of freedom that interact in a non-trivial way, as well as the requirement of 1-form gauge invariance on the ground states of $\mathbb T^2 \times [0,1]$.

Although we focused on the case of loop-like excitations, we could easily consider more complex excitations whose classifications correspond to the classifications of boundary conditions of higher-genus surfaces $\Sigma$. Such scenarios have been studied using the language of strict 2-groupoids \cite{Bullivantthesis}. In this case, the relevant 2-groupoid consists of objects given by boundary colourings of $\Sigma \times [0,1]$, 1-morphisms given by bulk colourings, and 2-morphisms that correspond to 1-form gauge transformations between bulk colourings. Within this context, tube algebras can be rephrased in terms of groupoid algebras, and the corresponding simple modules can be conveniently found using the technology of groupoid representations \cite{willerton2008twisted}.

The techniques introduced in this work admit several generalisations. Firstly, we could include a crossed module 4-cocycle as input of our model, where such an algebraic cocycle would be identified with a simplicial 4-cocycle in the cohomology of the crossed module classifying space. Secondly, we could replace the input strict 2-group by a weak 2-group defined as  a monoidal category whose objects are all weakly invertible and morphisms are all invertible. Isomorphism classes of weak 2-groups are classified by quadruples $(\Gamma_1, \Gamma_2, [\alpha], \act)$ where $\Gamma_1$ is a group, $\Gamma_2$ an abelian group, $\act: \Gamma_1 \to {\rm Aut}(\Gamma_2)$ a group action, and $[\alpha] \in H^3(\Gamma_1,\Gamma_2)$. In the present context, the 3-cocycle $\alpha$, which determines the monoidal associator, would appear in the definition of the 2-flatness constraint for a 3-simplex. As such, this scenario would require dealing with triangulations instead of cubulations. Thirdly, the strategy employed in this manuscript can be adapted to study the excitation content of gapped boundaries for higher gauge models. These generalizations will be reported in a forthcoming paper. 

Finally, it is tantalising  to study the fusion and the braiding statistics of the elementary loop-like excitations derived in this manuscript. Indeed, the authors showed in \cite{Bullivant:2019fmk} that for gauge models the tube algebra for torus boundary can be equipped with a comultiplication map and an $R$-matrix that encode the fusion and the braiding statistics of the excitations, respectively. Similarly, we could try to endow the algebra obtained in the present manuscript with the corresponding structures. However, in light of the complexity of the elementary excitations in higher gauge models, there is no straightforward way of generalizing these notions. A well-studied approach to understanding the braid statistics of loop-like excitations in (3+1)d is to consider the mapping class group representations of the three-torus ${\rm SL}(3,\mathbb{Z})$ induced from the three-torus ground state subspace \cite{Jiang:2014ksa,Wan:2014woa,Wang:2014oya,Wang:2014xba}. It is  expected that the fusion rules for loop-like excitations are related to a generalised Verlinde formula \cite{verlinde1988fusion} induced by representations of ${\rm SL}(2,\mathbb{Z})\subset {\rm SL}(3,\mathbb{Z})$. Such techniques will be applied to the higher gauge theory model in a subsequent work \cite{MCG}.

\begin{center}
	\textbf{Acknowledgments}
\end{center}
\noindent
CD would like to thank Apoorv Tiwari for numerous discussions on related topics.
This project has received funding from the European Research Council (ERC) under the European Union’s Horizon 2020 research and innovation programme through the ERC Starting Grant WASCOSYS (No. 636201) as well as the Deutsche Forschungsgemeinschaft (DFG, German Research Foundation) under Germany’s Excellence Strategy – EXC-2111 – 390814868.
AB would like to thank Yidun Wan for related discussions. AB is funded by the EPSRC doctoral prize fellowship.

\newpage
\appendix
\titleformat{name=\section}[display]
{\normalfont}
{\footnotesize\centering {APPENDIX \thesection}}
{0pt}
{\large\bfseries\centering}
\section{Strict 2-groups and 2-groupoids\label{sec:app_groupoid}}
\emph{In this appendix, we formulate strict 2-groups as one-object 2-groupoids. We then show that this can be used to define strict 2-group connections as functors from path 2-groupoids to strict 2-groups.}

\subsection{Crossed modules as 2-groupoids}
Given a crossed module $(G,H,\partial,\act)$, let us derive the corresponding strict 2-group $\mathcal{G}$, or more precisely its delooping. The strict 2-group $\mathcal{G}$ is a 2-groupoid whose single object is notated $\bul$, (1-)morphisms are elements in $G$ depicted as $\bul \xrightarrow{\, g \,} \bul$, and 2-morphisms are pairs $\lambda := (g,h) \in G \times H$ such that $(g,h)$ is the 2-morphisms from the \emph{source} 1-morphism $g$ to the \emph{target} 1-morphism $\partial(h)g$ depicted as
\begin{equation*}
	\begin{tikzpicture}[scale=1,baseline=0em]
	\matrix[matrix of math nodes,row sep =2em,column sep=3.7em, ampersand replacement=\&] (m) {
		\bul \& \bul
		\\
	};
	\path
	(m-1-1) edge[out = 60, in=120, looseness=0.8, ->, edge node={coordinate[pos=0.5] (sp1)}] node[pos=0.5, above] {${\sss g}$} (m-1-2)
	(m-1-1) edge[out = -60, in=-120, looseness=0.8, ->, edge node={coordinate[pos=0.5] (sp2)}] node[pos=0.5, below] {${\sss \partial(h)g}$} (m-1-2)
	(sp1) edge[double, double equal sign distance, -implies, line cap=butt, shab] node[pos=0.5, right] {${\sss h}$} (sp2) 
	;
	\end{tikzpicture} \!\! .
\end{equation*}
The 1-morphisms compose according to the group multiplication in $G$, i.e. $\bul \xrightarrow{\, g_1 \,} \bul \xrightarrow{\, g_2 \,} = \bul \xrightarrow{g_1g_2}\bul$, while the vertical composition of the 2-morphisms is provided by the group multiplication in $H$, i.e.
\begin{equation*}
	\begin{tikzpicture}[scale=1,baseline=-0.2em]
		\matrix[matrix of math nodes,row sep =2em,column sep=3.7em, ampersand replacement=\&] (m) {
			\bul \& \bul
			\\
		};
		\path
		(m-1-1) edge[out = 60, in=120, looseness =0.8, ->, edge node={coordinate[pos=0.5] (sp1)}] node[pos=0.5, above] {${\sss g}$} (m-1-2)
		(m-1-1) edge[out = -60, in=-120, looseness =0.8, ->, edge node={coordinate[pos=0.5] (sp3)}] node[pos=0.5, below] {${\sss \partial(h_2h_1)g}$} (m-1-2)
		(m-1-1) edge[->, edge node={coordinate[pos=0.5] (sp2)}] (m-1-2)
		(sp1) edge[double, double equal sign distance, -implies, line cap=butt, shab] node[pos=0.5, right] {${\sss h_1}$} (sp2) 
		(sp2) edge[double, double equal sign distance, -implies, line cap=butt, shab] node[pos=0.5, right] {${\sss h_2}$} (sp3) 
		;
	\end{tikzpicture} = 
	\begin{tikzpicture}[scale=1,baseline=-0.2em]
		\matrix[matrix of math nodes,row sep =2em,column sep=3.7em, ampersand replacement=\&] (m) {
			\bul \& \bul
			\\
		};
		\path
		(m-1-1) edge[out = 60, in=120, looseness =0.8, ->, edge node={coordinate[pos=0.5] (sp1)}] node[pos=0.5, above] {${\sss g}$} (m-1-2)
		(m-1-1) edge[out = -60, in=-120, looseness =0.8, ->, edge node={coordinate[pos=0.5] (sp3)}] node[pos=0.5, below] {${\sss \partial(h_2h_1)g}$} (m-1-2)
		(sp1) edge[double, double equal sign distance, -implies, line cap=butt, shab] node[pos=0.5, right] {${\sss h_2h_1}$} (sp3) 
		;
	\end{tikzpicture}\!\! .
\end{equation*}
It is also possible to compose the 2-morphisms horizontally
\begin{equation*}
	\begin{tikzpicture}[scale=1,baseline=0em]
		\matrix[matrix of math nodes,row sep =2em,column sep=3.7em, ampersand replacement=\&] (m) {
			\bul \& \bul \& \bul
			\\
		};
		\path
		(m-1-1) edge[out = 60, in=120, looseness=0.8, ->, edge node={coordinate[pos=0.5] (sp1)}] node[pos=0.5, above] {${\sss g_1}$} (m-1-2)
		(m-1-1) edge[out = -60, in=-120, looseness=0.8, ->, edge node={coordinate[pos=0.5] (sp2)}] node[pos=0.5, below] {${\sss \partial(h_1)g_1}$} (m-1-2)
		(sp1) edge[double, double equal sign distance, -implies, line cap=butt, shab] node[pos=0.5, right] {${\sss h_1}$} (sp2) 
		(m-1-2) edge[out = 60, in=120, looseness=0.8, ->, edge node={coordinate[pos=0.5] (sp3)}] node[pos=0.5, above] {${\sss g_2}$} (m-1-3)
		(m-1-2) edge[out = -60, in=-120, looseness=0.8, ->, edge node={coordinate[pos=0.5] (sp4)}] node[pos=0.5, below] {${\sss \partial(h_2)g_2}$} (m-1-3)
		(sp3) edge[double, double equal sign distance, -implies, line cap=butt, shab] node[pos=0.5, right] {${\sss h_2}$} (sp4)
		;
	\end{tikzpicture} = 
	\begin{tikzpicture}[scale=1,baseline=0em]
		\matrix[matrix of math nodes,row sep =2em,column sep=3.7em, ampersand replacement=\&] (m) {
			\bul \& \bul
			\\
		};
		\path
		(m-1-1) edge[out = 60, in=120, looseness=0.8, ->, edge node={coordinate[pos=0.5] (sp1)}] node[pos=0.5, above] {${\sss g_1g_2}$} (m-1-2)
		(m-1-1) edge[out = -60, in=-120, looseness=0.8, ->, edge node={coordinate[pos=0.5] (sp2)}] node[pos=0.5, below] {${\sss \partial(h)g_1g_2}$} (m-1-2)
		(sp1) edge[double, double equal sign distance, -implies, line cap=butt, shab] node[pos=0.5, right] {${\sss \tilde h}$} (sp2) 
		;
	\end{tikzpicture}
\end{equation*}
with $\tilde h := h_2(g_2 \act h_1)$ so that the set of 2-morphisms forms the semidirect product $G \ltimes H$. Notating $\lambda_i := (g_i,h_i)$, the multiplication rule $*$ in $G \ltimes H$ reads
\begin{equation*}
	\lambda_{h_1} * \lambda_{h_2} = (g_1,h_1) * (g_2,h_2)  = (g_1g_2, h_2(g_2 \act h_1)) \; .
\end{equation*}
Vertical and horizontal compositions can be checked to satisfy the interchange law
\begin{equation*}
	(\lambda_1 \circ \lambda_2) * (\lambda_1' \circ \lambda_2') = (\lambda_1 * \lambda_1') \circ (\lambda_2 * \lambda_2')	
\end{equation*}
such that there is a well-defined 2-morphism associated with the diagram
\begin{equation*}
	\begin{tikzpicture}[scale=1,baseline=0em]
		\matrix[matrix of math nodes,row sep =2em,column sep=3.7em, ampersand replacement=\&] (m) {
			\bul \& \bul \& \bul
			\\
		};
		\path
		(m-1-1) edge[out = 60, in=120, looseness=0.8, ->, edge node={coordinate[pos=0.5] (sp1)}] node[pos=0.5, above] {${\sss g_1}$} (m-1-2)
		(m-1-1) edge[out = -60, in=-120, looseness=0.8, ->, edge node={coordinate[pos=0.5] (sp2)}] node[pos=0.5, below] {${\sss \partial(h_2h_1)g_1}$} (m-1-2)
		(m-1-1) edge[->, edge node={coordinate[pos=0.5] (sp11)}] (m-1-2)
		(sp1) edge[double, double equal sign distance, -implies, line cap=butt, shab] node[pos=0.5, right] {${\sss h_1}$} (sp11) 
		(sp11) edge[double, double equal sign distance, -implies, line cap=butt, shab] node[pos=0.5, right] {${\sss h_2}$} (sp2) 
		(m-1-2) edge[out = 60, in=120, looseness=0.8, ->, edge node={coordinate[pos=0.5] (sp3)}] node[pos=0.5, above] {${\sss g_2}$} (m-1-3)
		(m-1-2) edge[out = -60, in=-120, looseness=0.8, ->, edge node={coordinate[pos=0.5] (sp4)}] node[pos=0.5, below] {${\sss \partial(h_2'h_1')g_2}$} (m-1-3)
		(m-1-2) edge[->, edge node={coordinate[pos=0.5] (sp33)}] (m-1-3)
		(sp3) edge[double, double equal sign distance, -implies, line cap=butt, shab] node[pos=0.5, right] {${\sss h_1'}$} (sp33)
		(sp33) edge[double, double equal sign distance, -implies, line cap=butt, shab] node[pos=0.5, right] {${\sss h_2'}$} (sp4)
		;
	\end{tikzpicture} \! \! ,
\end{equation*}
independent of the order of composition.
Conversely, we can define the crossed module associated with a given one-object 2-groupoid.

\subsection{Path groupoids and strict 2-group colourings}

Conventional gauge theories are built from (1-)connections on principle bundles, and, given a topologically trivial bundle, a 1-connection  can be completely determined by the holonomies of a 1-form gauge field valued in the Lie algebra of the gauge group. In the limiting case that $G$ is a finite group, the 1-connection is commonly replaced by a $G$-valued 1-cochain. In this limit, a systematic way of providing a local description of a connection on a manifold $\mathcal{M}$ is through the language of category theory in terms of path groupoids. The same language can then be used to describe strict 2-group connections. Here we provide only an overview of such ideas and suggest the following sources for a more comprehensive treatment: \cite{baez2004higher,schreiber2007parallel,bartels2004higher,Pfeiffer:2003je}.

Let us begin by describing the \emph{path groupoid} $\mathcal{P}(\mathcal{M})$ for a manifold $\mathcal{M}$. The path groupoid $\mathcal{P}(\mathcal{M})$ is a category whose object set is given by a finite set of points in $\mathcal{M}$ and morphisms are given by oriented paths connecting such points. Composition of morphisms then corresponds to the concatenation of paths. Given two points $\mathsf{v}, \mathsf{v}' \in \mathcal{M}$, we denote an oriented (1-)path between them as $\mathsf{v}\xrightarrow{\, \mathsf{e} \,}\mathsf{v}'$.\footnote{In order to facilitate the comparison, we use the same notation for points, 1-paths and 2-paths as the one for vertices, edges and plaquettes in sec.~\ref{sec:Ham} namely $\mathsf{v}$, $\mathsf{e}$ and $\mathsf{p}$, respectively.} For such data to define a category, we additionally require for each point $\mathsf{v} \in \mathcal{M}$ a `trivial path' $ \mathsf{v} \xrightarrow{\, 1_\mathsf{v} \,} \mathsf{v}$ whose support is the point $\mathsf{v}$, defining the identity morphism for the point $\mathsf{v}$. The groupoid structure is given by defining the inverse of an oriented path $\mathsf{v}\xrightarrow{\, \mathsf{e} \,}\mathsf{v}'$ as the orientation reversal of the path, notated via $\mathsf{v}'\xrightarrow{\, \mathsf{e}^{-1}}\mathsf{v}$ so that it satisfies the identities
\begin{align*}
	\mathsf{v}\xrightarrow{\, \mathsf{e}\, } \mathsf{v}' \xrightarrow{\, \mathsf{e}^{-1}}\mathsf{v}
	:=\mathsf{v}\xrightarrow{\, 1_{\mathsf{v}} \, }\mathsf{v}\q \text{and} \q
	\mathsf{v}'\xrightarrow{\, \mathsf{e}^{-1}} \mathsf{v} \xrightarrow{\, \mathsf{e} \, }\mathsf{v}'
	:=\mathsf{v}'\xrightarrow{ \, 1_{\mathsf{v}' \, }}\mathsf{v}'\,\,.
\end{align*}
Building on the path groupoid construction, a $G$-connection is expressed via the \emph{1-holonomy functor}
\begin{align*}
	{\rm hol}_{1}:\mathcal{P}(\mathcal{M})\rightarrow \overline{G} \; ,
\end{align*}
where $\overline{G}$ is the \emph{delooping} of $G$, i.e. the one object groupoid with morphisms labelled by elements of $G$ and composition given by multiplication in $G$. In particular, the functor ${\rm hol}_{1}$ assigns to each path $\mathsf{v} \xrightarrow{\, \mathsf{e} \,} \mathsf{v}'\in {\rm Hom}_{\mathcal{P}(\mathcal{M})}(\mathsf{v}, \mathsf{v}')$ an element ${\rm hol}_{1}(\mathsf{v}\xrightarrow{\, \mathsf{e} \, }\mathsf{v}')=\bul\xrightarrow{ \, g_{\mathsf{e}} \, }\bul\in G$ such that
\begin{align*}
	{\rm hol}_{1}(\mathsf{v}\xrightarrow{\mathsf{e}}\mathsf{v}'\xrightarrow{\mathsf{e}'}\mathsf{v}'')
	=
	{\rm hol}_{1}(\mathsf{v}\xrightarrow{\mathsf{e}}\mathsf{v}')
	{\rm hol}_{1}(\mathsf{v}'\xrightarrow{\mathsf{e}'}\mathsf{v}'') \; ,
\end{align*}
where the composition rule on the r.h.s is the multiplication in $G$. Furthermore, functorality implies the relations
\begin{align*}
	{\rm hol}_{1}(\mathsf{v}'\xrightarrow{\, \mathsf{e}^{-1}}\mathsf{v})
	=
	{\rm hol}_{1}(\mathsf{v}\xrightarrow{\, \mathsf{e} \,}\mathsf{v}')^{-1}
	\q \text{and} \q
	{\rm hol}_{1}(\mathsf{v}\xrightarrow{\, 1_{\mathsf{e}} \, }\mathsf{v})
	=\mathbb 1_{G} \; .
\end{align*}
In this way, the requirement that ${\rm hol}_{1}$ is a functor is equivalent to the condition that composition of holonomies is well-defined in the connection. In this manuscript, we are primarily interested in 2-connections arising from 2-bundles associated to finite 2-groups $\mG$, which we refer to as $\mG$-labellings. One key advantage of the category theoretical definition of a finite $G$-connection presented above is that it can be neatly extended to 2-connections for finite 2-bundles. Mimicking the group case, let us begin by defining the path 2-groupoid $\mathcal{P}_{2}(\mathcal{M})$. The path 2-groupoid $\mathcal{P}_{2}(\mathcal{M})$ is a strict 2-groupoid whose underlying 1-category is the path groupoid $\mathcal{P}(\mathcal{M})$ and 2-morphisms $\mathsf{e} \xRightarrow{ \; \mathsf{p} \;\; } \mathsf{e}'$ correspond to 2-paths in $\mathcal{M}$  with the topology of a bigon connecting pairs $(\mathsf{e}, \mathsf{e}')$ of 1-paths with the same source and target points, e.g.
\begin{align*}
	\begin{tikzpicture}[scale=1,baseline=0em]
	\matrix[matrix of math nodes,row sep =2em,column sep=3.7em, ampersand replacement=\&] (m) {
		\mathsf{v} \& \mathsf{v}\smash{'}
		\\
	};
	\path
	(m-1-1) edge[out = 60, in=120, looseness=0.8, ->, edge node={coordinate[pos=0.5] (sp1)}] node[pos=0.5, above] {${\sss \mathsf{e}}$} (m-1-2)
	(m-1-1) edge[out = -60, in=-120, looseness=0.8, ->, edge node={coordinate[pos=0.5] (sp2)}] node[pos=0.5, below] {${\sss \mathsf{e}'}$} (m-1-2)
	(sp1) edge[double, double equal sign distance, -implies, line cap=butt, shab] node[pos=0.5, right] {${\sss \mathsf{p}}$} (sp2) 
	;
	\end{tikzpicture} \!\! .
\end{align*}
To ensure this defines a strict 2-category, we require for each path $\mathsf{v} \xrightarrow{\, \mathsf{e} \,}\mathsf{v}'$ a `trivial 2-path' $\mathsf{e} \xRightarrow{\, 1_{\mathsf{e}} \;\;} \mathsf{e}'$ whose support is contained on the path $\mathsf{v} \xrightarrow{\, \mathsf{e} \,}\mathsf{v}'$, defining the identity 2-morphism  for the path $\mathsf{v} \xrightarrow{\, \mathsf{e} \,}\mathsf{v}'$. The 2-groupoid structure then follows by defining a vertical inverse $^{-1}$ and a horizontal inverse $^\dagger$, associated with the two ways in which we can reverse the orientation of a bigon, via
\begin{align*}
	\bigg( \!\!\!
	\begin{tikzpicture}[scale=1,baseline=0em]
	\matrix[matrix of math nodes,row sep =2em,column sep=3.7em, ampersand replacement=\&] (m) {
		\mathsf{v} \& \mathsf{v}\smash{'}
		\\
	};
	\path
	(m-1-1) edge[out = 60, in=120, looseness=0.8, ->, edge node={coordinate[pos=0.5] (sp1)}] node[pos=0.5, above] {${\sss \mathsf{e}}$} (m-1-2)
	(m-1-1) edge[out = -60, in=-120, looseness=0.8, ->, edge node={coordinate[pos=0.5] (sp2)}] node[pos=0.5, below] {${\sss \mathsf{e}'}$} (m-1-2)
	(sp1) edge[double, double equal sign distance, -implies, line cap=butt, shab] node[pos=0.5, right] {${\sss \mathsf{p}}$} (sp2) 
	;
	\end{tikzpicture} \!\!
	\!\! \bigg)^{-1} = 
	\begin{tikzpicture}[scale=1,baseline=0em]
	\matrix[matrix of math nodes,row sep =2em,column sep=3.7em, ampersand replacement=\&] (m) {
		\mathsf{v} \& \mathsf{v}\smash{'}
		\\
	};
	\path
	(m-1-1) edge[out = 60, in=120, looseness=0.8, ->, edge node={coordinate[pos=0.5] (sp1)}] node[pos=0.5, above] {${\sss \mathsf{e}}$} (m-1-2)
	(m-1-1) edge[out = -60, in=-120, looseness=0.8, ->, edge node={coordinate[pos=0.5] (sp2)}] node[pos=0.5, below] {${\sss \mathsf{e}'}$} (m-1-2)
	(sp1) edge[double, double equal sign distance, implies-, line cap=butt, shab] node[pos=0.5, right] {${\sss \mathsf{p}^{-1}}$} (sp2) 
	;
	\end{tikzpicture} 
	\q \text{and} \q
	\bigg( \!\!\!
	\begin{tikzpicture}[scale=1,baseline=0em]
	\matrix[matrix of math nodes,row sep =2em,column sep=3.7em, ampersand replacement=\&] (m) {
		\mathsf{v} \& \mathsf{v}\smash{'}
		\\
	};
	\path
	(m-1-1) edge[out = 60, in=120, looseness=0.8, ->, edge node={coordinate[pos=0.5] (sp1)}] node[pos=0.5, above] {${\sss \mathsf{e}}$} (m-1-2)
	(m-1-1) edge[out = -60, in=-120, looseness=0.8, ->, edge node={coordinate[pos=0.5] (sp2)}] node[pos=0.5, below] {${\sss \mathsf{e}'}$} (m-1-2)
	(sp1) edge[double, double equal sign distance, -implies, line cap=butt, shab] node[pos=0.5, right] {${\sss \mathsf{p}}$} (sp2) 
	;
	\end{tikzpicture} \!\!
	\!\! \bigg)^\dagger
	=
	\begin{tikzpicture}[scale=1,baseline=0em]
	\matrix[matrix of math nodes,row sep =2em,column sep=3.7em, ampersand replacement=\&] (m) {
		\mathsf{v}\smash{'}\, \& \mathsf{v}
		\\
	};
	\path
	(m-1-1) edge[out = 60, in=120, looseness=0.8, <-, edge node={coordinate[pos=0.5] (sp1)}] node[pos=0.5, above] {${\sss \mathsf{e}^{-1}}$} (m-1-2)
	(m-1-1) edge[out = -60, in=-120, looseness=0.8, <-, edge node={coordinate[pos=0.5] (sp2)}] node[pos=0.5, below] {${\sss \mathsf{e}'^{-1}}$} (m-1-2)
	(sp1) edge[double, double equal sign distance, -implies, line cap=butt, shab] node[pos=0.5, right] {${\sss \mathsf{p}^\dagger}$} (sp2) 
	;
	\end{tikzpicture} \!\! .
\end{align*}
Building on the path 2-groupoid construction, a $\mG$-connection is expressed via the holonomy strict 2-functor
\begin{align*}
	{\rm hol}_{2}:\mathcal{P}_{2}(\mathcal{M})\rightarrow \mathcal{G} \; .
\end{align*}
In particular, the requirement that such a map corresponds to a strict 2-functor states that there is a functor ${\rm hol}_{1}:\mathcal{P}(\mathcal{M})\rightarrow \overline{G}$
of the underlying 1-category assigning to each path $\mathsf{v} \xrightarrow{\, \mathsf{e} \,}\mathsf{v}'\in {\rm Hom}_{\mathcal{P}(\mathcal{M})}(\mathsf{v}, \mathsf{v}')$ a morphism $\bul\xrightarrow{\, g_{\mathsf{e}} \, }\bul\in \overline{G}$ and to each 2-path
\begin{align*}
	{\rm hol}_{2}:
	\begin{tikzpicture}[scale=1,baseline=0em]
	\matrix[matrix of math nodes,row sep =2em,column sep=3.7em, ampersand replacement=\&] (m) {
		\mathsf{v} \& \mathsf{v}\smash{'}
		\\
	};
	\path
	(m-1-1) edge[out = 60, in=120, looseness=0.8, ->, edge node={coordinate[pos=0.5] (sp1)}] node[pos=0.5, above] {${\sss \mathsf{e}}$} (m-1-2)
	(m-1-1) edge[out = -60, in=-120, looseness=0.8, ->, edge node={coordinate[pos=0.5] (sp2)}] node[pos=0.5, below] {${\sss \mathsf{e}'}$} (m-1-2)
	(sp1) edge[double, double equal sign distance, -implies, line cap=butt, shab] node[pos=0.5, right] {${\sss \mathsf{p}}$} (sp2) 
	;
	\end{tikzpicture}
	\mapsto
	\begin{tikzpicture}[scale=1,baseline=0em]
	\matrix[matrix of math nodes,row sep =2em,column sep=3.7em, ampersand replacement=\&] (m) {
		\bul \& \bul
		\\
	};
	\path
	(m-1-1) edge[out = 60, in=120, looseness=0.8, ->, edge node={coordinate[pos=0.5] (sp1)}] node[pos=0.5, above] {${\sss g_{\mathsf{e}}}$} (m-1-2)
	(m-1-1) edge[out = -60, in=-120, looseness=0.8, ->, edge node={coordinate[pos=0.5] (sp2)}] node[pos=0.5, below] {${\sss g_{\mathsf{e}'}=\partial(h_{\mathsf{p}})g_{\mathsf{e}}}$} (m-1-2)
	(sp1) edge[double, double equal sign distance, -implies, line cap=butt, shab] node[pos=0.5, right] {${\sss h_{\mathsf{p}}}$} (sp2) 
	;
	\end{tikzpicture}
	\in G\ltimes H \; ,
\end{align*}
which is compatible with the 1-functor ${\rm hol}_{1}$ by the requirement $g_{\mathsf{e}'}=\partial(h_{\mathsf{p}})g_{\mathsf{e}}$. This compatibility between 1-path and 2-path labellings corresponds to the fake-flatness condition expressed in \eqref{eq:trivHol1}. Additionally the functor requires
\begin{align*}
	{\rm hol}_{2}\bigg( \!\!\!
	\begin{tikzpicture}[scale=1,baseline=0em]
		\matrix[matrix of math nodes,row sep =2em,column sep=3.7em, ampersand replacement=\&] (m) {
			\mathsf{v} \& \mathsf{v}\smash{'}
			\\
		};
		\path
		(m-1-1) edge[out = 60, in=120, looseness=0.8, ->, edge node={coordinate[pos=0.5] (sp1)}] node[pos=0.5, above] {${\sss \mathsf{e}}$} (m-1-2)
		(m-1-1) edge[out = -60, in=-120, looseness=0.8, ->, edge node={coordinate[pos=0.5] (sp2)}] node[pos=0.5, below] {${\sss \mathsf{e}'}$} (m-1-2)
		(sp1) edge[double, double equal sign distance, -implies, line cap=butt, shab] node[pos=0.5, right] {${\sss \mathsf{p}^{-1}}$} (sp2) 
		;
	\end{tikzpicture} \!\!
	\!\! \bigg) = \!\!
	\begin{tikzpicture}[scale=1,baseline=0em]
		\matrix[matrix of math nodes,row sep =2em,column sep=3.7em, ampersand replacement=\&] (m) {
			\bul \& \bul
			\\
		};
		\path
		(m-1-1) edge[out = 60, in=120, looseness=0.8, ->, edge node={coordinate[pos=0.5] (sp1)}] node[pos=0.5, above] {${\sss \mathsf{e}}$} (m-1-2)
		(m-1-1) edge[out = -60, in=-120, looseness=0.8, ->, edge node={coordinate[pos=0.5] (sp2)}] node[pos=0.5, below] {${\sss \mathsf{e}'}$} (m-1-2)
		(sp1) edge[double, double equal sign distance, implies-, line cap=butt, shab] node[pos=0.5, right] {${\sss h_{\mathsf{p}}^{-1}}$} (sp2) 
		;
	\end{tikzpicture} \!\! \; , \q
	{\rm hol}_{2}\bigg( \!\!\!
	\begin{tikzpicture}[scale=1,baseline=0em]
		\matrix[matrix of math nodes,row sep =2em,column sep=3.7em, ampersand replacement=\&] (m) {
			\mathsf{v} \& \mathsf{v}\smash{'}
			\\
		};
		\path
		(m-1-1) edge[out = 60, in=120, looseness=0.8, <-, edge node={coordinate[pos=0.5] (sp1)}] node[pos=0.5, above] {${\sss \mathsf{e}^{-1}}$} (m-1-2)
		(m-1-1) edge[out = -60, in=-120, looseness=0.8, ->, edge node={coordinate[pos=0.5] (sp2)}] node[pos=0.5, below] {${\sss \mathsf{e}^{'-1}}$} (m-1-2)
		(sp1) edge[double, double equal sign distance, -implies, line cap=butt, shab] node[pos=0.5, right] {${\sss \mathsf{p}^\dagger}$} (sp2) 
		;
	\end{tikzpicture} \!\!
	\!\! \bigg)
	= \!\!
	\begin{tikzpicture}[scale=1,baseline=0em]
		\matrix[matrix of math nodes,row sep =2em,column sep=3.7em, ampersand replacement=\&] (m) {
			\bul \& \bul
			\\
		};
		\path
		(m-1-1) edge[out = 60, in=120, looseness=0.8, <-, edge node={coordinate[pos=0.5] (sp1)}] node[pos=0.5, above] {${\sss \mathsf{e}^{-1}}$} (m-1-2)
		(m-1-1) edge[out = -60, in=-120, looseness=0.8, <-, edge node={coordinate[pos=0.5] (sp2)}] node[pos=0.5, below] {${\sss \mathsf{e}^{'-1}}$} (m-1-2)
		(sp1) edge[double, double equal sign distance, -implies, line cap=butt, shab] node[pos=0.5,right] {${\sss \tilde{h}_{\mathsf{p}}}$} (sp2) 
		;
	\end{tikzpicture} \!\!
\end{align*}
with $\tilde{h}_{\mathsf{p}}=g_{\mathsf{e}}^{-1}\triangleright h_{\mathsf{p}}^{-1}$, and
\begin{align*}
	{\rm hol}_{2}\bigg( \!\!\!
	\begin{tikzpicture}[scale=1,baseline=0em]
		\matrix[matrix of math nodes,row sep =2em,column sep=3.7em, ampersand replacement=\&] (m) {
			\mathsf{v} \& \mathsf{v}\smash{'}
			\\
		};
		\path
		(m-1-1) edge[out = 60, in=120, looseness=0.8, ->, edge node={coordinate[pos=0.5] (sp1)}] node[pos=0.5, above] {${\sss \mathsf{e}}$} (m-1-2)
		(m-1-1) edge[out = -60, in=-120, looseness=0.8, ->, edge node={coordinate[pos=0.5] (sp2)}] node[pos=0.5, below] {${\sss \mathsf{e}}$} (m-1-2)
		(sp1) edge[double, double equal sign distance, -implies, line cap=butt, shab] node[pos=0.5, right] {${\sss 1_{\mathsf{e}}}$} (sp2) 
		;
	\end{tikzpicture} \!\!
	\!\! \bigg)
	= \!\!
	\begin{tikzpicture}[scale=1,baseline=0em]
		\matrix[matrix of math nodes,row sep =2em,column sep=3.7em, ampersand replacement=\&] (m) {
			\mathsf{v} \& \mathsf{v}\smash{'}
			\\
		};
		\path
		(m-1-1) edge[out = 60, in=120, looseness=0.8, ->, edge node={coordinate[pos=0.5] (sp1)}] node[pos=0.5, above] {${\sss g_\mathsf{e}}$} (m-1-2)
		(m-1-1) edge[out = -60, in=-120, looseness=0.8, ->, edge node={coordinate[pos=0.5] (sp2)}] node[pos=0.5, below] {${\sss g_\mathsf{e}}$} (m-1-2)
		(sp1) edge[double, double equal sign distance, -implies, line cap=butt, shab] node[pos=0.5, right] {${\sss \mathbb 1_{H}}$} (sp2) 
		;
	\end{tikzpicture} \!\!
\end{align*}
ensuring that composition of 2-holonomies is well defined. This definition is very general. It can thus be used to redefine $\mathcal{G}$-labellings on cubulations as presented in the main text, but also to define $\mathcal{G}$-labellings of cell decompositions of a manifold given by any CW-complex as illustrated in \cite{Bullivant:2017sjz}.

\section{Properties of the representations matrices\label{sec:app_prop}}
\noindent
\emph{In this appendix, we collect the proofs of several properties satisfied by the representation matrices.}

\subsection{Proof of \eqref{eq:repStab}\label{sec:app_proofStab}}
Given definitions \eqref{eq:defQ}, \eqref{eq:irreps} and \eqref{eq:defEstab}, we confirm in this appendix that $[\mE_\mC^{\rm stab.}]_{i,j} \in Z_\mC$ for every $\mE_{\gy,\gz} \in \mathfrak{B}_{\gy,\gz}$.
Following the definition \eqref{eq:defZC} of $Z_\mathcal{C}$, we first compute
\begin{align}
	\nn
	c_{{\rm y},1}^{\pxi^{-1}\exOne \pxj\, ; \,\pxi^{-1} \act [q_{\hat{\rm z},i}^{-1} \fzOne(\exOne \act \qzj)]} 
	&\stackrel{\mathmakebox[\widthof{\;=\;}]{{\rm def}}}{=} (\pxj^{-1}\exOne^{-1}\pxi)\, \partial\big(\pxi^{-1} \act \big[(\exOne \act \qzj^{-1})\fzOne^{-1}\qzi \big]\big)\, c_{{\rm y},1} \, (\pxi^{-1}\exOne \pxj) \\
	\nn
	&\stackrel{\mathmakebox[\widthof{\;=\;}]{\eqref{eq:Peiffer1}}}{=}
	\pxj^{-1}\exOne^{-1}\,  
	\partial\big( (\exOne \act \qzj^{-1})\fzOne^{-1}\qzi\big)\, \pxi \, c_{{\rm y},1} \, \pxi^{-1}\exOne \pxj \\
	\nn
	&\stackrel{\mathmakebox[\widthof{\;=\;}]{\eqref{eq:Peiffer1}}}{=}
	\pxj^{-1}\, \partial(\qzj^{-1})\exOne^{-1}\partial(\fzOne^{-1}) \partial(\qzi)\, \pxi \, c_{{\rm y},1} \, \pxi^{-1} \exOne \pxj \\
	&\stackrel{\mathmakebox[\widthof{\;=\;}]{\eqref{eq:defQ}}}{=} \pxj^{-1}\, \partial(\qzj^{-1})\exOne^{-1}\partial(\fzOne^{-1}) \, c_{{\rm y},i} \, \exOne \pxj
	\nn
	\\ 
	& \stackrel{\mathmakebox[\widthof{\;=\;}]{\eqref{eq:irreps}}}{=} \pxj^{-1}\partial(\qzj^{-1})c_{{\rm y},j}\pxj \stackrel{\mathmakebox[\widthof{\;=\;}]{\eqref{eq:defQ}}}{=} c_{{\rm y },1}
	\label{eq:check1}
\end{align}
where we used between the second and third lines the fact that $\partial$ is a group homomorphism in addition to \eqref{eq:Peiffer1}. Following exactly the same steps, we similarly find that
\begin{align}
	\label{eq:check2}
	c_{{\rm z},1}^{\pxi^{-1}g_{\rm x}\pxj \,;\,\pxi^{-1} \act [q_{\hat{\rm y},i}^{-1} \fyOne(\exOne \act \qyj)]}  = c_{{\rm z },1} \; .
\end{align}
We are left to check the final identity, namely
\begin{equation}
	\label{eq:check3}
	d_{\hat{\rm x},1}^{\pxi^{-1}\exOne\pxj, c_{{\rm y},1},c_{{\rm z},1}\, ; \,\pxi^{-1} \act [q_{\hat{\rm y},i}^{-1} \fyOne(\exOne \act \qyj)],\pxi^{-1} \act [q_{\hat{\rm z},i}^{-1} \fzOne(\exOne \act \qzj)]}  = d_{\hat{\rm x},1} \; .
\end{equation}
By definition of the notation, the l.h.s is equal to
\begin{align*}
	{\rm l.h.s}\eqref{eq:check3} = (\pxj^{-1}\exOne^{-1}\pxi)\act
	\Big[
	&\big(\pxi^{-1}\act[(\exOne \act \qyj^{-1})\fyOne^{-1}\qyi ]\big)
	\big(( c_{{\rm z},1}\pxi^{-1})\act [(\exOne \act \qzj^{-1})\fzOne^{-1}\qzi] \big) 
	\\
	&d_{\hat{\rm x},1}
	\big((c_{{\rm y},1}\pxi^{-1}) \act [\qyi^{-1}\fyOne(\exOne \act \qyj)] \big)
	\big(\pxi^{-1}\act [\qzi^{-1}\fzOne(\exOne \act \qzj)] \big) \Big] 
\end{align*}
which in virtue of the axioms \eqref{eq:axioms} can be rewritten
\begin{align*}
	{\rm l.h.s}\eqref{eq:check3} = (\pxj^{-1}\exOne^{-1})\act
	\Big[
	&\big((\exOne \act \qyj^{-1})\fyOne^{-1}\qyi\big)
	\big((\pxi c_{{\rm z},1}\pxi^{-1})\act [(\exOne \act \qzj^{-1})\fzOne^{-1}\qzi] \big)
	\\
	&\big(\pxi \act d_{\hat{\rm x},1} \big)
	\big((\pxi c_{{\rm y},1}\pxi^{-1}) \act [ \qyi^{-1}\fyOne(\exOne \act \qyj) ]\big)
	\big(\qzi^{-1}\fzOne(\exOne \act \qzj) \big) \Big] \; .
\end{align*}
But according to \eqref{eq:defQ} we have $\pxi c_{{\rm z},1}\pxi^{-1} = \partial(\qyi^{-1})c_{{\rm z},i}$ and $\pxi c_{{\rm y},1}\pxi^{-1} = \partial(\qzi^{-1})c_{{\rm y},i}$. Moreover, using the second Peiffer identity \eqref{eq:Peiffer2} together with the axioms \eqref{eq:axioms}, the expression above becomes
\begin{align*}
	{\rm l.h.s}\eqref{eq:check3} = (\pxj^{-1}\exOne^{-1})\act
	\Big[
	&\big((\exOne \act \qyj^{-1})\fyOne^{-1}\qyi\big)
	\qyi^{-1}\big(c_{{\rm z},i}\act [(\exOne \act \qzj^{-1})\fzOne^{-1}\qzi ]\big)\qyi
	\\
	&\big(\pxi \act d_{\hat{\rm x},1}\big)
	\qzi^{-1}\big(c_{{\rm y},i} \act [\qyi^{-1}\fyOne(\exOne \act \qyj)] \big)\qzi
	\big(\qzi^{-1}\fzOne(\exOne \act \qzj) \big) \Big] \; .
\end{align*}
Removing superfluous brackets, applying the axioms \eqref{eq:axioms}, and using the fact that according to \eqref{eq:defQ} we have $d_{\hat{\rm x},1} = \pxi^{-1}\act [\qyi^{-1}(c_{{\rm z},i}\act \qzi^{-1})d_{\hat{\rm x},i}(c_{{\rm y},i} \act \qyi)\qzi]$ yields
\begin{align*}
	{\rm l.h.s}\eqref{eq:check3} & \; = \; (\pxj^{-1}\exOne^{-1})\act
	\Big[
	(\exOne \act \qyj^{-1})\fyOne^{-1}
	\big(c_{{\rm z},i}\act [(\exOne \act \qzj^{-1})\fzOne^{-1} ]\big)
	d_{\hat{\rm x},i} \\
	& \hspace{7.3em}
	\big(c_{{\rm y},i} \act [\fyOne(\exOne \act \qyj)] \big)
	\fzOne(\exOne \act \qzj)\Big]
	\\
	&  \stackrel{\mathmakebox[\widthof{\;=\;}]{\eqref{eq:axioms}}}{=}
	\big(\pxj^{-1}\exOne^{-1} \big)\act
	\Big[
	(\exOne \act \qyj^{-1})\fyOne^{-1}
	(c_{{\rm z},i}\exOne \act \qzj^{-1})(c_{{\rm z},i} \act \fzOne^{-1})
	d_{\hat{\rm x},i} \\
	&\hspace{7.3em}
	(c_{{\rm y},i} \act \fyOne)(c_{{\rm y},i} \exOne \act \qyj) 
	\fzOne(\exOne \act \qzj) \Big] \; .
\end{align*}
Making use of the delta functions entering the definition \eqref{eq:irreps} of the representation matrices, we finally obtain
\begin{align*}
	{\rm l.h.s}\eqref{eq:check3} & \; = \; 
	(\pxj^{-1}\exOne^{-1})\act
	\Big[
	(\exOne \act \qyj^{-1})\fyOne^{-1}
	\big(\partial(\fyOne)\exOne c_{{\rm z},j} \act \qzj^{-1}\big)(c_{{\rm z},i} \act \fzOne^{-1})
	d_{\hat{\rm x},i}\\
	& \hspace{7.3em}
	(c_{{\rm y},i} \act \fyOne)\big(\partial(\fzOne)\exOne c_{{\rm y},i}  \act \qyj\big)
	\fzOne(\exOne \act \qzj)\Big]	\\
	\nn
	& \stackrel{\mathmakebox[\widthof{\;=\;}]{\eqref{eq:Peiffer2}}}{=}
	(\pxj^{-1}\exOne^{-1})\act
	\Big[
	(\exOne \act \qyj^{-1})
	(\exOne c_{{\rm z},j} \act \qzj^{-1})\fyOne^{-1}(c_{{\rm z},i} \act \fzOne^{-1}) 
	d_{\hat{\rm x},i} \\
	& \hspace{7.3em}
	(c_{{\rm y},i} \act \fyOne)\fzOne(\exOne c_{{\rm y},i}  \act \qyj)
	(\exOne \act \qzj)  \Big]	\\
	& \stackrel{\mathmakebox[\widthof{\;=\;}]{\eqref{eq:axioms}}}{=}
	\pxj^{-1}\act
	\Big[
	\qyj^{-1}
	\big(c_{{\rm z},j} \act \qzj^{-1})
	\big( \exOne^{-1} \act [\fyOne^{-1}(c_{{\rm z},i} \act \fzOne^{-1}) 
	d_{\hat{\rm x},i} 
	(c_{{\rm y},i} \act \fyOne)\fzOne]\big)( c_{{\rm y},i}  \act \qyj)
	\qzj  \Big] \\
	&  \stackrel{\mathmakebox[\widthof{\;=\;}]{\eqref{eq:irreps}}}{=}
	\pxj^{-1}\act
	\Big[
	\qyj^{-1}
	(c_{{\rm z},j} \act \qzj^{-1})
	d_{\hat{\rm x},j}( c_{{\rm y},i}  \act \qyj)
	 \qzj  \Big] \stackrel{\mathmakebox[\widthof{\;=\;}]{\eqref{eq:defQ}}}{=} d_{\hat{\rm x},1} \; ,
\end{align*}
hence the identity \eqref{eq:check3}.
Putting (\ref{eq:check1}--\ref{eq:check3}) together, we checked that \eqref{eq:repStab} is true, hence confirming definition \eqref{eq:irreps}.

\subsection{Proof of the linearity property \eqref{eq:Lin}\label{sec:app_proofLin}}
In this appendix, we check that the representation matrices as defined in \eqref{eq:irreps} indeed realise an algebra homomorphism:
\begin{align*}
	&	\sum_{k=1}^{|\mathcal C|}\sum_{o=1}^{{\rm dim}(V_R)} \!\!\!
	D^{\mathcal{C},R}_{im,ko}\Big( 	\Big| \, \tubeSavg{\hx}{\gy}{\gz}{\mE_{\gy,\gz}}{0.75}{1.25}
	\Big\ra\Big)
	D^{\mathcal{C},R}_{ko,jn} \Big(
	\Big| \, \tubeSavg{\hx'}{\gy'}{\gz'}{\mE'_{\gy',\gz'}}{0.75}{1.25}
	\Big\ra\Big) 
	\\[-0.3em]
	& \q = \sum_{k=1}^{|\mathcal C|}
	\delta(\gy , c_{{\rm y},i}) \, 
	\delta(\gz , c_{{\rm z},i}) \, 
	\delta(\hx , d_{\hat{\rm x},i}) \, 
	\delta(\gy^{\exOne \, ; \, \fzOne} , c_{{\rm y},k}) \, 
	\delta(\gz^{\exOne \, ; \, \fyOne} , c_{{\rm z},k}) \, \delta(\hx^{\exOne,\gy,\gz \, ; \, \fyOne,\fzOne} , d_{\hat{\rm x},k})
	\\[-0.5em]
	& \hspace{2.6em} 
	\times \delta(\gy' , c_{{\rm y},k}) \, 
	\delta(\gz' , c_{{\rm z},k}) \, 
	\delta(\hx' , d_{\hat{\rm x},k}) \, 
	\delta(\gy'^{\exOne'\, ; \, \fzOne'} , c_{{\rm y},j}) \, 
	\delta(\gz'^{\exOne'\, ; \, \fyOne'} , c_{{\rm z},j}) \, \delta(\hx'^{\exOne',\gy',\gz' \,;\, \fyOne',\fzOne'} , d_{\hat{\rm x},j})
	\\[-0.2em]
	& \q\times  \!\!\!\!
	\sum_{o= 1}^{{\rm dim(V_R)}} 
	D^R_{mo}\big( 	\big| \xrightarrow{\; [\mE_\mC^{\rm stab.}]_{i,k} \;}\big\ra \big) 
	D^R_{on}\big( 	\big| \xrightarrow{\; [\mE'^{\rm stab.}_\mC]_{k,j} \;}\big\ra \big)
	\\
	& \q =
	\delta(\gy , c_{{\rm y},i}) \, 
	\delta(\gz , c_{{\rm z},i}) \,
	\delta(\hx , d_{\hat{\rm x},i}) \, 
	\delta(\gy'^{\exOne' \, ; \, \fzOne'} , c_{{\rm y},j}) \, 
	\delta(\gz'^{\exOne' \,;\, \fyOne'} , c_{{\rm z},j}) \, \delta(\hx'^{\exOne',\gy',\gz' \,;\, \fyOne',\fzOne'} , d_{\hat{\rm x},j})
	\\[0.3em]
	& \q \times
	\delta(\gy^{\exOne \,;\, \fzOne} , \gy') \, 
	\delta(\gz^{\exOne \,;\, \fyOne} , \gz') \, 
	\delta(\hx^{\exOne,\gy,\gz \,;\, \fyOne,\fzOne}, \hx')  D^R_{mn}\big( \big| \xrightarrow{\; [(\mE \cdot \mE')^{\rm stab.}_\mC]_{i,j} \;}\big)
	\\[-0.5em]
	& \q =
	D^{\mathcal{C},R}_{im,jn}\Big( 	\Big| \,	\tubeSavg{\hx}{\gy}{\gz}{\mE_{\gy,\gz}}{0.75}{1.25}
	\Big\ra \! \star
	\Big| \, \tubeSavg{\hx'}{\gy'}{\gz'}{\mE'_{\gy',\gz'}}{0.75}{1.25}
	\Big\ra \Big)  \; ,
\end{align*}
where we used in particular the fact that
\begin{align*}
	&\pxi^{-1} \act [\qzi^{-1} \fzOne(\exOne \act \qzk)] (\pxi^{-1}\exOne \pxk)\act \pxk^{-1} \act [\qzk^{-1} \fzOne'(\exOne' \act \qzj)] \\
	& \q = 
	\pxi^{-1} \act [\qzi^{-1} \fzOne(\exOne \act \fzOne')(\exOne \exOne' \act \qzj)] \; ,
\end{align*}
which simply follows from repeated use of the axioms \eqref{eq:axioms}, so as to define the equivalence class $[(\mE \cdot \mE')^{\rm stab.}_\mC]_{i,j} \in \mathfrak{B}_\mathcal{C}$  whose representative element is provided by the triplet
\begin{align*}
	\big (\pxi^{-1}\exOne \exOne' \pxj \,,\, \pxi^{-1} \act [\qzi^{-1} \fzOne(\exOne \act \fzOne')(\exOne \exOne' \act \qzj)] \,,\, \pxi^{-1} \act [\qyi^{-1} \fyOne(\exOne \act \fyOne')(\exOne \exOne' \act \qyj)]\big) \; .
\end{align*}
Furthermore, between the second and the third steps, we used the linearity of the irreducible representation on $Z_\mathcal{C}$.

\subsection{Proof of the orthogonality relation \eqref{eq:ortho}\label{sec:app_proofOrtho}}
In this appendix, we prove the orthogonality relation \eqref{eq:ortho}. It suffices to write down explicitly the definition \eqref{eq:irreps} for the irreducible representations of ${\rm Tube}^{\mG}[\mathbb{T}^{2}_{\boxempty}]$ and use the orthogonality of the representations in the stabiliser group, i.e.
\begin{align*}
	&\!\!\!\!	\summand \!\!\!\!
	D^{\mC,R}_{im,jn}\Big( 	\Big| \, 	\tubeSavg{\hx}{\gy}{\gz}{\mE_{\gy,\gz}}{0.75}{1.25}
	\Big\ra \Big)
	\overline{D^{\mC',R'}_{i'm',j'n'}\Big( 	\Big| \, 	\tubeSavg{\hx}{\gy}{\gz}{\mE_{\gy,\gz}}{0.75}{1.25}
	\Big\ra \Big)} 
	\\[0.2em]
	& \q = \!\!\!\!\!\!	\summand \q \q
	\delta(\gy , c_{{\rm y},i}) \, 
	\delta(\gz , c_{{\rm z},i}) \, 
	\delta(\hx , d_{\hat{\rm x},i}) 
	\\[-2.8em]
	& \hspace{6.8em}  \q \times \delta(\gy^{\exOne \, ; \, \fzOne} \! ,c_{{\rm y},j}) \, 
	\delta(\gz^{\exOne \, ; \, \fyOne} \! ,c_{{\rm z},j}) \, \delta(\hx^{\exOne,\gy,\gz \, ; \, \fyOne,\fzOne} , d_{\hat{\rm x},j})
	\\[0.1em]
	& \hspace{6.8em}  \q \times  	\delta(\gy , c_{{\rm y},i'}) \, 
	\delta(\gz ,c_{{\rm z},i'}) \, 
	\delta(\hx , d_{\hat{\rm x},i'}) 
	\\
	&\hspace{6.8em}  \q \times 
	\delta(\gy^{\exOne \, ; \, \fzOne}, c_{{\rm y},j'}) \,
	\delta(\gz^{\exOne \, ; \, \fyOne} \! ,c_{{\rm z},j'}) \, \delta(\hx^{\exOne,\gy,\gz \, ; \, \fyOne,\fzOne} \! ,d_{\hat{\rm x},j'})
	\\
	& \hspace{6.8em}  \q \times 
	D^R_{mn}\big( 	\big| \xrightarrow{\; [\mE_\mC^{\rm stab.}]_{i,j} \;}\big\ra \big) 
	\overline{D^R_{m'n'}\big( 	\big| \xrightarrow{\; [\mE^{\rm stab.}_{\mC'}]_{i',j'} \;}\big\ra \big) }
	\\[0.7em]
	& \q = \!\!\!\!\!\!\summand \q \q
	\delta_{\mathcal C, \mathcal C'}\delta_{i,i'}\delta_{j,j'}
	\delta(\gy^{\exOne \, ; \, \fzOne} \! ,c_{{\rm y},j}) \, 
	\delta(\gz^{\exOne \, ; \, \fyOne} \! ,c_{{\rm z},j}) \, \delta(\hx^{\exOne,\gy,\gz \, ; \, \fyOne,\fzOne} \! ,d_{\hat{\rm x},j})
	\\[-3em]
	& \hspace{6.8em}  \q \times 
	D^R_{mn}\big( 	\big| \xrightarrow{\; [\mE_\mC^{\rm stab.}]_{i,j} \;}\big\ra \big) 
	\overline{D^R_{m'n'}\big( 	\big| \xrightarrow{\; [\mE^{\rm stab.}_{\mC}]_{i,j} \;}\big\ra \big) }
	\\[1em]
	& \q = \frac{|\mathfrak{B}_{\mC}|\delta_{\mathcal C, \mathcal C'}\delta_{R,R'}}{|\mathcal C|{\rm dim}(V_R)}\, \delta_{i,i'}\delta_{j,j'}\delta_{m,m'}\delta_{n,n'} \; ,
\end{align*}
where we used in the third step the orthogonality of the irreducible representations of $Z_{\mathcal C}$  as well as the orbit-stabiliser theorem which states that $|\mathfrak{B}_{\mC}| = |\mathcal C| \cdot |Z_{\mathcal C}|$.

\subsection{Proof of the completeness relation \eqref{eq:complete}\label{sec:app_proofComplete}}
In this appendix, we prove the completeness relation \eqref{eq:complete}. Let us first write down explicitly the l.h.s using definition \eqref{eq:irreps}:
\begin{align*}
	&	\frac{1}{| \mathfrak{B}_{\gy,\gz}|}\sum_{\mC,R}\sum_{\substack{i,j \\ m ,n}}
	|\mC|{\rm dim}(V_R)
	D^{\mC,R}_{im,jn}\Big( 	\Big| \, 	\tubeSavg{\hx}{\gy}{\gz}{\mE_{\gy,\gz}}{0.75}{1.25}
	\Big\ra \Big)
	\overline{D^{\mC,R}_{im,jn}\Big( \Big| \, \tubeSavg{\thx}{\tgy}{\tgz}{\tilde{\mE}_{\tgy,\tgz}}{0.75}{1.25}
		\Big\ra \Big)}
	\\	
	& \q = 
	\sum_{\mathcal C,R}\sum_{\substack{i,j \\ m,n}}
	\delta(\gy , c_{{\rm y},i}) \, 
	\delta(\gz , c_{{\rm z},i}) \, 
	\delta(\hx , d_{\hat{\rm x},i}) \, 
	\delta(\gy^{\exOne \, ; \, \fzOne} , c_{{\rm y},j}) \, 
	\delta(\gz^{\exOne \, ; \, \fyOne} , c_{{\rm z},j}) \, \delta(\hx^{\exOne,\gy,\gz \, ; \, \fyOne,\fzOne} , d_{\hat{\rm x},j})
	\\[-0.7em]
	& \hspace{4.3em} 
	\times 	\delta(\tgy , c_{{\rm y},i}) \, 
	\delta(\tgz , c_{{\rm z},i}) \, 
	\delta(\thx , d_{\hat{\rm x},i}) \, 
	\delta(\tgy^{\texOne \, ; \, \tfzOne} , c_{{\rm y},j}) \, 
	\delta(\tgz^{\texOne \, ; \, \tfyOne} , c_{{\rm z},j}) \, \delta(\thx^{\texOne,\tgy,\tgz \, ; \, \tfyOne,\tfzOne} , d_{\hat{\rm x},j})
	\\
	& \hspace{4.3em} \times 
	\frac{|\mathcal C|{\rm dim}(V_R)}{|\mathfrak{B}_{\gy,\gz}|}
	D^R_{mn}\big( 	\big| \xrightarrow{\; [\mE_\mC^{\rm stab.}]_{i,j} \;}\big\ra \big) 
	\overline{	D^R_{mn}\big( 	\big| \xrightarrow{\; [\tilde{\mE}_\mC^{\rm stab.}]_{i,j} \;}\big\ra \big) } \; .
\end{align*}
But, given an equivalence class $\mE_\mC \in Z_C$ whose representative element is $(\exOne,\fyOne , \fzOne)$, the representation matrices of $Z_{\mathcal C}$ satisfy the identity
\begin{equation*}
	\overline{	D^R_{mn}\big( 	\big| \xrightarrow{\; \mE_\mC \;}\big\ra \big) }
	= 	D^R_{nm}\big( 	\big| \xrightarrow{\; \mE_\mC^{-1} \;}\big\ra \big) \; ,
\end{equation*}
such that $\mE_\mC^{-1}$ is the equivalence class with representative element $(\exOne^{-1}, \exOne^{-1} \act \fyOne^{-1}, \exOne^{-1} \act \fzOne^{-1})$. Inserting this identity in the computation above, we can then use the linearity of the representations of $Z_\mathcal{C}$ together with \eqref{eq:algZC} and the fundamental property
\begin{equation*}
	\frac{1}{|Z_{\mathcal C}|}\sum_{\{R\}} {\rm dim}(V_R)\chi^R \big( 	\big| \xrightarrow{\; \mE_\mC \;}\big\ra \big)  =
		\delta\big( \mE_\mC, \mE_\mC^{\rm triv.}\big)
\end{equation*}
so as to obtain
\begin{align*}
	\frac{1}{| \mathfrak{B}_{\gy,\gz}|}\sum_{\mC,R}\sum_{\substack{i,j \\ m ,n}}
	|\mC|{\rm dim}(V_R)
	D^{\mC,R}_{im,jn}\Big( 	\Big| \, 	\tubeSavg{\hx}{\gy}{\gz}{\mE_{\gy,\gz}}{0.75}{1.25}
	\Big\ra \Big)
	\overline{D^{\mC,R}_{im,jn}\Big( \Big| \, \tubeSavg{\thx}{\tgy}{\tgz}{\tilde{\mE}_{\tgy,\tgz}}{0.75}{1.25}
		\Big\ra \Big)} = 	
	\Big\la \; 	
	\tubeSavg{\hx}{\gy}{\gz}{\mE_{\gy,\gz}}{0.75}{1.25}\; \Big| \, \tubeSavg{\thx}{\tgy}{\tgz}{\tilde{\mE}_{\tgy,\tgz}}{0.75}{1.25}
	\Big\ra\; .
\end{align*}

\section{Properties of the canonical basis}
\emph{In this appendix, we collect the proofs of several important properties satisfied by the canonical basis defined in sec.~\ref{sec:torigs}}

\subsection{Proof of the orthonormality relation \eqref{eq:canonicalortho}}\label{sec:app_canonicalortho}

In this appendix, we prove the orthonormality of the canonical basis stated in equation \eqref{eq:canonicalortho}. Utilising the definition of the canonical basis in equation \eqref{eq:canonicaltrans} and the inner product for basis elements of ${\rm Tube}^{\mG}[\mathbb{T}^{2}_{\boxempty}]$, it follows that
\begin{align*}
	\la \mC',R';M'N'\, | \, \mC,R \, ;MN\ra &= 
	\bigg(\frac{d_{\mC,R}d_{\mC',R'}}{|\mathfrak{B}_{\mC}||\mathfrak{B}_{\mC'}|}\bigg)^\frac{1}{2} \!\!\!\!\!\!
	\summand \!\!\!\!
	{D^{\mC',R'}_{M'N'}\Big( \Big|\; \tubeSavg{\hx}{\gy}{\gz}{\mE_{\gy,\gz}}{0.75}{1.25} \Big\ra \Big)}
	\overline{D^{\mC,R}_{MN}\Big( \Big| \, \tubeSavg{\hx}{\gy}{\gz}{\mE_{\gy,\gz}}{0.75}{1.25} \Big\ra \Big)}
	\\
	&=\bigg(\frac{d_{\mC,R}d_{\mC',R'}}{|\mathfrak{B}_{\mC}||\mathfrak{B}_{\mC'}|}\bigg)^\frac{1}{2}
	\frac{|\mathfrak{B}_{\mC}|}{d_{\mC,R}}
	\delta_{\mC,\mC'} \, \delta_{R,R'} \, \delta_{M,M'} \,\delta_{N,N'}
	=
	\delta_{\mC,\mC'} \, \delta_{R,R'} \, \delta_{M,M'} \,\delta_{N,N'} \; ,
\end{align*}
where we made use of eq.~\eqref{eq:ortho}.

\subsection{Proof of the completeness relation \eqref{eq:canonicalcomp}}\label{sec:app_canonicalcomplete}
In this appendix, we prove that the canonical basis defined in equation \eqref{eq:canonicaltrans} provides a complete basis for ${\rm Tube}^{\mG}[\mathbb{T}^{2}_{\boxempty}]$.
To confirm this statement, it is enough to verify that the dimension of the canonical basis is equal to the dimension of ${\rm Tube}^{\mG}[\mathbb{T}^{2}_{\boxempty}]$, i.e.
\begin{align*}
	\sum_{\mC,R}\sum_{M,N}\la \mC,R;MN\, | \, \mC,R \, ;MN\ra
	&=
	\sum_{\mC,R }\sum_{M,N}	\frac{d_{\mC,R}}{|\mathfrak{B}_{\mC}|}
	\!\!\!\!	\summand \!\!\!\!
	{D^{\mC,R}_{MN}\Big( \Big| \, \g \Big\ra \Big)}\overline{D^{\mC,R}_{MN}\Big( \Big| \, \g \Big\ra \Big)}
	\nonumber\\
	&=\summand 1=\big| {\rm Tube}^{\mG}[\mathbb{T}^{2}_{\boxempty}] \big| \; ,
\end{align*}
where we made use of eq.~\eqref{eq:complete}.

\subsection{Proof of the canonical algebra product \eqref{eq:canonicalproduct}}\label{sec:app_canonicalalgebraprod}
In this appendix, we prove relation \eqref{eq:canonicalproduct} that provides the tube algebra product for the canonical basis elements defined in \eqref{eq:canonicaltrans}. By definition of the canonical basis elements we have:

\begin{align*}
	&\ket{\mC,R \, ;MN}\star\ket{\mC',R';M'N'}
	\\
	& \q =	\bigg(\frac{d_{\mC,R}d_{\mC',R'}}{|\mathfrak{B}_{\mC}||\mathfrak{B}_{\mC'}|}\bigg)^\frac{1}{2} \!\!\!\!
	\summand \,\,\,
	\summandh \;
	\overline{D^{\mC,R}_{MN}\Big( \Big| \, \tubeSavg{\hx}{\gy}{\gz}{\mE_{\gy,\gz}}{0.75}{1.25} \Big\ra \Big)}\;
	\overline{D^{\mC',R'}_{M'N'}\Big( \Big| \, \h \Big\ra \Big)}
	\\[-1.5em]
	&\hspace{19.7em}\times \Big| \, 
	\tubeSavg{\hx}{\gy}{\gz}{\mE_{\gy,\gz}}{0.75}{1.25}
	\Big\ra \!
	\star 
	\Big| \,
	\tubeSavg{\hx'}{\gy'}{\gz'}{\mE'_{\gy',\gz'}}{0.75}{1.25}
	\Big\ra \; .
\end{align*}
The tube algebra product was defined to be
\begin{align*}
	\Big| \,
	\tubeSavg{\hx}{\gy}{\gz}{\mE_{\gy,\gz}}{0.75}{1.25}
	\Big\ra \!
	\star 
	\Big| \,
	\tubeSavg{\hx'}{\gy'}{\gz'}{\mE'_{\gy',\gz'}}{0.75}{1.25}
	\Big\ra
	=  
	\frac{\delta \big(\hx', \hx^{\exOne,\gy,\gz \,;\, \fyOne,\fzOne} \big) 
		\delta \big(\gy',\gy^{\exOne \,;\, \fzOne} \big) 
		\delta \big(\gz' , \gz^{\exOne \,;\, \fyOne} \big)}{|\mathfrak{B}_{\gy,\gz}|^\frac{1}{2}}
	\Big| \, \tubeSavg{\hx}{\gy}{\gz}{(\mE \cdot \mE')_{\gy,\gz}}{1.05}{1.85} \Big\ra 
\end{align*}
where the tube algebra element on the r.h.s can be decomposed into canonical basis states as
\begin{align*}
	\Big| \, \tubeSavg{\hx}{\gy}{\gz}{(\mE \cdot \mE')_{\gy,\gz}}{1.05}{1.85} \Big\ra 
	=\frac{1}{|\mathfrak{B}_{\gy,\gz}|^\frac{1}{2}}\sum_{\mC'',R''}d_{\mC'',R''}^\frac{1}{2}
	\sum_{M'',N''}
	D^{\mC'',R''}_{M''N''}\Big(\Big| \, \tubeSavg{\hx}{\gy}{\gz}{(\mE \cdot \mE')_{\gy,\gz}}{1.05}{1.85} \Big\ra  \Big) \; |\mC'',R'' \, ;M''N''\ra \; .
\end{align*}
But in virtue of \eqref{eq:Lin}, we have 
\begin{align*}
	&
	\delta \big(\hx', \hx^{\exOne,\gy,\gz \,;\, \fyOne,\fzOne} \big) 
	\delta \big(\gy',\gy^{\exOne \,;\, \fzOne} \big) 
	\big(\gz' , \gz^{\exOne \,;\, \fyOne} \big)
	D^{\mC'',R''}_{M''N''}\Big(\Big| \, \tubeSavg{\hx}{\gy}{\gz}{(\mE \cdot \mE')_{\gy,\gz}}{1.05}{1.85} \Big\ra  \Big) \\
	& \q =
	\sum_O
	D^{\mC'',R''}_{M''O}\Big( 	\Big| \, \tubeSavg{\hx}{\gy}{\gz}{\mE_{\gy,\gz}}{0.75}{1.25}
	\Big\ra \Big)
	D^{\mC'',R''}_{ON''} \Big(
	\Big| \, \tubeSavg{\hx'}{\gy'}{\gz'}{\mE'_{\gy,\gz}}{0.75}{1.25}
	\Big\ra\Big) \; .
\end{align*}
The orthogonality relations
\begin{align*}
	\!\!\!\!	\summand \!\!\!\!
	D^{\mC'',R''}_{M''O}\Big( 	\Big| \, 	\tubeSavg{\hx}{\gy}{\gz}{\mE_{\gy,\gz}}{0.75}{1.25}
	\Big\ra \Big)
	\overline{D^{\mC,R}_{MN}\Big( 	\Big| \, 	\tubeSavg{\hx}{\gy}{\gz}{\mE_{\gy,\gz}}{0.75}{1.25}
	\Big\ra \Big)} 
	&= \frac{|\mathfrak{B}_\mC |\delta_{\mC'', \mC}\delta_{R'',R}}{d_{\mC,R}}\delta_{M'',M} \delta_{O,N} 
	\\
	\!\!\!\!	\summandh \!\!\!\!
	D^{\mC'',R''}_{ON''}\Big( 	\Big| \, 	\tubeSavg{\hx'}{\gy'}{\gz'}{\mE'_{\gy',\gz'}}{0.75}{1.25}
	\Big\ra \Big)
	\overline{D^{\mC',R'}_{M'N'}\Big( 	\Big| \, 	\tubeSavg{\hx'}{\gy'}{\gz'}{\mE'_{\gy',\gz'}}{0.75}{1.25}
		\Big\ra \Big)} 
	&= \frac{|\mathfrak{B}_\mC |\delta_{\mC'', \mC'}\delta_{R'',R'}}{d_{\mC',R'}}\delta_{O,M'} \delta_{N'',N'} 
\end{align*}
finally yield the desired result, namely
\begin{equation*}
	\ket{\mC,R \, ;MN}\star\ket{\mC',R';M'N'} = 	\frac{\delta_{N,M'} \, \delta_{\mC,\mC'} \, \delta_{R,R'}}{d_{\mC,R}^\frac{1}{2}}
	\, \ket{\mC,R \, ;MN'} \; .
\end{equation*}

\subsection{Ground state projector}\label{sec:app_gsprojector}

In this appendix, we demonstrate the equality between the two expressions for the three-torus ground state projector $\mathbb{P}^{\mG}[\mathbb{T}^{3}_{\boxempty}]$ given in equations \eqref{eq:tori-gsproj} and \eqref{eq:tori-gsprojalt}. Let us consider
\begin{align}
	\label{eq:interTube0}
	\summand \;\; \summandh \!\!\!\!
	\bigg(
	\Big| \, 
	\tubeSavg{\hx'}{\gy'}{\gz'}{{\mE}'_{\gy',\gz'}}{0.75}{1.25}
	\Big\ra^{\!\! -1} \!\!\!\!\!
	\star
	\Big| \,
	\tubeSavg{\hx}{\gy}{\gz}{\mE_{\gy,\gz}}{0.75}{1.25}
	\Big\ra \!
	\star
	\Big| \,  
	\tubeSavg{\hx'}{\gy'}{\gz'}{\mE'_{\gy',\gz'}}{0.75}{1.25}
	\Big\ra
	\bigg)
	\Big\la \,
	\tubeSavg{\hx}{\gy}{\gz}{\mE_{\gy,\gz}}{0.75}{1.25} \;
	\Big| \, .
\end{align}
By definition of the tube algebra, we have
\begin{align*}
	\Big| \,
	\tubeSavg{\hx}{\gy}{\gz}{\mE_{\gy,\gz}}{0.75}{1.25}
	\Big\ra \!
	\star 
	\Big| \,
	\tubeSavg{\hx'}{\gy'}{\gz'}{\mE'_{\gy',\gz'}}{0.75}{1.25}
	\Big\ra
	=  
	\frac{\delta \big(\hx', \hx^{\exOne,\gy,\gz \,;\, \fyOne,\fzOne} \big) 
		\delta \big(\gy',\gy^{\exOne \,;\, \fzOne} \big) 
		\delta \big(\gz' , \gz^{\exOne \,;\, \fyOne} \big)}{|\mathfrak{B}_{\gy,\gz}|^\frac{1}{2}}
	\Big| \, \tubeSavg{\hx}{\gy}{\gz}{(\mE \cdot \mE')_{\gy,\gz}}{1.05}{1.85} \Big\ra 
\end{align*}
and it follows from def.~\eqref{eq:inverse} of the inverse that
\begin{align}
	\label{eq:interTube2}
	\Big| \, 
	\tubeSavg{\hx'}{\gy'}{\gz'}{{\mE}'_{\gy',\gz'}}{0.75}{1.25}
	\Big\ra^{\!\! -1} \!\!\!\!\!
	\star
	\Big| \, \tubeSavg{\hx}{\gy}{\gz}{(\mE \cdot \mE')_{\gy,\gz}}{1.05}{1.85} \Big\ra 
	=
	\frac{\delta \big(\hx', \hx \big) 
	\delta \big(\gy',\gy\big) 
	\delta \big(\gz' , \gz \big)}{|\mathfrak{B}_{\gy,\gz}|^\frac{1}{2}}
	\Big| \, \tubeSavg{\bhx'}{\bgy'}{\bgz'}{(\mE'^{-1} \cdot\mE \cdot \mE')_{\bgy,\bgz}}{1.35}{2.45} \Big\ra 
\end{align}
with $\bgy'={\gy'}^{\exOne'\,;\, \fzOne'}, \bgz'={\gz'}^{\exOne' \,;\, \fyOne'}$, $\bhx'={\hx'}^{\exOne',\gy',\gz' \,;\, \fyOne',\fzOne'}$ and such that $({\mE'}^{-1}\cdot \mE \cdot \mE')_{\bgy',\bgz'}\in\mathfrak{B}_{\bgy',\bgz'}$ is the equivalence class with representative element
\begin{equation*}
	\big(\exOne'^{-1} \exOne \exOne' \, , \, 
	\exOne'^{-1} \act [\fyOne'^{-1} \fyOne(\exOne \act \fyOne')] \, , \, 
	\exOne'^{-1} \act [\fzOne'^{-1} \fzOne(\exOne \act \fzOne')] \big) \; .
\end{equation*}
In order to obtain the delta functions in \eqref{eq:interTube2}, we used the fact that
\begin{gather*}
	\bgy' \phantom{|}^{\exOne'^{-1} \, ; \, \exOne'^{-1}\act \fzOne'^{-1}} = \gy' \; , \q
	\bgz' \phantom{|}^{\exOne'^{-1} \, ; \, \exOne'^{-1}\act \fyOne'^{-1}} = \gz' \\ \text{and} \;\;
	\bhx' \phantom{|}^{\exOne'^{-1},{\gy'}^{\exOne' \, ; \, \fzOne'},{\gz'}^{\exOne' \, ; \, \fyOne'} \, ; \, \exOne'^{-1}\act \fyOne'^{-1}, \exOne'^{-1} \act \fzOne'^{-1}} = \hx' \; . 
\end{gather*}
Expanding the resulting tube algebra element in \eqref{eq:interTube2} according to \eqref{eq:avgTube}, we can rewrite this equation in terms of the gauge operators as
\begin{align*}
	{\rm l.h.s}\eqref{eq:interTube2} 
	&= 
	\frac{\delta \big(\hx', \hx \big) 
		\delta \big(\gy',\gy\big) 
		\delta \big(\gz' , \gz \big)}{|\mathfrak{B}_{\gy,\gz}|^\frac{1}{2}
		|H||Z_{\mE_{\gy,\gz}}|}
	\!\! \sum_{\lambda,\lambda' \in H} \!\!
	\mathbb{A}_{\rm x}^{\exOne'^{-1}\act [\lambda'^{-1}\lambda(\exOne \act \lambda')]}
	\mathbb{A}^{\exOne'^{-1}}_{(0)}
	\mathbb{A}^{\fyOne'^{-1}}_{\rm z}
	\mathbb{A}^{\fzOne'^{-1}}_{\rm y}
	\Big| \,
	\tubeSext{\hx}{\gy}{\gz}{\exOne}{\fzOne}{\fyOne}{0.75}{1.25}
	\Big\ra
	\\
	&=
	\frac{\delta \big(\hx', \hx \big) 
	\delta \big(\gy',\gy\big) 
	\delta \big(\gz' , \gz \big)}{|\mathfrak{B}_{\gy,\gz}|^\frac{1}{2}
	|H||Z_{\mE_{\gy,\gz}}|}
	\!\! \sum_{\lambda,\lambda' \in H} \!\!
	\mathbb{A}^{\exOne'^{-1}\partial(\lambda'^{-1})}_{(0)}
	\mathbb{A}^{\lambda'\fyOne'^{-1}(\gz \act \lambda'^{-1})}_{\rm z}
	\mathbb{A}^{\lambda'\fzOne'^{-1}(\gy \act \lambda'^{-1})}_{\rm y}
	\mathbb{A}^{\lambda}_{\rm x}
	\Big| \,
	\tubeSext{\hx}{\gy}{\gz}{\exOne}{\fzOne}{\fyOne}{0.75}{1.25}
	\Big\ra .
\end{align*}
Moreover, we can introduce a new summation variable via
\begin{equation*}
	\sum_{\lambda \in H}\mathbb{A}^\lambda_{\rm x} = \frac{1}{|H|}\sum_{\lambda, \lambda_{\rm x} \in H}\mathbb A^{\lambda_{\rm x}}_{\rm x}\mathbb A^\lambda_{\rm x} \; .
\end{equation*}
Putting everything together so far, and using the fact that $|\mathfrak{B}_{\gy,\gz}| = |G| |H| |Z_{\mE_{\gy,\gz}}|$, we have obtained that
\begin{align*}
	\eqref{eq:interTube0} &= 
	\frac{1}{|G||H|^3} \!\!\!\!\!\!
	\sum_{\substack{\gy,\gz\in G\\[-0.2em] \hx\in H\,|\,\gz=\gz^{\gy;\hx} \\[0.05em] \mE_{\gy,\gz}, \mE'_{\gy,\gz,} \in \mathfrak{B}_{\gy,\gz}}} \!\!\!\!\!\!
	\frac{1}{|Z_{\mE_{\gy,\gz}}|^{2}}
	\delta(\hx,\hx^{\exOne,\gy,\gz \,;\, \fyOne,\fzOne})
	\delta(\gy,\gy^{\exOne\,;\, \fzOne })
	\delta(\gz,\gz^{\exOne\,;\, \fyOne })
	\\[-0.8em] 
	&\times  
	\sum_{\lambda,\lambda', \lambda_{\rm x} \in H} \!
	\mathbb{A}^{\exOne'^{-1}\partial(\lambda'^{-1})}_{(0)}
	\mathbb{A}^{\lambda'\fyOne'^{-1}(\gz \act \lambda'^{-1})}_{\rm z}
	\mathbb{A}^{\lambda'\fzOne'^{-1}(\gy \act \lambda'^{-1})}_{\rm y}
	\mathbb{A}^{\lambda_{\rm x}}_{\rm x}
	\Big| \, \tubeSext{\hx}{\gy}{\gz}{\partial(\lambda)\exOne}{(\gy \act \lambda)\fzOne \lambda^{-1}}{(\gz \act \lambda)\fyOne \lambda^{-1}}{1.2}{2.1} \Big\ra
	\Big\la \,
	\tubeSext{\hx}{\gy}{\gz}{\partial(\lambda)\exOne}
	{(\gy\act \lambda)\fzOne\lambda^{-1}}
	{(\gz\act \lambda)\fyOne\lambda^{-1}}
	{1.2}{2.1}
	\Big| \; .
\end{align*}
Performing the changes of variables
\begin{equation*}
	\sum_{\mE_{\gy,\gz}\in\mathfrak{B}_{\gy,\gz}}
	\frac{1}{|Z_{\mE_{\gy,\gz}}|}
	\sum_{\lambda \in H}=\sum_{\gx\in G}\sum_{\hy,\hz\in H}  \q \text{and} \q
	\sum_{\mE'_{\gy,\gz}\in\mathfrak{B}_{\gy,\gz}}
	\frac{1}{|Z_{\mE'_{\gy,\gz}}|}
	\sum_{\lambda' \in H}=\sum_{k\in G}\sum_{\lambda_{\rm y},\lambda_{\rm z}\in H} \; ,
\end{equation*}
where the factors $|Z_{\mE_{\gy,\gz}}|$ and $|Z_{\mE'_{\gy,\gz}}|$ account for the possible overcounting of $\mG$-colourings, we finally obtain that
\begin{align*}
	\eqref{eq:interTube0}
	=
	\frac{1}{|G||H|^{3}} \! \!
	\sum_{
		\substack{
			\gx,\gy,\gz\in G\\
			\hx,\hy,\hz\in H
		}
	}
	\sum_{
		\substack{
			k \in G\\
			\lambda_{\rm x},\lambda_{\rm y},\lambda_{\rm z} \in H
		}
	} \!\!
	&\mathbb{A}^{k}_{(0)}
	\mathbb{A}^{\lambda_{\rm z}}_{\rm z}
	\mathbb{A}^{\lambda_{\rm y}}_{\rm y}
	\mathbb{A}^{\lambda_{\rm x}}_{\rm x}
	\Big| \,
	\tubeS{\hx}{\gy}{\gz}{\gx}{\hz}{\hy}
	\Big\ra
	\Big\la \;
	\tubeS{\hx}{\gy}{\gz}{\gx}{\hz}{\hy} \;
	\Big|
	\\[-0.6em] \nn
	\times \; &\delta(\hx,\hx^{\gx,\gy,\gz \,;\, \hy,\hz}) \,
	\delta(\gy,\gy^{\gx\,;\, \hz }) \,
	\delta(\gz,\gz^{\gx\,;\, \hy }) \,
	\delta(\gz,\gz^{\gy\,;\,\hx}) \; ,
\end{align*}
which is the definition of $\mathbb{P}^{\mG}[\mathbb{T}^{3}_{\boxempty}]$ given in \eqref{eq:tori-gsproj}, as expected.

\bibliographystyle{JHEP}
\bibliography{ref_cat}

\end{document}